\documentclass[%
 aip,
 amsmath,amssymb,
 reprint,%
]{revtex4-1}

\usepackage{graphicx}
\usepackage{dcolumn}
\usepackage{bm}

\usepackage[utf8]{inputenc}
\usepackage[T1]{fontenc}
\usepackage{mathptmx}
\usepackage{url}  

\begin{document}

\preprint{AIP/123-QED}

\title[Im$\{\chi^{(3)}\}$ spectra of 110-cut GaAs, GaP, and Si near the two-photon absorption band edge]{$\mathrm{Im}\{\chi^{(3)}\}$ spectra of 110-cut GaAs, GaP, and Si near the two-photon absorption band edge}

\author{Brandon J. Furey}
\email{furey@utexas.edu.}
\affiliation{Physics Department, University of Texas at Austin, 2515 Speedway, C1600, Austin, TX, USA 78712}
\author{Rodrigo M. Barba-Barba}
\affiliation{Centro de Investigaciones en \'{O}ptica, A.C., Loma del Bosque 115, Colonia Lomas del Campestre, Le\'{o}n, Gto., M\'{e}xico 37150}
\author{Ramon Carriles}
\email{ramon@cio.mx.}
\affiliation{Centro de Investigaciones en \'{O}ptica, A.C., Loma del Bosque 115, Colonia Lomas del Campestre, Le\'{o}n, Gto., M\'{e}xico 37150}
\author{Alan Bernal}
\affiliation{Centro de Investigaciones en \'{O}ptica, A.C., Loma del Bosque 115, Colonia Lomas del Campestre, Le\'{o}n, Gto., M\'{e}xico 37150}
\author{Bernardo S. Mendoza}
\affiliation{Centro de Investigaciones en \'{O}ptica, A.C., Loma del Bosque 115, Colonia Lomas del Campestre, Le\'{o}n, Gto., M\'{e}xico 37150}
\author{Michael C. Downer}
\email{downer@physics.utexas.edu.}
\affiliation{Physics Department, University of Texas at Austin, 2515 Speedway, C1600, Austin, TX, USA 78712}

\date{\today}

\begin{abstract}
Spectra of the degenerate two-photon absorption coefficient $\beta(\omega)$, anisotropy parameter $\sigma(\omega)$, and dichroism parameter $\delta(\omega) = \big[\sigma(\omega) + 2\eta(\omega)\big]/2$ of crystalline 110-cut GaAs, GaP, and Si at 300 K were measured using femtosecond pump-probe modulation spectroscopy over an excitation range in the vicinity of each material's half-band gap $E_g/2$ (overall $0.62 < \hbar \omega < 1.91$ eV, or $2000 > \lambda > 650$ nm). Together these three parameters completely characterize the three independent components of the imaginary part of the degenerate third-order nonlinear optical susceptibility tensor $\mathrm{Im}\{\chi^{(3)}_{abcd}(\omega)\}$. In direct-gap GaAs, these components peak at $\hbar \omega \approx 0.78 E_g$ which is close to the peak at $\hbar \omega = 0.71 E_g$ predicted by the Jones-Reiss phenomenological model. The dispersion is comparable to ab initio calculations. In indirect-gap GaP and Si, these components tend to increase with $\hbar \omega$ over our tuning range. In Si, the dispersion differs significantly from predictions of semi-empirical models and ab initio calculations do not account for transitions below the two-photon direct band gap, motivating further investigation. Kleinman symmetry was observed to be broken in all three materials. We also note anomalies observed and their possible origins, emphasizing the advantages of a 2-beam experiment in identifying the contribution of various nonlinear effects.
\end{abstract}

\maketitle
%

\section{\label{sec:intro}Introduction}
Column III-V and IV semiconductors are not only building blocks of modern microelectronics, but have become important \textit{photonic} materials for telecommunications and integrated light-wave systems.\cite{asghari} Direct-gap gallium arsenide (GaAs) is used in photovoltaic\cite{yoon,bett} and optoelectronic\cite{yoon,wada} devices. Among indirect-gap semiconductors, gallium phosphide (GaP) is a key component of green LEDs\cite{pilkuhn} and nanophotonic devices,\cite{wilson} and an emerging platform for integrated nonlinear photonics,\cite{wilson} while silicon (Si) has important applications in all-optical switches,\cite{euser} photonic crystals,\cite{vlasov} lasers,\cite{rong} and optical waveguides.\cite{tsang} In photonic systems utilizing ultrashort light pulses, long propagation lengths, or high light intensities ($\geq$ MW/cm$^2$), \textit{nonlinear} optical properties of these semiconductors come into play. In particular, two-photon absorption (2PA), which is proportional to the imaginary part of the third-order nonlinear optical susceptibility tensor $\chi^{(3)}(\omega)$, limits optical switching energy\cite{dvorak,blanco} and affects long-haul pulse transmission at telecommunication wavelengths $\lambda = 1300$ and $1500$ nm.\cite{bristow,tsang} Nonlinear refraction, which is proportional to $\mathrm{Re}\{\chi^{(3)}(\omega)\}$, but depends on the dispersion of the degenerate $\mathrm{Im}\{\chi^{(3)}(\omega)\}$ as well as the nondegenerate $\mathrm{Im}\{\chi^{(3)}(\omega;\omega, \omega',-\omega')\}$ via nonlinear Kramers-Kronig relations,\cite{bassani,caspers,hutchings2,kogan,nussenzweig,price,ridener,sheikbahae,sheikbahae2,toll} underlies all-optical switching\cite{gibbs,stegeman} and long-haul light pulse propagation.\cite{reintjes} Accurate measurement and calculation of $\mathrm{Im}\{ \chi^{(3)}(\omega)\}$ is essential to understanding the operation and limits of diverse photonic systems.

Early characterization of $\mathrm{Im}\{\chi^{(3)}(\omega)\}$ in GaAs, GaP, and Si consisted of single-wavelength measurements of degenerate 2PA amplitude using pulsed lasers with picosecond durations.\cite{reintjes,bechtel,bosacchi} Reported values of the extracted 2PA coefficient $\beta$ at a given wavelength often varied over an order of magnitude or more.\cite{vanstryland,sheikbahae3} With the advent of tunable optical parametric amplifiers (OPAs) providing femtosecond pulses at $\sim 1$ kHz repetition rate, 2PA \textit{spectra} of Si were measured with improved accuracy using \textit{single-beam} excitation over a range $0.56 < \hbar \omega < 1.46$ eV at peak intensities $\sim 10$ GW cm$^{-2}$.\cite{bristow} This intensity induced nonlinear aborption as large as $\sim 10\%$, but can also introduce competing nonlinearities [e.g. three-photon absorption (3PA) and free-carrier absorption (FCA)\cite{vanstryland}] that can be challenging to distinguish from 2PA.\cite{sheikbahae3} Above-gap 2PA spectra of Si were also measured over a range of $2.00 < \hbar \omega < 2.31$ eV.\cite{reitze} 2PA spectra of GaAs were also measured using single-beam excitation over a limited range of $0.73 < \hbar \omega < 0.95$ eV,\cite{hurlbut} and a degenerate pump-probe experiment over a limited range of $0.74 < \hbar \omega < 0.87$ eV.\cite{fishman} To our knowledge two-beam 2PA anisotropy in these three semiconductors, i.e. the dependence of 2PA on the angle between incident light polarization and crystollographic axes, has only been completely characterized at a single wavelength in GaAs.\cite{dvorak}

In this paper, we report \textit{spectroscopic} measurements of the amplitude \textit{and} anisotropy of degenerate 2PA in GaAs, GaP, and Si using \textit{two}-beam pump-probe modulation spectroscopy.\cite{shank,dvorak} In this technique the 1 kHz pulse train from a tunable fs OPA is split into a pump beam that is chopped at a sub-harmonic of its repetition rate and focused loosely into the samples and a weak sychronized probe beam of the same wavelength that intersects the central portion of the pump beam path at a small angle inside the sample. Although alignment is more complicated, 2-beam 2PA offers two advantages over single-beam measurements. First, by detecting pump-induced probe transmission changes with lock-in techniques, the setup becomes sensitive to sub-1\% absorption, excited by pump intensity $< 1$ GW cm$^{-2}$. The magnitude of 2PA in co-polarized beams is also doubled relative to the single-beam case.\cite{dvorak, sheikbahae3} Competing higher-order nonlinearities are therby more effectively avoided than in single-beam measurements at higher intensity. Second, by scanning the time delay $\Delta t$ between pump and probe pulses, the shape of the resulting cross-correlation signal reveals signatures of competing nonlinear processes such as FCA, 3PA,\cite{benis} and 2-beam pump-probe coupling, when present. Absence of such signatures, i.e. an instantaneous or nearly-instantaneous cross-correlation response, provides strong \textit{in-situ} evidence that data is being acquired in a pure 2PA regime. From combined measurements of 2PA amplitude and anistropy spectra, we obtain not only degenerate $\beta(\omega)$, as in past work, but anisotropy parameter $\sigma(\omega)$ and $\eta(\omega)$ as well, enabling us to extract complete $\mathrm{Im}\{\chi^{(3)}(\omega)\}$ spectra of GaAs, GaP, and Si near their respective 2PA band edges for the first time. $\mathrm{Im}\{\chi^{(3)}(\omega)\}$ spectra are ultimately better suited than $\beta(\omega)$ alone for quantitative comparison with first-principles calculations.

The paper is organized as follows: Section \ref{sec:theory} adapts the standard theory of single-beam 2PA to the current 2-beam context, expands it to include 2PA anisotropy, and defines approximations used in our data analysis. Section \ref{sec:experiment} describes our 2-beam experimental setup, presents and discusses experimental results, and examines anomalies and limitations of this technique. Section \ref{sec:discussion} summarizes our main findings. 

\section{\label{sec:theory}Theory}

\subsection{\label{sec:theorypp2pa}Pump-probe two-photon absorption}
Up to 2-photon interactions, loss processes in a 2-beam open-aperture experiment can be characterized by the differential equations\cite{cirloganu}
\begin{align}
\label{eqdifeqprobe}
\frac{\mathrm{d}I_1}{\mathrm{d}z} &=  -\beta_{12}I_1 I_2 - \sigma_{fca}N_e I_1,\\
\label{eqdifeqpump}
\frac{\mathrm{d}I_2}{\mathrm{d}z} &= -\beta_{22} I_2^2,\\
\label{eqdifeqfca}
\frac{\mathrm{d}N_e}{\mathrm{d}t} &= \frac{\beta_{22} I_2^2}{2\hbar \omega},
\end{align}
where $(I_1,I_2)$ is the (probe, pump) intensity, $(\beta_{12}, \beta_{22})$ is the (pump-probe, pump-pump) 2PA coefficient, $\sigma_{fca}$ is the FCA cross section, and $N_e$ is the free-carrier density. For $I_2 \gg I_1$, probe-probe 2PA is neglected and both probe-probe and probe-pump excitations of free-carriers can also be neglected. We also only consider depletion of the pump due to pump-pump 2PA, since that term dominates over all other terms. For excitation below the band gap $\hbar \omega < E_g$, linear absorption is negligible. We also assume that the free-carrier recombination rate is $f_{rep} \ll \Gamma \ll 1/\tau_g$, much greater than the laser reptition rate and much less than the inverse of the pulse duration, and thus relaxation on the timescale of a pulse is neglected but the system is assumed to be fully relaxed before the arrival of the subsequent pulse. Exceptional cases in which other 2-beam coupling processes (e.g. photorefractive phase gratings, free carrier gratings, nonlinear refractive index gratings) influence the results are discussed in Section \ref{sec:experimentresultstimedelay}. Eqs. \ref{eqdifeqprobe} -- \ref{eqdifeqfca} are strictly valid for co-propagating pump and probe beams, but accurately approximate experimental geometries, such as those used here, with small intersection angles (here $\sim 5^{\circ}$), large pump/probe beam-size ratios (here $\sim 5)$, and interaction lengths much shorter than the pump beam size.

The incident pulse intensity profiles (at $z = 0$) can be expressed as
\begin{equation}
\label{eqintensity}
I_j(r,0,t) = I_j^0 e^{-\frac{2r^2}{w_j^2}}e^{-\frac{2(t + \Delta t_j)^2}{\tau_{g}^2}},
\end{equation}
where $j = 1,2$ labels the beams, the on-axis peak intensity of beam $j$ is
\begin{equation}
I_j^0 = \bigg(\frac{2}{\pi}\bigg)^{3/2} \frac{\varepsilon_j}{w_j^2\tau_g},
\end{equation}
$w_j$ is the beam radius, $\tau_g = \tau_{FWHM}/\sqrt{2\mathrm{ln}2}$ is the full-width at half-maximum (FWHM) Gaussian pulse duration, $\varepsilon_j = \varepsilon_j(0) (1 - R_j)$ is the incident pulse energy, and $R_j$ is the Fresnel power reflection coefficient. The relative time delay $\Delta t$ between the pump and probe pulses is defined as
\begin{equation}
\Delta t_j = \delta_{j2}\ \Delta t,
\end{equation}
where $\delta_{j2}$ is the Kronecker delta. If the sample thickness $L$ is shorter than the Rayleigh range $z_R$, then beam radii can be considered constant with respect to $z$. In the limit that FCA is weak and only considered to first-order, solving Eqs. \ref{eqdifeqprobe} - \ref{eqdifeqfca} for the intensity profile in Eq. \ref{eqintensity}, and integrating over $\mathrm{d}z$ from $z = 0$ to $L$, $2\pi r\ \mathrm{d}r$ from $r = 0$ to $\infty$, and $\mathrm{d}t$ from $t = -\infty$ to $\infty$ and expanding about $\beta_{12} I_2^0 L \approx 0$ gives an analytical expression for the change in probe transmission when the pump is on from when the pump is off, normalized to the probe transmission when the pump is off, $\Delta T_1/T_1 = \big[\varepsilon_1(L)/\varepsilon_1(0)\big] - 1$. The result is
\begin{equation}
\label{eqtransmission}
\frac{\Delta T_1^{\parallel,\perp}}{T_1} =\lim_{N \rightarrow \infty} \sum_{m=1}^{N} a^{\parallel,\perp}_m e^{\frac{-2m\Delta t^2}{(m+1)\tau_g^2}} - b^{\parallel,\perp} \bigg[ 1 + \mathrm{erf}\Big(\frac{2\Delta t}{\sqrt{3}\tau_g}\Big)\bigg],
\end{equation}
where 
\begin{align}
a^{\parallel}_m &= \frac{\Big(-\frac{1}{2} \beta_{12}^{\parallel} I_2^0 L\Big)^m \sqrt{m+1}}{1 + m w_r^2},\\
a^{\perp}_m &= \frac{ c^{\perp}_m(\Phi) \Big( \Phi \beta_{12}^{\perp} I_2^0 L \Big)^m}{(1 + m w_r^2) \sqrt{m+1}}
\end{align}
characterize the magnitude of pump-probe 2PA and 
\begin{align}
b^{\parallel} &= \bigg[\frac{\sqrt{\pi}\sigma_{fca}\tau_g}{16 \hbar \omega \big(1 + 2w_r^2\big)}\bigg]\beta_{12}^{\parallel}(I_2^0)^2 L,\\
b^{\perp} &= \bigg[\frac{\sqrt{\pi}\sigma_{fca}\tau_g \Phi}{8 \hbar \omega \big(1 + 2w_r^2\big)}\bigg]\beta_{12}^{\perp}(I_2^0)^2 L
\end{align}
characterize the magnitude of FCA. The symbols $(\parallel,\perp)$ refer to the cases of co-polarized and cross-polarized beams, resepectively. The generalized binomial coefficients in the cross-polarized case are
\begin{equation}
c^{\perp}_m(\Phi) = \frac{\Gamma(-1/\Phi+1)}{\Gamma(m+1) \Gamma(-1/\Phi - m +1)}
\end{equation}
which are valid so long as $1 - 1/\Phi - m \notin \mathbb{Z}^{0-}$ for $m \in \mathbb{Z}^{0+}$ and $\Phi \in \mathbb{R}$. The quantity $\Phi(\theta_0,\sigma,\eta)$ relates the pump-pump 2PA coefficient to the cross-polarized pump-probe 2PA coefficient by
\begin{equation}
\Phi(\theta_0,\sigma,\eta) = \frac{\beta_{22}^{\perp}(-\theta_0)}{\beta_{12}^{\perp}(-\theta_0)} = \frac{\beta_{12}^{\parallel}(\pi/2 - \theta_0)}{2 \beta_{12}^{\perp}(-\theta_0)}.
\end{equation}
The quantity $w_r^2 = w_1^2/w_2^2$ is the ratio of the probe to pump beam areas. Pump-probe 2PA produces a characteristic 2nd-order autocorrelation dip in probe transmission while FCA causes an approximately step-function decrease in probe transmission for positive delays. The undepleted pump regime is the case of $N=1$ and higher-order terms capture the effect of pump depletion by pump-pump 2PA. 

\subsection{\label{sec:theoryanisotropy}2PA anisotropy}
2PA is ``anisotropic'' when it depends on the relative orientations of incident fields $\vec{E}_i = E^a (\omega, \vec{k}_i)$ and crystallographic axes. For energetically degenerate, nearly collinear beams co-propagating along the laboratory $Z$-axis, the probe and pump fields in laboratory $XYZ$ coordinates are
\begin{align}
\vec{E}_1&=\big[E^X_1(\omega,\vec{k}_1) + c.c.\big] \hat{X}= A_1(Z) e^{i(k_1 Z - \omega t)} \hat{X} + c.c.,\\
\begin{split}
\vec{E}_2&=\big[E^X_2(\omega,\vec{k}_2) + c.c.\big] \hat{X} + \big[E^Y_2(\omega,\vec{k}_2) + c.c.\big] \hat{Y}\\
&= \big[A_2^X (Z) \hat{X} + A_2^Y(Z) \hat{Y}\big] e^{i(k_2 Z - \omega t + \gamma)} + c.c.,
\end{split}
\end{align}
where $A_i = |A_i|e^{i \phi_i}$ are the complex field amplitudes, $A_2^X(Z) = A_2(Z) \cos \zeta$, $A_2^Y(Z) = A_2(Z) \sin \zeta$, $\zeta$ is the angle between linearly-polarized probe and pump fields, $\phi_i$ are the phases of the complex field amplitudes, and $\gamma$ is an arbitrary phase shift between the pump and probe fields (assumed to not vary in time over the duration of the pulse such that the two pulses are mutually coherent). The field experienced by the nonlinear medium is just the sum field $\vec{E} = \vec{E}_1 + \vec{E}_2$, and the intensity of the fields as defined is\cite{boyd}
\begin{equation}
I_i(Z) = 2\epsilon_0 n c |A_i(Z)|^2,
\end{equation}
where $\epsilon_0$ is the vacuum permittivity, $n$ the refractive index, and $c$ the speed of light. 2PA is a third-order nonlinear optical process. If the beams are co-polarized, i.e. $\zeta = 0$, then the Fourier component $\vec{P}^{(3)}(\pm \omega, \pm \vec{k}_1)$ is
\begin{multline}
P_X^{(3)}(\pm \omega,\pm \vec{k}_1) = 3 \epsilon_0 \chi^{(3)}_{XXXX}(\omega) \big[ A_1(Z) |A_1(Z)|^2\\
+ 2 A_1(Z) |A_2(Z)|^2 \big] e^{i(k_1 Z - \omega t)} + c.c.,
\end{multline}
which is the nonlinear polarization density induced by the incident field $\vec{E}$ that governs nonlinear propagation of probe field $\vec{E}_1$. Here, the factor of 3 arises from the sum over allowed frequency permutations and $\chi^{(3)}_{abcd}(\omega; \omega, \omega, -\omega) = \chi^{(3)}_{abcd}(\omega)$ is the degenerate third-order nonlinear optical susceptibility tensor describing a polarization density response at the same frequency as the incident frequency.

Similarly, the expression for the Fourier component $\vec{P}^{(3)}(\pm\omega, \pm \vec{k}_1)$ for cross-polarized beams, i.e. $\zeta = \pi/2$, is
\begin{multline}
P_X^{(3)}(\pm \omega,\pm\vec{k}_1) = 3 \epsilon_0 \big[\chi^{(3)}_{XXXX}(\omega) A_1(Z) |A_1(Z)|^2\\
+ 2 \chi^{(3)}_{XXYY}(\omega) A_1(Z) |A_2(Z)|^2\big] e^{i(k_1 Z - \omega t)} + c.c..
\end{multline}

Solving the nonlinear wave propagation equation for these fields in the plane-wave basis and approximating slowly-varying field amplitudes gives an expression relating $\beta_{12}$ to the imaginary part of the third-order nonlinear optical susceptiblity tensor by
\begin{align}
\label{eqbetachi}
\beta_{12}^{\parallel}(\omega) &= \frac{3 \omega}{\epsilon_0 \big[n(\omega)\big]^2 c^2}\mathrm{Im}\{\chi^{(3)}_{XXXX}(\omega)\} = 2 \beta_{11} (\omega),\\
\label{eqbetachi2}
\beta_{12}^{\perp}(\omega) &= \frac{3 \omega}{\epsilon_0 \big[n(\omega)\big]^2 c^2}\mathrm{Im}\{\chi^{(3)}_{XXYY}(\omega)\},
\end{align}
where $n(\omega)$ is the refractive index, $c$ the speed of light, and $\beta_{12}^{\parallel}(\omega)$ refers to the two-beam degenerate 2PA coefficient for co-polarized beams ($\zeta = 0$) which is twice the 2PA coefficient for a single beam with the same polarization and propagation direction, $\beta_{11}(\omega)$. $\beta_{12}^{\perp}(\omega)$ is the two-beam degenerate 2PA coefficient for cross-polarized beams ($\zeta = \pi/2$). See the Supplementary Material for the complete derivation of these relationships.

In our experiments, the beams propagate along $Z=110$ in crystallographic coordinates, with $X=\overline{1}10$ and $Y=001$ defining a basis of linear polarization directions. We therefore rotationally transform $\chi^{(3)}$ using the relation
\begin{equation}
\chi_{ABCD} = R_A^a R_B^b R_C^c R_D^d\ \chi_{abcd},
\end{equation}
using rotation matrix
\begin{gather}
R=\frac{1}{\sqrt{2}}
 \begin{bmatrix}
   -\cos{\theta}      & \cos{\theta} &  \sqrt{2}\sin{\theta} \\
    \sin{\theta}      & -\sin{\theta} & \sqrt{2}\cos{\theta} \\
  1   &1 & 0
\end{bmatrix},
\end{gather}
where $\theta$ is the probe polarization angle relative to the $X'$-axis, or equivalently, the sample rotation angle, in order to express Eqs. \ref{eqbetachi} -- \ref{eqbetachi2} in terms of $\chi^{(3)}$ tensor components in crystallographic coordinates. This enables us to take advantage of the $\overline{4}3m$ (GaP and GaAs) or $m3m$ (Si) crystal symmetry for which there are only three independent, non-vanishing components; $\chi_{aaaa}$, $\chi_{aabb} = \chi_{abab}$, and $\chi_{abba}$, where $a \neq b$.\cite{popov} Here, intrinsic permutation (but not Kleinman)\cite{boyd,kleinman1} symmetry has been applied. $\beta_{12}^{\parallel, \perp}$ can then be written
\begin{widetext}
\begin{align}
\beta_{12}^{\parallel}(\omega,\theta) &= \frac{3 \omega\ \mathrm{Im}\{\chi^{(3)}_{xxxx}(\omega)\}}{16 \epsilon_0 \big[n(\omega)\big]^2 c^2} \bigg\{\Big[3\cos{(4\theta)} - 4\cos{(2\theta)} - 7\Big] \sigma(\omega) + 16\bigg\},\\
\beta_{12}^{\perp}(\omega,\theta) &= \frac{3 \omega\ \mathrm{Im}\{\chi^{(3)}_{xxxx}(\omega)\}}{16 \epsilon_0 \big[n(\omega)\big]^2 c^2} \bigg\{8\Big[1 - \eta(\omega)\Big]- \Big[ 3 \cos{(4\theta)} + 5 \Big]\sigma(\omega)\bigg\},
\end{align}
\end{widetext}
where $\theta \rightarrow \theta - \theta_0$ to allow for an arbitrary initial orientation, and
\begin{equation}
\sigma(\omega)= 1 - \Bigg[\frac{2\ \mathrm{Im}\{\chi_{xxyy}^{(3)}(\omega)\} + \mathrm{Im}\{\chi_{xyyx}^{(3)}(\omega)\}}{\mathrm{Im}\{\chi_{xxxx}^{(3)}(\omega)\}}\Bigg]
\end{equation}
is the anisotropy parameter and 
\begin{equation}
\eta(\omega) = \frac{\mathrm{Im}\{\chi_{xyyx}^{(3)}(\omega)\}}{\mathrm{Im}\{\chi_{xxxx}^{(3)}(\omega)\}} = \frac{2\delta(\omega) - \sigma(\omega)}{2}
\end{equation}
is the indicated ratio of tensor components. The difference in magnitude of 2PA between linear and circular polarizations in single-beam experiments is related to the dichroism parameter, $\delta$. Under the usual weak 2PA conditions there is no population inversion, which constrains $\sigma \leq 1 - \eta$ and $0 \leq \eta \leq 1$. See the Supplementary Material for the derivation of these constraints. Thus the independent components of $\mathrm{Im}\{\chi^{(3)}_{abcd}(\omega)\}$ of a $\overline{4}3m$ or $m3m$ crystal can be completely characterized by measuring 2-beam 2PA while rotating the sample about $Z=110$ in both co-polarized ($\zeta = 0$) and cross-polarized ($\zeta = \pi/2$) geometries.

\subsection{\label{sec:theorydispersionanisomodels}Dispersion and anisotropy models}
Ab initio calculations using density functional theory (DFT) have been performed by Murayama and Nakayama for $\mathrm{Im}\{\chi^{(3)}_{abcd}(\omega)\}$ in GaAs and Si.\cite{murayama} The calculations are applicable for GaAs over the entire spectral range, but for Si only for direct transitions above the two-photon direct-gap. 2PA in Si between the two-photon indirect-gap $E_g/2 = 0.56$ eV and the two-photon direct-gap $E_0'/2 = 1.75$ eV is most likely due to phonon-mediated transitions, as in linear absorption, but could also be affected by the presence of defect states or dopants.\cite{noffsinger} 

We can model the dispersion in $\beta(\omega)$ using the phenomenological models developed by Jones and Reiss for direct band gap semiconductors\cite{jones,wherrett} and Garcia and Kalyanaraman for indirect band gap semiconductors.\cite{garcia1} The Jones-Reiss model for direct-gap semiconductors (GaAs) has the form
\begin{equation}
\label{eqjrmodel}
\beta^{JR}(x) = B \frac{(2x - 1)^{3/2}}{(2x)^5},
\end{equation}
where $x = \hbar \omega / E_g$ and $B$ is a material-dependent constant which functions as a fit parameter. This function peaks at a value of $\beta^{JR}_{max}(0.71) \approx 0.0472\ B$. The Garcia-Kalyanaraman model for indirect-gap semiconductors (GaP, Si) considers parabolic electron and hole bands with indirect matrix elements and also accounts for ``allowed-allowed'' (a-a), ``allowed-forbidden'' (a-f), and ``forbidden-forbidden'' (f-f) transitions. 2PA dispersion takes the form
\begin{multline}
\label{eqgkmodel}
\beta^{GK}(x) = C \frac{(2x - 1)^2}{(2x)^5}\\
\times \bigg[ \underbrace{\frac{\pi}{8}}_\text{a-a} + \underbrace{\frac{\pi}{16}(2x - 1)}_\text{a-f} + \underbrace{\frac{5\pi}{128}(2x - 1)^2}_\text{f-f} \bigg]
\end{multline}
in the limit where the phonon energy $\hbar \nu_Q \ll E_g$ and in the absence of a DC-field. $C$ is a material-dependent constant which functions as a fit parameter. This function peaks at $\beta^{GK}_{max}(1.10) \approx 0.0225\ C$. The 2PA dispersion in the presence of a DC-field for both direct-gap\cite{garcia2} and indirect-gap\cite{garcia1} semiconductors acquires an exponential tail for $\hbar \omega < E_g/2$ due to tunneling-assisted 2PA. While these models are not based on full band-structure calculations, they have achieved some success in reproducing the dispersion observed in published results using single-beam techniques.

The anisotropy parameter $\sigma$ was estimated using a 3-band model where the dominant anisotropy is due to a-f transitions giving\cite{dvorak}
\begin{equation}
\sigma^{theory} \approx - \frac{2 Q^2 E_{g}}{P^2 E_{g}'},
\end{equation}
where ($P,Q$) are momentum matrix elements for ($\Gamma_{15}^V \rightarrow \Gamma_1^C,\Gamma_{15}^V \rightarrow \Gamma_{15}^C$) transitions and ($E_{g},E_{g}'$) are the direct band gaps between the valence and the (conduction, higher conduction) bands. The values of $P$ and $Q$ for $\overline{4}3m$ semiconductors are not reliably known\cite{dvorak} but we still apply this model to GaAs for predicting the 2PA anisotropy near the $\Gamma$-point which corresponds to the two-photon band edge. The model is still reasonable for GaP for transitions near the direct-gap energy of 2.78 eV at the $\Gamma$-point, which is only 25\% higher than the indirect-gap, as there are few available indirect transitions. It is not straightforward to apply this model to Si due to the nearly parallel band structure, necessitating a calculation of the mean contribution of anisotropy over the $k$-space whose direct transitions and available indirect transitions at 300 K are comparable in energy within the bandwidth of the excitation source. We report the theoretical values for the anisotropy near the two-photon band edge for a range of values for the momentum matrix elements from literature for GaAs and GaP in Table \ref{tab:aniso}.

\begin{table}
 \caption{\label{tab:aniso}Theoretical calculations of 2PA anisotropy near the $\Gamma$-point using Dvorak et al. model.}
 \begin{ruledtabular}
  \begin{tabular}{@{}lll@{}}
    Parameter & GaAs & GaP\\
    \hline
    $(Q/P)^2$ &  0.60,\cite{aspnes1} (0.71, 1.06)\cite{moss} & 0.57,\cite{aspnes1} (0.71, 1.06)\cite{moss}\\
    $E_{g-d}$ (eV) & 1.44\cite{dvorak} & 2.78\cite{zallen}\\
    $E_{g-d}'$ (eV) & 4.5\cite{dvorak} & 3.71\cite{zallen}\\
    $\sigma^{theory}$ & -0.38, (-0.45, -0.68) & -0.85, (-1.06, -1.59)
  \end{tabular}
  \end{ruledtabular}
\end{table}

\section{\label{sec:experiment}Experiment}

\subsection{\label{sec:experimenttechnique}Experimental technique}

\subsubsection{\label{sec:experimenttechniqueppmssetup}Pump-probe modulation spectroscopy setup}
\begin{figure*}[htbp]
  \includegraphics[width=\textwidth]{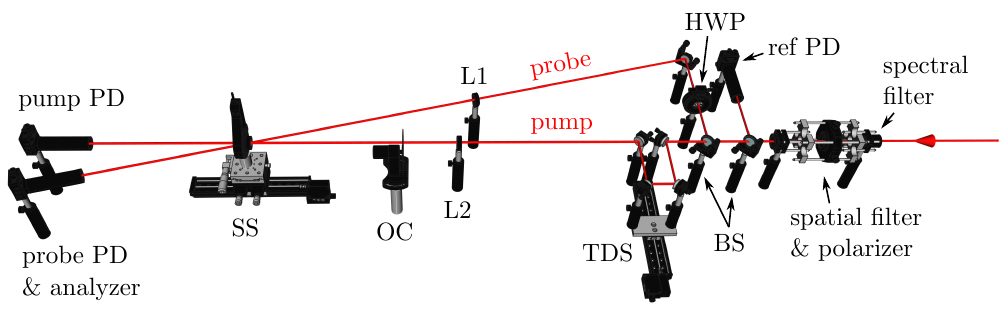}
  \caption{\label{fig:setup}PPMS experimental setup. The beam passes through spectral and spatial filters and a thin-film polarizer. The first of two beam splitters (BS) picks off a reference beam by reflecting 4\% to the reference photodiode detector (ref PD). A second pellicle beam splitter (BS) transmits 90\% (pump) and reflects 10\% (probe). A time delay stage (TDS) controls the relative time delay between the pump and probe pulses, and a half-wave plate (HWP) controls the relative probe polarization. The beams are focused by lenses (L1 and L2), and the pump beam is modulated with an optical chopper (OC). The beams overlap at an angle of $\sim 15^{\circ}$ on the sample which is mounted in the sample stage (SS) having fine $X$,$Z$-translation, 2-axis rotation, and tip-tilt control. The transmitted beams are detected with photodiode detectors (PD). Image includes CAD models courtesy of Thorlabs Inc., Zaber Technologies Inc., and Edmund Optics Inc. All rights reserved.}
\end{figure*}

Fig. \ref{fig:setup} shows the pump-probe modulation spectroscopy (PPMS) experimental setup. The light source is a Light Conversion TOPAS-C optical parametric amplifier (tuning range $240 < \lambda < 2600$ nm) pumped by a 1 kHz Coherent Libra HE USP titanium-doped sapphire regenerative amplifier (nominally $\lambda = 800$ nm, $\varepsilon_P = 4.0$ mJ, $\tau_{FWHM} < 50$ fs). Spectral filters attenuate undesired frequency components. A spatial filter consisting of two plano-convex $f = 100$ mm lenses and a pinhole (of diameter selected based on beam parameters) homogenizes and shapes the transverse beam profile to approximate a Gaussian. These filters were configured for each excitation wavelength. A linear thin-film polarizer ensures definite polarization. An achromatic half-wave plate (HWP) controls probe polarization relative to the pump. Plano-convex lenses L1 and L2 (with same focal length $f = 250$ or 500 mm, depending on beam parameters) focus pump and probe beams to a common spot on the sample. Temporal and spatial overlap was optimized by monitoring two-beam second-harmonic generation (SHG) from a barium borate crystal in the sample position. The optical chopper (OC) was sychronized to the fifth subharmonic of the laser repetition rate ($f_{mod} = 200$ Hz) where spectral noise was relatively low, and could be configured to modulate only the pump beam (PPMS experiment) or both beams (probe absorption calibration). The outputs of all photodiode detectors (Thorlabs DET100A for 325 - 1100 nm, Thorlabs PDA30G for 1200 - 2000 nm) were integrated and held for each pulse until the next trigger by Stanford Research Systems SR250 gated integrators. The probe and pump integrated signals were the inputs for two Stanford Research Systems SR830 DSP lock-in amplifiers referenced to the optical chopper TTL output. A Tektronix DPO 2014 oscilloscope recorded the outputs of the lock-in amplifiers, which were digitally recorded by a computer. The computer controlled pump-probe time delay, sample Z-scan to optimize pump-probe spatial overlap, and sample rotation.

\subsubsection{\label{sec:experimenttechniqueprofiles}Pump and probe spatiotemporal profiles}
The pulse duration was measured by performing a SHG second-order autocorrelation. Pump power was measured with a Coherent FieldMate Si photodiode (PD) head (650 - 1100nm) or with a reference PD calibrated with a thermal head (1200 - 2000 nm). In order to reduce sensitivity on overlap position and geometric effects, the probe radius was kept $5 \times$ smaller than the pump\cite{diels} (the beam waists were not at the sample plane) which necessitated attenuating the probe pulse energy with neutral density filters $\sim 250 \times$ lower than the pump to keep the fluence ratio $F_2/F_1 > 10$. Pump and probe beams impinged on each sample at $\approx 7.5^{\circ}$ from the surface normal, and at $\approx 15^{\circ}$ with respect to each other. The latter angle became $\approx 5^{\circ}$ (i.e. approximately collinear) inside each sample after taking refraction into account. The beam radii at the overlap position were measured using an automated knife-edge technique, and measurements at $Z = \pm 10$ and $\pm 20$ mm from the overlap were used to measure the beam angles and Rayleigh ranges.

\subsubsection{\label{sec:experimenttechniquesamples}Samples}
The samples were semi-insulating undoped n-type GaAs (110) with resistivity $\rho = 1.9$ -- $4.4$ E7 $\Omega$ cm, intrinsic carrier concentration $N_c = 3.2$ -- $11$ E7 cm$^{-3}$, carrier mobility $\mu = 3.7$ -- $4.4$ E3 cm$^2$ V$^{-1}$ s$^{-1}$, etch pitch density $EPD < 6000$ cm$^{-2}$, and thickness $L = 0.538 \pm 0.005$ mm; undoped GaP (110) with $L = 0.536 \pm 0.005$ mm; and undoped Si (110) with $\rho = 1$ -- $10$ $\Omega$ cm and $L = 0.485 \pm 0.005$ mm. The spatial and temporal overlap of the beams was fine-tuned for each sample using 2PA prior to all data acquisition scans. The linear optical properties of the samples were measured using a J.A. Wollam M2000 variable-angle spectroscopic ellipsometer (VASE) and are reported in the Supplementary Material. Linear optical properties for $\lambda > 1690$ nm were obtained from literature.\cite{skauli,li}

\subsection{\label{sec:experimentprocedure}Data collection and analysis procedure}

\subsubsection{\label{sec:experimentproceduretimedelay}Time delay scans}
The 2PA coefficient was determined by performing a time delay scan and fitting the normalized probe transmission to Eq. \ref{eqtransmission}. The fitting algorithm compared the variation in the fitting parameter $\beta_{12}^{\parallel,\perp} I_2^0 L$ for the $N=1$ and $N=2$ cases, and if the variation was below a threshold of $1\%$, then the result for the $N=1$ case was reported; if the variation was larger, the process was repeated, comparing successively higher-order terms. The normalized change in probe transmission was calculated by 
\begin{equation}
\frac{\Delta T_1(\Delta t)}{T_1} = \frac{V_1(\Delta t)}{\overline{V}_1^{cal}},
\end{equation}
where $V_1$ is the probe lock-in amplitude during the time delay scan and $\overline{V}_1^{cal}$ is the mean lock-in amplitude corresponding to $100\%$ absorption when the probe is modulated by the optical chopper during a calibration scan. After fitting, $\Delta T_1(\Delta t)/T_1$ was adjusted by the vertical offset parameter which removes the effect of any scattered pump light. When multiple scans were performed at a single excitation energy, the mean value of $\beta_{12}$ was reported and the uncertainty was conservatively quantified by the maximum of the set including the standard error of the mean, the propagated uncertainty of the mean, and the propagated uncertainty in the individual measurements.

\subsubsection{\label{sec:experimentprocedurerotation}Rotation scans}
The values of $\sigma$ and $\eta$ were determined by performing rotation scans of the sample fixed at the spatial overlap position and $\Delta t = 0$ in both co-polarized and cross-polarized geometries. Power fluctations can sometimes complicate the signal, and so the magnitude of the pump-induced absorption was corrected according to
\begin{equation}
\beta_{12}(\theta) \sim -\bigg(\frac{\Delta T_1(\theta)}{T_1}\bigg) \bigg(\frac{1}{V_{ref}(\theta)}\bigg)
\end{equation}
and
\begin{equation}
\frac{\Delta T_1(\theta)}{T_1} = \frac{V_1(\Delta t = 0, \theta) - V_1(\Delta t \rightarrow -\infty, \theta)}{\overline{(V_1^{cal}/V_{ref}^{cal})}V_{ref}(\theta) - V_1(\Delta t \rightarrow - \infty, \theta)},
\end{equation}
where $V_{ref}(\theta)$ is a reference voltage monitoring the incident power, $V_1(\Delta t \rightarrow - \infty)$ is the lock-in amplitude of the probe at large negative time delay (and thus negligible 2PA, capturing any offset), and $\overline{(V_1^{cal}/V_{ref}^{cal})}$ is the mean of the probe lock-in amplitude normalized to the reference amplitude during the calibration scan. 

Zincblende semiconductors can exhibit the photorefractive effect which can lead to transient two-beam coupling.\cite{dvorak} Transient energy transfer between beams propagating along the $110$-direction due to the photorefractive effect is antisymmetric about rotations of $R_Z(\pi)$, while 2PA is symmetric.\cite{dvorak} Analyzing only the symmetric part removes any contribution from the photorefractive effect. In addition, random fluctuations are also reduced. The symmetric and antisymmetric parts of the change in probe transmission are
\begin{equation}
\frac{\Delta T_1^{\pm} (\theta)}{T_1} = \frac{1}{2}\bigg[\frac{\Delta T_1(\theta)}{T_1} \pm \frac{\Delta T_1(\theta + \pi)}{T_1}\bigg],
\end{equation}
where $\Delta T_1^+ (\theta)/T_1$ and $\Delta T_1^- (\theta)/T_1$ are the symmetric and antisymmetric parts, respectively, of the change in probe transmission upon rotations of $R_Z(\pi)$.

\subsection{\label{sec:experimentresults}Results}

\subsubsection{\label{sec:experimentresultstimedelay}Time delay scans}
\begin{figure}[htbp]
  \includegraphics[width=\linewidth]{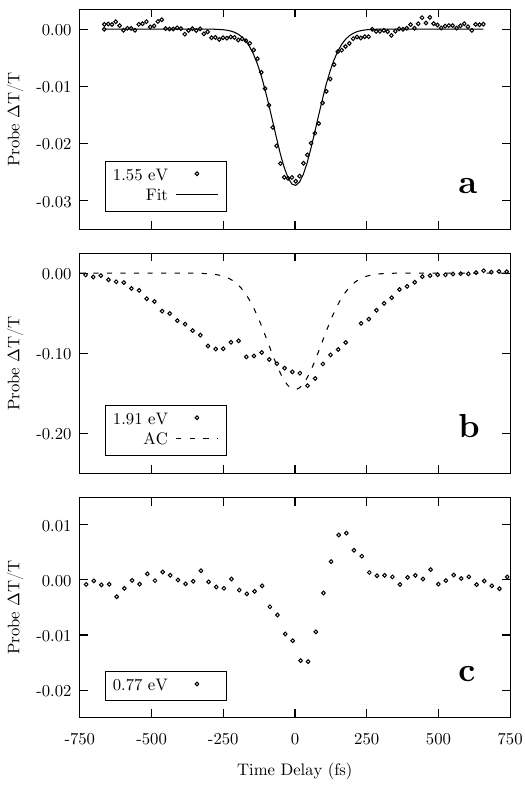}
  \caption{\label{fig:tdscan}a) $\Delta T_1/T_1$ vs. $\Delta t$ for GaP at $\hbar \omega =1.55$ eV measured by PPMS in co-polarized geometry (circles) representative of a typical time delay scan. The on-axis peak pump intensity was $I_2^0 = 0.236 \pm 0.012$ GW cm$^{-2}$ with pulse duration $\tau_{FWHM} = 125 \pm 2$ fs. Fitting the data (solid curve) gave a 2PA coefficient of $\beta_{12}^{\parallel} = 4.7 \pm 0.3$ cm GW$^{-1}$ for $N = 2$ terms and FCA was negligible at $b^{\parallel} = 0.0000 \pm 0.0001$. b) $\Delta T_1/T_1$ vs. $\Delta t$ for GaP at $\hbar \omega =1.91$ eV measured by PPMS in co-polarized geometry (circles) showing the temporally-broadened asymmetric anomaly as $\hbar \omega \rightarrow E_g$. The second-order autocorrelation signal expected for 2PA is superimposed (dashed curve) for $\tau_{FWHM} = 154 \pm 6$ fs. c) $\Delta T_1/T_1$ vs. $\Delta t$ for GaP at $\hbar \omega =0.77$ eV measured by PPMS in co-polarized geometry (circles) showing the coherent artifact with an increase in transmission for positive time delays likely due to transient energy transfer from a nonlinear refractive index phase grating as $\hbar \omega \rightarrow E_g/2$.}
\end{figure}

Fig. \ref{fig:tdscan} (a) shows an example of a time delay scan for GaP at $\hbar \omega =1.55$ eV in co-polarized geometry. Baseline probe absorption $\Delta T/T$ returned to zero at both positive and negative delays $|\Delta t| > 250$ fs, indicating negligible FCA. We observed the same symmetry in all $\Delta t$ scans presented in this study.

For $|\Delta t| < 250$ fs, the width of the measured 2PA response (data points) fits very closely to the calculated response (solid black curve) with the pulse duration independently measured by SHG second-order autocorrelation. We obtained a similar match for most data presented here. For a small number of exceptional cases, we observed a 2PA response slightly wider than the SHG autocorrelation. For these cases, we considered the possiblity of saturation effect and this led to a widening of error bars in reported spectra (see Supplementary Material). 

In addition to revealing complete information regarding the $\mathrm{Im}\{\chi^{(3)}(\omega)\}$ tensor structure, the 2-beam PPMS technique offers another advantage over single-beam techniques in its ability to discern anomalous effects from 2PA. Such effects include FCA, transient energy transfer originating from photorefractive phase gratings, nonlinear refractive index phase gratings, 2PA amplitude gratings, free-carrier gratings, and saturable absorption which can couple the two beams.\cite{smirl,valley}

The temporal width of the nonlinear loss was anomalously broadened in GaAs and GaP as the excitation neared their respective band gaps in both polarization geometries. Fig. \ref{fig:tdscan} (b) shows this anomaly in GaP at $\hbar \omega = 1.91$ eV (0.85$E_g$) and it was also observed in GaAs at $\hbar \omega = 1.38$ eV (0.96$E_g$). Pump depletion by pump-pump 2PA cannot account for this effect. The observed temporal FWHM in the nonlinear loss was a factor of 2.3 -- 3 greater than expected for 2PA alone. Asymmetry in the time delay scans also supports the presence of other effects not accounted for in our model. Both of these features are obvious against the second-order autocorrelation signal expected for 2PA (as measured by SHG) which is superimposed on the data.

A fs-scale time delay-dependent FCA cross-section could produce such an anomaly as free carriers excited predominantly by pump-pump 2PA diffuse through $k$-space in the conduction band. Zero time delay in this experiment was determined by a fit parameter, but the peak could be shifted towards positive time delays as a result of this effect. Another possibility is that a photorefractive index grating could induce transient energy transfer and generate the observed anomaly. Both of these effects are encompassed more generally by any non-instantaneous response or memory effect as the 2PA process approaches resonance. If such an effect is present it can induce a polarization density responsive not only to the instantaneous value of the electric field but also to its history on fs timescales.\cite{diels} Measurements at different crystal orientations, pulse energies, pulse durations, and a nondegenerate two-photon absorption experiment\cite{fishman,hutchings,bolger,hannes} could reveal more information about the nature of such anomalies.

Below the two-photon absorption band edge of GaAs and GaP, nonlinear loss was very small except at high intensities (a factor of 1 -- 20 greater than intensities at excitation energies $\hbar \omega > E_g/2$). This is consistent with 2PA being energetically disallowed but a transient coherent artifact\cite{lebedev} was still observed near zero time delay in both polarization geometries. A transient coherent artifact was also observed in Si near to but slightly above $E_g/2$ at $\hbar \omega = 0.62$ eV, where the magnitude of 2PA was also quite low. This coherent artifact was asymmetric and notably included an antisymmetric component (about $\Delta t =0$) where $\Delta T_1/T_1 > 0$ briefly for positive time delays. While $\mathrm{Im}\{\chi^{(3)}(\omega)\} \rightarrow 0$ rapidly at the two-photon band edge, $\mathrm{Re}\{\chi^{(3)}(\omega)\}$ in general does not. $\mathrm{Re}\{\chi^{(3)}(\omega)\}$ encodes the nonlinear refractive index which is related to $\mathrm{Im}\{\chi^{(3)}(\omega)\}$ by nonlinear Kramers-Kronig relations.\cite{boyd} Below resonance, $\mathrm{Re}\{\chi^{(3)}(\omega)\} \sim \omega^{-3}$ in semiconductors and thus is the dominant third-order nonlinearity at low excitation energies.\cite{boyd}

This coherent artifact shown in Fig. \ref{fig:tdscan} (c) likely arises from a nonlinear refractive index phase grating, which can transfer energy between the probe and pump and produce such antisymmetric time delay profiles.\cite{palfrey} This effect is a background effect that likely occurs in all measurements, but is so weak that it is only visible when 2PA is no longer the dominant third-order nonlinearity.

\subsubsection{\label{sec:experimentresults2paspectra}2PA spectra}
\begin{figure}[htbp]
  \includegraphics[width=\linewidth]{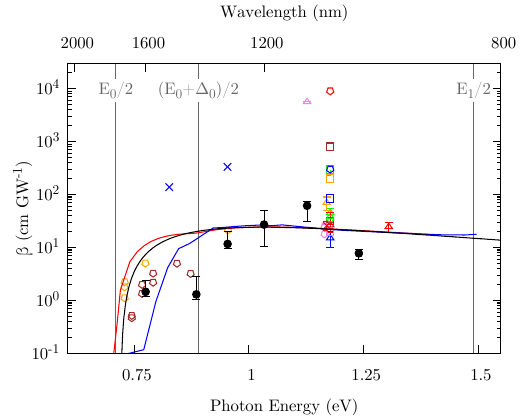}
  \caption{\label{fig:betagaas}2PA coefficient ($\beta = \frac{1}{2}\beta_{12}^{\parallel}$) spectrum of GaAs measured by PPMS in co-polarized geometry for $E_g/2 < \hbar \omega < E_g$ (solid black circles). The best fit of this data to the Jones-Reiss model is shown with the black curve. Data from a two-beam experiment is shown with a red triangle\cite{dvorak}. Data from other sources using single-beam experiments are also shown; blue triangle\cite{bosacchi}, green triangle\cite{penzkofer}, orange triangle\cite{saissy}, violet triangle\cite{jayaraman}, brown triangle\cite{bechtel}, red square\cite{kleinman2}, blue square\cite{bepko}, green square\cite{lee}, orange square\cite{oksman}, violet square\cite{ralston}, brown square\cite{arsenev}, red pentagon\cite{basov}, blue pentagon\cite{grasyuk}, green pentagon\cite{zubarev}, orange pentagons\cite{hurlbut}, violet pentagons\cite{desalvo}, brown pentagons\cite{fishman}, red X\cite{said}, and blue X's\cite{zhihui}. Theoretical calculations from Hutchings, et al.\cite{hutchings} are shown with the red curve, and ab initio calculations from Murayama, et al.\cite{murayama} are shown with the blue curve. Critical point energies are marked with gray vertical lines.\cite{lautenschlager1}}
\end{figure}

The 2PA spectrum for GaAs measured by PPMS in co-polarized geometry is shown in Fig. \ref{fig:betagaas} along with the fit using the Jones-Reiss model in Eq. \ref{eqjrmodel}, published values using mostly single-beam techniques, and ab initio calculations. The best fit of this data to the Jones-Reiss model gives $B = 510 \pm 210$ cm GW$^{-1}$ and a 2PA peak of $\beta^{JR}_{max} = 24 \pm 10$ cm GW$^{-1}$. The 2PA coefficient reported in literature has a large spread, mainly from single-beam experiments clustered around 1053 nm using picosecond lasers.

\begin{figure}[htbp]
  \includegraphics[width=\linewidth]{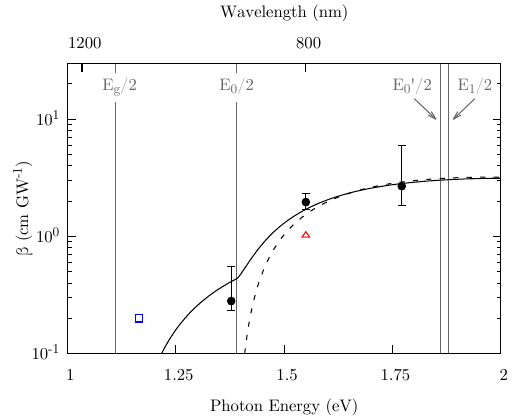}
  \caption{\label{fig:betagap}2PA coefficient ($\beta = \frac{1}{2}\beta_{12}^{\parallel}$) spectrum of GaP measured by PPMS in co-polarized geometry for $E_g/2 < \hbar \omega < E_g$ (solid black circles). The best fit of this data to the cumulative model is shown with the solid curve and the Jones-Reiss model for direct transitions with the dashed curve. Data from other sources using single-beam experiments are also shown; red triangle\cite{hoffman} and blue square\cite{bechtel}. Critical point energies are marked with gray vertical lines.\cite{zallen}}
\end{figure}

\begin{figure}[htbp]
  \includegraphics[width=\linewidth]{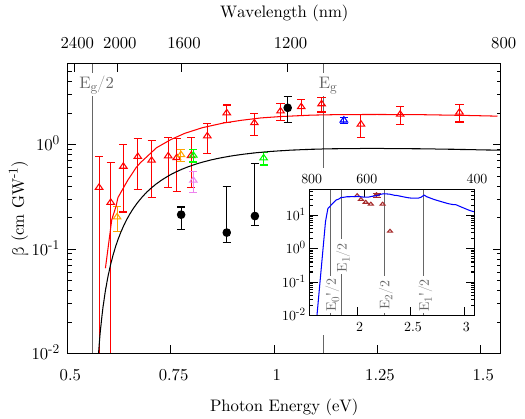}
  \caption{\label{fig:betasi}2PA coefficient ($\beta = \frac{1}{2}\beta_{12}^{\parallel}$) spectrum of Si measured by PPMS in co-polarized geometry for $E_g/2 < \hbar \omega < E_g$ (solid black circles). The best fit of this data to the Garcia-Kalyanaraman model is shown with the black curve and the fit to the data from Bristow, et al.\cite{bristow} is shown with the red curve and red triangles, respectively. Data from other sources using single-beam experiments are also shown; blue triangle\cite{reintjes}, green triangles\cite{dinu}, orange triangles\cite{euser}, and violet triangle\cite{tsang}. Inset: Data from a two-beam experiment above the direct two-photon band gap is shown with brown triangles\cite{reitze} and ab initio calculations from Murayama, et al.\cite{murayama} are shown with the blue curve. Critical point energies are marked with gray vertical lines.\cite{murayama,lautenschlager2}}
\end{figure}

The 2PA spectrum for GaP measured by PPMS in co-polarized geometry is shown in Fig. \ref{fig:betagap} along with a fit using a cumulative model for direct and indirect transitions, a fit using the Jones-Reiss model for direct transitions near $E_0/2$, and a comparison to published values using single-beam techniques. The cumulative model is defined as using the Garcia-Kalyanaraman model for $E_g / 2 \leq \hbar \omega < E_0 / 2$, and a sum of this function and the Jones-Reiss model above $E_0 / 2$. The best fit to this model gives $B = 44 \pm 18$ cm GW$^{-1}$ and $C = 49 \pm 35$ cm GW$^{-1}$, with 2PA contributions peaking at $\beta^{GK}_{max} = 1.1 \pm 0.8$ cm GW$^{-1}$ for indirect transitions and $\beta^{JR}_{max} = 2.1 \pm 0.9$ cm GW$^{-1}$ for direct transitions.

The 2PA spectrum for Si measured by PPMS in co-polarized geometry is shown in Fig. \ref{fig:betasi} along with the fit using the Garcia-Kalyanaraman model and a comparison to published values using single-beam techniques. The best fit of this data to the Garcia-Kalyanaraman model gives $C = 41 \pm 25$ cm GW$^{-1}$, peaking at $\beta^{GK}_{max} =  0.9 \pm 0.6$ cm GW$^{-1}$, which is considerably lower than the fit to the data from Bristow, et al.\cite{bristow} of $C = 86$ cm GW$^{-1}$, peaking at $\beta^{GK}_{max} = 1.93$ cm GW$^{-1}$. The measured dispersion differs from the phenomenological model. These results are tabulated in the Supplementary Material. An inset in Fig. \ref{fig:betasi} shows the degenerate 2PA spectrum predicted by ab initio calculations valid above $E_0'/2 = 1.75$ eV and published values using a two-beam technique.

\subsubsection{\label{sec:experimentresultsrotation}Rotation scans}
\begin{figure}[htbp]
  \includegraphics[width=\linewidth]{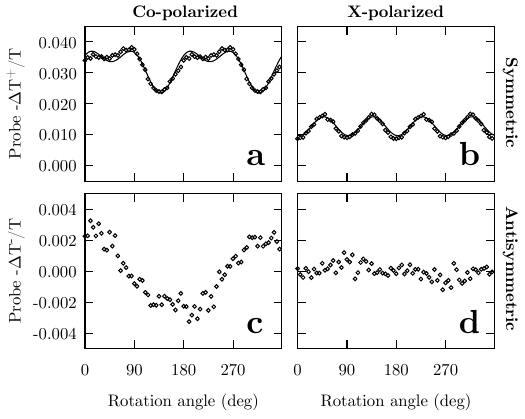}
  \caption{\label{fig:rotscan}$-\Delta T_1^+ /T_1$ vs. $\theta$ for GaP at $\hbar \omega = 1.55$ eV in a) co-polarized and b) cross-polarized geometries. The resulting fits (solid curves) are shown and give $\sigma = - 0.86 \pm 0.07$, $\eta = 0.18 \pm 0.18$, and $\theta_0 = 132.2^{\circ} \pm 1.0^{\circ}$ (describing the relative orientation of lab $X$-axis from $\overline{1}10$ crystal direction). The antisymmetric part $-\Delta T_1^- /T_1$ vs. $\theta$ in c) co-polarized and d) cross-polarized geometries.}
\end{figure}

An example of a rotation scan for GaP at $\hbar \omega = 1.55$ eV showing symmetric and antisymmetric parts of the normalized $- \Delta T_1(\theta)/T_1$ in co- and cross-polarized geometries is shown in Fig. \ref{fig:rotscan}. The fits of the co- and cross-polarized symmetric data share global parameters $\sigma$ and $\theta_0$, while the other parameters are independent. The antisymmetric part is typically $<10\%$ of the magnitude of the symmetric part.

\subsubsection{\label{sec:experimentresultsanisospectra}Anisotropy spectra}
The spectra of anisotropy and dichroism parameters for GaAs, GaP, and Si are reported in Table \ref{tab:anisosigma}, along with the calculations of the relative amplitudes of $\mathrm{Im}\{\chi^{(3)}_{xxyy}\}$ and $\mathrm{Im}\{\chi^{(3)}_{xyyx}\}$ to $\mathrm{Im}\{\chi^{(3)}_{xxxx}\}$. The values of $\sigma$ from this work are compared to the range of theoretical predictions of $\sigma$ for GaAs and GaP from Table \ref{tab:aniso} as well as results reported in literature for GaAs and Si. $|\sigma|$ is largest for GaP, which is consistent with theoretical predictions at the two-photon band edge. The values of $\eta$ from this work are compared to a previously reported value in literature for GaAs, and ab initio calculations for anisotropy in GaAs are also shown. See the Supplementary Material for additional information on comparing values between different sources in literature.

\begin{table*}
 \caption{\label{tab:anisosigma}Measured spectra of 2PA anisotropy and dichroism parameters for GaAs, GaP, and Si. The relative values of $\mathrm{Im}\{\chi^{(3)}_{xxyy}\} / \mathrm{Im}\{\chi^{(3)}_{xxxx}\} = \big[ 1 - ( \sigma + \eta ) \big]/2 = \big[2 - (\sigma + 2 \delta)\big]/2$ and $\mathrm{Im}\{\chi^{(3)}_{xyyx}\} / \mathrm{Im}\{\chi^{(3)}_{xxxx}\} = \eta = (2\delta - \sigma)/2$ are also tabulated. Literature and theoretical values are included and referenced in Notes. See Table \ref{tab:aniso} for derivations of theoretical values.}
 \begin{ruledtabular}
  \begin{tabular}{ l l l l l l l l l l l } 
   Sample & $\hbar \omega$ (eV) & $\sigma$ & $\delta \sigma$ & $\delta$ & $\delta (\delta)$ & $\frac{\mathrm{Im}\{\chi^{(3)}_{xxyy}\}}{\mathrm{Im}\{\chi^{(3)}_{xxxx}\}}$ & $\delta \frac{\mathrm{Im}\{\chi^{(3)}_{xxyy}\}}{\mathrm{Im}\{\chi^{(3)}_{xxxx}\}}$ & $\frac{\mathrm{Im}\{\chi^{(3)}_{xyyx}\}}{\mathrm{Im}\{\chi^{(3)}_{xxxx}\}}$ & $\delta \frac{\mathrm{Im}\{\chi^{(3)}_{xyyx}\}}{\mathrm{Im}\{\chi^{(3)}_{xxxx}\}}$ & Notes\\
    \hline
GaAs & 0.72 & -0.38 & & & & & & & & \textit{Theoretical}: see Table \ref{tab:aniso}\\
          & 0.72 & -0.45 & & & & & & & & \textit{Theoretical}: see Table \ref{tab:aniso}\\
          & 0.72 & -0.68 & & & & & & & & \textit{Theoretical}: see Table \ref{tab:aniso}\\
          & 0.77 & -0.13 & 0.08 & 0.68 & 0.17 & 0.20 & 0.13 & 0.74 & 0.17 & \\
          & 0.88 & -0.65 & & -0.50 & & 0.91 & & -0.17 & & \textit{Ab initio}: Murayama, et al.\cite{murayama}\\
          & 0.89 & -0.36 & 0.06 & 0.0 & 0.3 & 0.58 & 0.18 & 0.2 & 0.3 & \\
          & 0.93 & -0.46 & & -0.01 & & 0.62 & & 0.22 & & \textit{Ab initio}: Murayama, et al.\cite{murayama}\\
          & 0.95 & -0.29 & 0.05 & 0.47 & 0.11 & 0.34 & 0.08 & 0.61 & 0.10 & \\
          & 0.95 & -0.56 & & 0.02 & & 0.63 & & 0.29 & & \textit{Ab initio}: Murayama, et al.\cite{murayama}\\
          & 1.03 & -0.27 & 0.08 & 0.4 & 1.6 & 0.4 & 0.8 & 0.5 & 1.6 & \\
          & 1.03 & -0.47 & & 0.15 & & 0.54 & & 0.39 & & \textit{Ab initio}: Murayama, et al.\cite{murayama}\\
          & 1.13 & -0.40 & 0.05 & 0.42 & 0.12 & 0.39 & 0.07 & 0.62 & 0.09 & \\
          & 1.13 & -0.58 & & 0.26 & & 0.51 & & 0.55 & & \textit{Ab initio}: Murayama, et al.\cite{murayama}\\
          & 1.17 & -0.224 & 0.011 & & & & & & & Bepko, et al.\cite{bepko}\\
          & 1.17 & -0.3 & & & & & & & & Bechtel, et al.\cite{bechtel}\\
          & 1.17 & -0.74 & 0.18 & & & & & & & DeSalvo, et al.\cite{desalvo}\\
          & 1.17 & 0.00 & 0.15 & & & & & & & Said, et al.\cite{said}\\
          & 1.17 & -0.57 & & 0.28 & & 0.50 & & 0.55 & & \textit{Ab initio}: Murayama, et al.\cite{murayama}\\
          & 1.23 & -0.64 & & 0.29 & & 0.51 & & 0.61 & & \textit{Ab initio}: Murayama, et al.\cite{murayama}\\
          & 1.24 & -0.23 & 0.05 & 0.64 & 0.01 & 0.24 & 0.06 & 0.75 & 0.07 & \\
          & 1.31 & -0.76 & 0.08 & 0.22 & 0.24 & 0.58 & 0.08 & 0.60 & 0.20 & Dvorak, et al.\cite{dvorak}\\
          & 1.31 & -0.64 & & 0.34 & & 0.49 & & 0.66 & & \textit{Ab initio}: Murayama, et al.\cite{murayama}\\
          & 1.48 & -0.66 & & 0.34 & & 0.49 & & 0.67 & & \textit{Ab initio}: Murayama, et al.\cite{murayama}\\
          \hline
GaP    & 1.38 & -0.53 & 0.06 & 0.09 & 0.17 & 0.59 & 0.10 & 0.35 & 0.14 & \\
          & 1.39 & -0.85 & & & & & & & & \textit{Theoretical}: see Table \ref{tab:aniso}\\
          & 1.39 & -1.06 & & & & & & & & \textit{Theoretical}: see Table \ref{tab:aniso}\\
          & 1.39 & -1.59 & & & & & & & & \textit{Theoretical}: see Table \ref{tab:aniso}\\
          & 1.55 & -0.88 & 0.10 & -0.13 & 0.23 & 0.79 & 0.14 & 0.31 & 0.18 & \\
          & 1.77 & -0.61 & 0.09 & 0.1 & 0.7 & 0.6 & 0.3 & 0.4 & 0.6 & \\
          \hline
Si       & 0.58 & 0.00 & 0.05 & & & & & & & Bristow, et al.\cite{bristow}\\
          & 0.77 & -0.55 & 0.09 & 0.1 & 0.4 & 0.60 & 0.20 & 0.4 & 0.3 & \\
          & 0.89 & -0.03 & 0.04 & 0.95 & 0.08 & 0.04 & 0.05 & 0.96 & 0.06 & \\
          & 0.89 & 0.00 & 0.05 & & & & & & & Bristow, et al.\cite{bristow}\\
          & 0.95 & -0.15 & 0.14 & 0.77 & 0.23 & 0.16 & 0.15 & 0.84 & 0.16 & \\
          & 1.03 & -0.29 & 0.08 & 0.2 & 0.6 & 0.5 & 0.3 & 0.3 & 0.6 & \\
          & 1.65 & -0.08 & & 0.50 & & 0.18 & & 0.54 & & \textit{Ab initio}: Murayama, et al.\cite{murayama}
  \end{tabular}
 \end{ruledtabular}
\end{table*}           

\section{\label{sec:discussion}Discussion}
\begin{figure*}[htbp]
  \includegraphics[width=\textwidth]{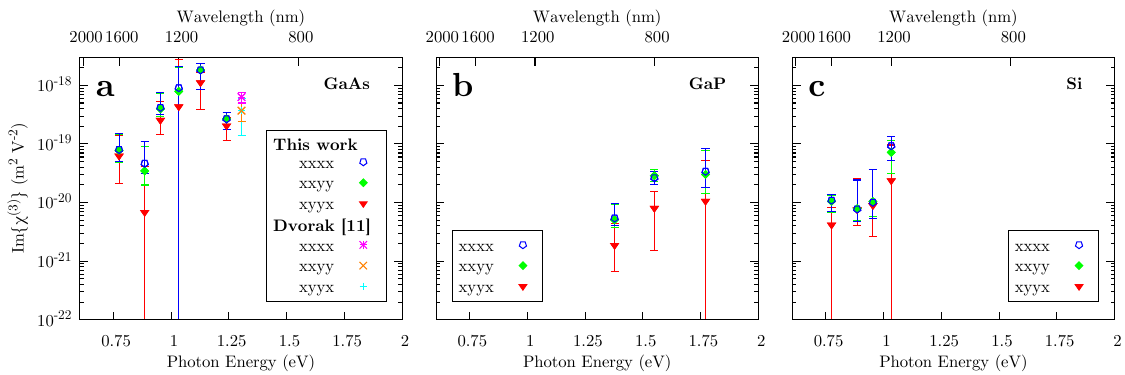}
  \caption{\label{fig:chi3}$\mathrm{Im}\{\chi^{(3)}(\omega)\}$ tensor spectrum obtained by 2PA PPMS time delay and anisotropy measurements for a) GaAs, b) GaP, and c) Si. Previously reported values for GaAs at $\hbar \omega = 1.31$ eV by Dvorak, et al.\cite{dvorak} are indicated. The values are tabulated in the Supplementary Material.}
\end{figure*}

The complete set of measurements performed ranged from 0.62 -- 1.38 eV for GaAs, 0.77 -- 1.91 eV for GaP, and 0.62 -- 1.03 eV for Si. Long-lifetime time-independent FCA was observed to be negligible in the regime studied in this work as evidenced by the complete recovery of the transmission of the probe at sufficiently large positive time delays. The antisymmetric part of $\Delta T_1/T_1$ in the rotation scans was generally small in comparison to the symmetric part, and thus the photorefractive effect did not significantly affect the anisotropy measurements in the regime studied. However, other anomalies were observed within this range for which the model for 2PA considered in this work was unable to explain, as detailed in Section \ref{sec:experimentresultstimedelay}. The anomaly observed in time delay scans characterized by asymmetric temporal broadening and strong nonlinear loss precluded an estimation of $\beta$ at 1.38 eV for GaAs and 1.91 eV for GaP, near their respective band gaps. This anomaly indicated the presence of a transient effect of the pump on the probe which depended on the history of the incident field. A distinctly different anomaly was observed in time delay scans characterized by an antisymmetric component with an increase in probe transmission for positive delays for which the likely source was a nonlinear refractive index phase grating. This precluded an estimation of $\beta$ at 0.62 -- 0.69 eV for GaAs, 0.77 -- 1.24 eV for GaP,  and 0.62 -- 0.69 eV for Si, near their respective two-photon band edges. The co-polarized rotation scans often did not follow the expected form of 2PA for $\overline{4}3m$ or $m3m$ crystal symmetry when the antisymmetric time delay scan anomaly was observed. However, some of the rotation scans did still match the expected form even with the presence of a time delay scan anomaly which suggests that 2PA may still be dominant at zero time delays. The complete set of measurements is outlined in the Supplementary Material. The 2PA spectra presented here reflect only the data for which no anomaly was observed.

The dispersion observed in GaAs in Fig. \ref{fig:betagaas} indicates that the spectral structure of 2PA is more complex than predicted by the Jones-Reiss model. The peak in 2PA measured at 1.13 eV is close to that expected from the Jones-Reiss model of 1.03 eV. The measured spectra is comparable in magnitude to ab initio calculations. The number of surviving spectral measurements of $\beta$ in GaP and Si are too few to make compelling statements about the agreement with phenomenological models, but this also illustrates the importance of identifying anomalous responses not attributable to 2PA which is often only possible with a 2-beam experiment. Measurements of 2PA in GaP at increased spectral density near $E_0/2$ could better constrain the relative strengths of direct and indirect two-photon transitions. 2PA in Si between $E_g/2 = 0.56$ eV and $E_0'/2 = 1.75$ eV is due to either phonon-mediated transitions or transitions involving defect or dopant states. Measurements of the degenerate 2PA spectrum dependence on temperature in this spectral range could verify the relative contributions to 2PA by these two mechanisms. The ab initio calculations of 2PA in Si by Murayama, et al. consider only direct transitions and thus are applicable in describing 2PA above $E_0'/2$.\cite{murayama}

The parameters $\beta_{12}(\omega)$, $\sigma(\omega)$, and $\eta(\omega)$ characterize the three independent components $\mathrm{Im}\{\chi^{(3)}_{xxxx}(\omega)\}$, $\mathrm{Im}\{\chi^{(3)}_{xxyy}(\omega)\}$, and $\mathrm{Im}\{\chi^{(3)}_{xyyx}(\omega)\}$. The relative amplitudes of the off-diagonal tensor components to the $\mathrm{Im}\{\chi^{(3)}_{xxxx}\}$ component are shown in Table \ref{tab:anisosigma}, and the absolute amplitude of the $\mathrm{Im}\{\chi^{(3)}_{abcd}(\omega)\}$ tensor spectra for GaAs, GaP, and Si are shown in Fig. \ref{fig:chi3} and compared with results from literature for GaAs.\cite{dvorak} See the Supplementary Material for more information on comparing values of the tensor components to literature results. The relative amplitudes of the tensor components $\mathrm{Im}\{\chi^{(3)}_{xxyy}(\omega)\}/\mathrm{Im}\{\chi^{(3)}_{xxxx}(\omega)\}$ differ by more than their uncertainties for all samples and excitation energies except for GaAs at $\hbar \omega = 1.03$ eV, where the anisotropy was only weakly constrained. The relative amplitudes of $\mathrm{Im}\{\chi^{(3)}_{xyyx}(\omega)\}/\mathrm{Im}\{\chi^{(3)}_{xxxx}\}$ also differ by more than their uncertainties for all samples and excitation energies except for GaAs at $\hbar \omega = 1.03$ eV again, and also for GaP at $\hbar \omega = 1.77$ eV, and Si at $\hbar \omega = 0.89$ and 0.95 eV. The relative amplitudes $\mathrm{Im}\{\chi^{(3)}_{xyyx}(\omega)\}/\mathrm{Im}\{\chi^{(3)}_{xxyy}(\omega)\}$ differ by more than their uncertainties for GaAs at $\hbar \omega = 0.77$, 0.95, 1.13, and 1.24 eV, for GaP at $\hbar \omega = 1.55$ eV, and for Si at $\hbar \omega = 0.89$ and 0.95 eV. Thus Kleinman symmetry is broken, a feature not discerned in the result from Dvorak, et al. for GaAs at 1.31 eV.\cite{dvorak} Indeed, Kleinman symmetry should be broken in the presence of dispersion in $\mathrm{Im}\{\chi^{(3)}(\omega)\}$.\cite{kleinman1,boyd}

\section{\label{sec:conclusions}Conclusions}
We characterized the $\mathrm{Im}\{\chi^{(3)}_{abcd}(\omega)\}$ tensor spectrum by measuring the degenerate two-photon absorption coefficient and its anisotropy using a 2-beam PPMS experiment for bulk 110-cut GaAs, GaP, and Si at 300 K over an overall excitation range $0.62 < \hbar \omega < 1.91$ eV ($650 < \lambda < 2000$ nm). The phenomenological 2PA dispersion models we used qualitatitvely agree with our experimental results for GaAs and GaP but there remain significant discrepancies, especially in Si. These could be due to the models lacking consideration of the full band structure of these materials including the effect of anisotropy or additional nonlinear effects. Ab initio calculations in GaAs roughly agree with the observed dispersion. We observed an anomalous asymmetric temporally-broadened nonlinear loss in GaAs and GaP when $\hbar \omega$ approached their respective band gaps which could be due to a variety of noninstantaneous nonlinear responses. We also observed a coherent artifact in all three samples near $E_g/2$ which was likely due to a nonlinear refractive index phase grating. Kleinman symmetry was observed to be broken in all three materials.

Our results motivate further development of ab initio theories\cite{vanstryland2,bechstedt,attaccalite,gruening,murayama} such as those which use DFT in the length gauge,\cite{anderson,anderson2,anderson3,anderson4} and experimental measurements of the degenerate 2PA spectrum dependence on temperature in Si and GaP which could help elucidate the potential contributions to 2PA below the two-photon direct-gap. Experiments at higher spectral density, using nondegenerate beams, and investigating a larger range of pulse energy- or duration-dependence could identify other nonlinear effects that may be present in these materials at femtosecond timescales. If phonon-mediated transitions are indeed dominant below the two-photon direct-gap in indirect band gap semiconductors, this would motivate further development of ab initio calculations of the $\mathrm{Im}\{\chi^{(3)}_{abcd}(\omega)\}$ tensor spectra which account for such transitions.\cite{noffsinger}

\begin{acknowledgments}
This work was funded by Robert A. Welch Foundation Grant F-1038 and the majority of experimental work performed at the Laboratorio de \'{O}ptica Ultrarr\'{a}pida at Centro de Investigaciones en \'{O}ptica, A.C. in Le\'{o}n, M\'{e}xico. B.S.M. acknowledges support from Consejo Nacional de Ciencia y Tecnolog\'{i}a, M\'{e}xico, Grant A1-S-9410. The authors would like to thank Enrique No\'{e}-Arias (Centro de Investigaciones en \'{O}ptica) for data acquisition program development, Dr. Sean Anderson (Wake Forest University) and Dr. Christopher Reilly (\'{E}cole Polytechnique F\'{e}d\'{e}rale de Lausanne) for assistance with data analysis, and Jack Clifford (University of Texas at Austin) and the Centro de Investigaciones en \'{O}ptica Machine Shop for machining assistance and part fabrication.
\end{acknowledgments}

\section*{\label{sec:dataavailability}Data Availability}
The data that supports the findings of this study are available within the article and its supplementary material, and additional information that support the findings of this study are available from the corresponding author upon reasonable request.

\section*{References}
\bibliography{ms}

\begin{thebibliography}{85}%
\makeatletter
\providecommand \@ifxundefined [1]{%
 \@ifx{#1\undefined}
}%
\providecommand \@ifnum [1]{%
 \ifnum #1\expandafter \@firstoftwo
 \else \expandafter \@secondoftwo
 \fi
}%
\providecommand \@ifx [1]{%
 \ifx #1\expandafter \@firstoftwo
 \else \expandafter \@secondoftwo
 \fi
}%
\providecommand \natexlab [1]{#1}%
\providecommand \enquote  [1]{``#1''}%
\providecommand \bibnamefont  [1]{#1}%
\providecommand \bibfnamefont [1]{#1}%
\providecommand \citenamefont [1]{#1}%
\providecommand \href@noop [0]{\@secondoftwo}%
\providecommand \href [0]{\begingroup \@sanitize@url \@href}%
\providecommand \@href[1]{\@@startlink{#1}\@@href}%
\providecommand \@@href[1]{\endgroup#1\@@endlink}%
\providecommand \@sanitize@url [0]{\catcode `\\12\catcode `\$12\catcode
  `\&12\catcode `\#12\catcode `\^12\catcode `\_12\catcode `\%12\relax}%
\providecommand \@@startlink[1]{}%
\providecommand \@@endlink[0]{}%
\providecommand \url  [0]{\begingroup\@sanitize@url \@url }%
\providecommand \@url [1]{\endgroup\@href {#1}{\urlprefix }}%
\providecommand \urlprefix  [0]{URL }%
\providecommand \Eprint [0]{\href }%
\providecommand \doibase [0]{http://dx.doi.org/}%
\providecommand \selectlanguage [0]{\@gobble}%
\providecommand \bibinfo  [0]{\@secondoftwo}%
\providecommand \bibfield  [0]{\@secondoftwo}%
\providecommand \translation [1]{[#1]}%
\providecommand \BibitemOpen [0]{}%
\providecommand \bibitemStop [0]{}%
\providecommand \bibitemNoStop [0]{.\EOS\space}%
\providecommand \EOS [0]{\spacefactor3000\relax}%
\providecommand \BibitemShut  [1]{\csname bibitem#1\endcsname}%
\let\auto@bib@innerbib\@empty
\bibitem [{\citenamefont {Asghari}\ and\ \citenamefont
  {Krishnamoorthy}(2011)}]{asghari}%
  \BibitemOpen
  \bibfield  {author} {\bibinfo {author} {\bibfnamefont {M.}~\bibnamefont
  {Asghari}}\ and\ \bibinfo {author} {\bibfnamefont {A.}~\bibnamefont
  {Krishnamoorthy}},\ }\href@noop {} {\bibfield  {journal} {\bibinfo  {journal}
  {Nat. Photonics}\ }\textbf {\bibinfo {volume} {5}},\ \bibinfo {pages}
  {268--270} (\bibinfo {year} {2011})}\BibitemShut {NoStop}%
\bibitem [{\citenamefont {Yoon}\ \emph {et~al.}(2010)\citenamefont {Yoon},
  \citenamefont {Jo}, \citenamefont {Chun}, \citenamefont {Jung}, \citenamefont
  {Kim}, \citenamefont {Meitl}, \citenamefont {Menard}, \citenamefont {Li},
  \citenamefont {Coleman}, \citenamefont {Paik},\ and\ \citenamefont
  {Rogers}}]{yoon}%
  \BibitemOpen
  \bibfield  {author} {\bibinfo {author} {\bibfnamefont {J.}~\bibnamefont
  {Yoon}}, \bibinfo {author} {\bibfnamefont {S.}~\bibnamefont {Jo}}, \bibinfo
  {author} {\bibfnamefont {I.}~\bibnamefont {Chun}}, \bibinfo {author}
  {\bibfnamefont {I.}~\bibnamefont {Jung}}, \bibinfo {author} {\bibfnamefont
  {H.}~\bibnamefont {Kim}}, \bibinfo {author} {\bibfnamefont {M.}~\bibnamefont
  {Meitl}}, \bibinfo {author} {\bibfnamefont {E.}~\bibnamefont {Menard}},
  \bibinfo {author} {\bibfnamefont {X.}~\bibnamefont {Li}}, \bibinfo {author}
  {\bibfnamefont {J.}~\bibnamefont {Coleman}}, \bibinfo {author} {\bibfnamefont
  {U.}~\bibnamefont {Paik}}, \ and\ \bibinfo {author} {\bibfnamefont
  {J.}~\bibnamefont {Rogers}},\ }\href@noop {} {\bibfield  {journal} {\bibinfo
  {journal} {Nature}\ }\textbf {\bibinfo {volume} {465}},\ \bibinfo {pages}
  {329--333} (\bibinfo {year} {2010})}\BibitemShut {NoStop}%
\bibitem [{\citenamefont {Bett}\ \emph {et~al.}(1999)\citenamefont {Bett},
  \citenamefont {Dimroth}, \citenamefont {Stollwerck},\ and\ \citenamefont
  {Sulima}}]{bett}%
  \BibitemOpen
  \bibfield  {author} {\bibinfo {author} {\bibfnamefont {A.}~\bibnamefont
  {Bett}}, \bibinfo {author} {\bibfnamefont {F.}~\bibnamefont {Dimroth}},
  \bibinfo {author} {\bibfnamefont {G.}~\bibnamefont {Stollwerck}}, \ and\
  \bibinfo {author} {\bibfnamefont {O.}~\bibnamefont {Sulima}},\ }\href@noop {}
  {\bibfield  {journal} {\bibinfo  {journal} {Appl. Phys. A}\ }\textbf
  {\bibinfo {volume} {69}},\ \bibinfo {pages} {119--129} (\bibinfo {year}
  {1999})}\BibitemShut {NoStop}%
\bibitem [{\citenamefont {Wada}(1988)}]{wada}%
  \BibitemOpen
  \bibfield  {author} {\bibinfo {author} {\bibfnamefont {O.}~\bibnamefont
  {Wada}},\ }\href@noop {} {\bibfield  {journal} {\bibinfo  {journal} {Opt.
  Quantum Electron.}\ }\textbf {\bibinfo {volume} {20}},\ \bibinfo {pages}
  {441--474} (\bibinfo {year} {1988})}\BibitemShut {NoStop}%
\bibitem [{\citenamefont {Pilkuhn}\ and\ \citenamefont
  {Foster}(1966)}]{pilkuhn}%
  \BibitemOpen
  \bibfield  {author} {\bibinfo {author} {\bibfnamefont {M.}~\bibnamefont
  {Pilkuhn}}\ and\ \bibinfo {author} {\bibfnamefont {L.}~\bibnamefont
  {Foster}},\ }\href@noop {} {\bibfield  {journal} {\bibinfo  {journal} {IBM J.
  Res. Dev.}\ }\textbf {\bibinfo {volume} {10}},\ \bibinfo {pages} {122--129}
  (\bibinfo {year} {1966})}\BibitemShut {NoStop}%
\bibitem [{\citenamefont {Wilson}\ \emph {et~al.}(2020)\citenamefont {Wilson},
  \citenamefont {Schneider}, \citenamefont {H\"{o}nl}, \citenamefont
  {Anderson}, \citenamefont {Baumgartner}, \citenamefont {Czornomaz},
  \citenamefont {Kippenberg},\ and\ \citenamefont {Seidler}}]{wilson}%
  \BibitemOpen
  \bibfield  {author} {\bibinfo {author} {\bibfnamefont {D.}~\bibnamefont
  {Wilson}}, \bibinfo {author} {\bibfnamefont {K.}~\bibnamefont {Schneider}},
  \bibinfo {author} {\bibfnamefont {S.}~\bibnamefont {H\"{o}nl}}, \bibinfo
  {author} {\bibfnamefont {M.}~\bibnamefont {Anderson}}, \bibinfo {author}
  {\bibfnamefont {Y.}~\bibnamefont {Baumgartner}}, \bibinfo {author}
  {\bibfnamefont {L.}~\bibnamefont {Czornomaz}}, \bibinfo {author}
  {\bibfnamefont {T.}~\bibnamefont {Kippenberg}}, \ and\ \bibinfo {author}
  {\bibfnamefont {P.}~\bibnamefont {Seidler}},\ }\href@noop {} {\bibfield
  {journal} {\bibinfo  {journal} {Nat. Photonics}\ }\textbf {\bibinfo {volume}
  {14}},\ \bibinfo {pages} {57--62} (\bibinfo {year} {2020})}\BibitemShut
  {NoStop}%
\bibitem [{\citenamefont {Euser}\ and\ \citenamefont {Vos}(2005)}]{euser}%
  \BibitemOpen
  \bibfield  {author} {\bibinfo {author} {\bibfnamefont {T.}~\bibnamefont
  {Euser}}\ and\ \bibinfo {author} {\bibfnamefont {W.}~\bibnamefont {Vos}},\
  }\href@noop {} {\bibfield  {journal} {\bibinfo  {journal} {J. Appl. Phys.}\
  }\textbf {\bibinfo {volume} {97}},\ \bibinfo {pages} {043102} (\bibinfo
  {year} {2005})}\BibitemShut {NoStop}%
\bibitem [{\citenamefont {Vlasov}\ \emph {et~al.}(2001)\citenamefont {Vlasov},
  \citenamefont {Bo}, \citenamefont {Sturm},\ and\ \citenamefont
  {Norris}}]{vlasov}%
  \BibitemOpen
  \bibfield  {author} {\bibinfo {author} {\bibfnamefont {Y.}~\bibnamefont
  {Vlasov}}, \bibinfo {author} {\bibfnamefont {X.}~\bibnamefont {Bo}}, \bibinfo
  {author} {\bibfnamefont {J.}~\bibnamefont {Sturm}}, \ and\ \bibinfo {author}
  {\bibfnamefont {D.}~\bibnamefont {Norris}},\ }\href@noop {} {\bibfield
  {journal} {\bibinfo  {journal} {Nature}\ }\textbf {\bibinfo {volume} {414}},\
  \bibinfo {pages} {289--293} (\bibinfo {year} {2001})}\BibitemShut {NoStop}%
\bibitem [{\citenamefont {Rong}\ \emph {et~al.}(2005)\citenamefont {Rong},
  \citenamefont {Liu}, \citenamefont {Jones}, \citenamefont {Cohen},
  \citenamefont {Hak}, \citenamefont {Nicolescu}, \citenamefont {Fang},\ and\
  \citenamefont {Paniccia}}]{rong}%
  \BibitemOpen
  \bibfield  {author} {\bibinfo {author} {\bibfnamefont {H.}~\bibnamefont
  {Rong}}, \bibinfo {author} {\bibfnamefont {A.}~\bibnamefont {Liu}}, \bibinfo
  {author} {\bibfnamefont {R.}~\bibnamefont {Jones}}, \bibinfo {author}
  {\bibfnamefont {O.}~\bibnamefont {Cohen}}, \bibinfo {author} {\bibfnamefont
  {D.}~\bibnamefont {Hak}}, \bibinfo {author} {\bibfnamefont {R.}~\bibnamefont
  {Nicolescu}}, \bibinfo {author} {\bibfnamefont {A.}~\bibnamefont {Fang}}, \
  and\ \bibinfo {author} {\bibfnamefont {M.}~\bibnamefont {Paniccia}},\
  }\href@noop {} {\bibfield  {journal} {\bibinfo  {journal} {Nature}\ }\textbf
  {\bibinfo {volume} {433}},\ \bibinfo {pages} {725--728} (\bibinfo {year}
  {2005})}\BibitemShut {NoStop}%
\bibitem [{\citenamefont {Tsang}\ \emph {et~al.}(2002)\citenamefont {Tsang},
  \citenamefont {Wong}, \citenamefont {Liang}, \citenamefont {Day},
  \citenamefont {Roberts}, \citenamefont {Harpin}, \citenamefont {Drake},\ and\
  \citenamefont {Asghari}}]{tsang}%
  \BibitemOpen
  \bibfield  {author} {\bibinfo {author} {\bibfnamefont {H.}~\bibnamefont
  {Tsang}}, \bibinfo {author} {\bibfnamefont {C.}~\bibnamefont {Wong}},
  \bibinfo {author} {\bibfnamefont {T.}~\bibnamefont {Liang}}, \bibinfo
  {author} {\bibfnamefont {I.}~\bibnamefont {Day}}, \bibinfo {author}
  {\bibfnamefont {S.}~\bibnamefont {Roberts}}, \bibinfo {author} {\bibfnamefont
  {A.}~\bibnamefont {Harpin}}, \bibinfo {author} {\bibfnamefont
  {J.}~\bibnamefont {Drake}}, \ and\ \bibinfo {author} {\bibfnamefont
  {M.}~\bibnamefont {Asghari}},\ }\href@noop {} {\bibfield  {journal} {\bibinfo
   {journal} {Appl. Phys. Lett.}\ }\textbf {\bibinfo {volume} {80}},\ \bibinfo
  {pages} {416--418} (\bibinfo {year} {2002})}\BibitemShut {NoStop}%
\bibitem [{\citenamefont {Dvorak}\ \emph {et~al.}(1994)\citenamefont {Dvorak},
  \citenamefont {Schroeder}, \citenamefont {Andersen}, \citenamefont {Smirl},\
  and\ \citenamefont {Wherrett}}]{dvorak}%
  \BibitemOpen
  \bibfield  {author} {\bibinfo {author} {\bibfnamefont {M.}~\bibnamefont
  {Dvorak}}, \bibinfo {author} {\bibfnamefont {W.}~\bibnamefont {Schroeder}},
  \bibinfo {author} {\bibfnamefont {D.}~\bibnamefont {Andersen}}, \bibinfo
  {author} {\bibfnamefont {A.}~\bibnamefont {Smirl}}, \ and\ \bibinfo {author}
  {\bibfnamefont {B.}~\bibnamefont {Wherrett}},\ }\href@noop {} {\bibfield
  {journal} {\bibinfo  {journal} {IEEE J. Quantum Electron.}\ }\textbf
  {\bibinfo {volume} {30}},\ \bibinfo {pages} {256--268} (\bibinfo {year}
  {1994})}\BibitemShut {NoStop}%
\bibitem [{\citenamefont {Blanco}\ \emph {et~al.}(2000)\citenamefont {Blanco},
  \citenamefont {Chomski}, \citenamefont {Grabtchak}, \citenamefont {Ibisate},
  \citenamefont {John}, \citenamefont {Leonard}, \citenamefont {Lopez},
  \citenamefont {Meseguer}, \citenamefont {Miguez}, \citenamefont {Mondia},
  \citenamefont {Ozin}, \citenamefont {Toader},\ and\ \citenamefont {van
  Driel}}]{blanco}%
  \BibitemOpen
  \bibfield  {author} {\bibinfo {author} {\bibfnamefont {A.}~\bibnamefont
  {Blanco}}, \bibinfo {author} {\bibfnamefont {E.}~\bibnamefont {Chomski}},
  \bibinfo {author} {\bibfnamefont {S.}~\bibnamefont {Grabtchak}}, \bibinfo
  {author} {\bibfnamefont {M.}~\bibnamefont {Ibisate}}, \bibinfo {author}
  {\bibfnamefont {S.}~\bibnamefont {John}}, \bibinfo {author} {\bibfnamefont
  {S.}~\bibnamefont {Leonard}}, \bibinfo {author} {\bibfnamefont
  {C.}~\bibnamefont {Lopez}}, \bibinfo {author} {\bibfnamefont
  {F.}~\bibnamefont {Meseguer}}, \bibinfo {author} {\bibfnamefont
  {H.}~\bibnamefont {Miguez}}, \bibinfo {author} {\bibfnamefont
  {J.}~\bibnamefont {Mondia}}, \bibinfo {author} {\bibfnamefont
  {G.}~\bibnamefont {Ozin}}, \bibinfo {author} {\bibfnamefont {O.}~\bibnamefont
  {Toader}}, \ and\ \bibinfo {author} {\bibfnamefont {H.}~\bibnamefont {van
  Driel}},\ }\href@noop {} {\bibfield  {journal} {\bibinfo  {journal} {Nature}\
  }\textbf {\bibinfo {volume} {405}},\ \bibinfo {pages} {437--440} (\bibinfo
  {year} {2000})}\BibitemShut {NoStop}%
\bibitem [{\citenamefont {Bristow}, \citenamefont {Rotenberg},\ and\
  \citenamefont {van Driel}(2007)}]{bristow}%
  \BibitemOpen
  \bibfield  {author} {\bibinfo {author} {\bibfnamefont {A.}~\bibnamefont
  {Bristow}}, \bibinfo {author} {\bibfnamefont {N.}~\bibnamefont {Rotenberg}},
  \ and\ \bibinfo {author} {\bibfnamefont {H.}~\bibnamefont {van Driel}},\
  }\href@noop {} {\bibfield  {journal} {\bibinfo  {journal} {Appl. Phys.
  Lett.}\ }\textbf {\bibinfo {volume} {90}},\ \bibinfo {pages} {191104}
  (\bibinfo {year} {2007})}\BibitemShut {NoStop}%
\bibitem [{\citenamefont {Bassani}\ and\ \citenamefont
  {Scandolo}(1991)}]{bassani}%
  \BibitemOpen
  \bibfield  {author} {\bibinfo {author} {\bibfnamefont {F.}~\bibnamefont
  {Bassani}}\ and\ \bibinfo {author} {\bibfnamefont {S.}~\bibnamefont
  {Scandolo}},\ }\href@noop {} {\bibfield  {journal} {\bibinfo  {journal}
  {Phys. Rev. B}\ }\textbf {\bibinfo {volume} {44}},\ \bibinfo {pages}
  {8446--8453} (\bibinfo {year} {1991})}\BibitemShut {NoStop}%
\bibitem [{\citenamefont {Caspers}(1964)}]{caspers}%
  \BibitemOpen
  \bibfield  {author} {\bibinfo {author} {\bibfnamefont {P.}~\bibnamefont
  {Caspers}},\ }\href@noop {} {\bibfield  {journal} {\bibinfo  {journal} {Phys.
  Rev. A}\ }\textbf {\bibinfo {volume} {133}},\ \bibinfo {pages} {1249}
  (\bibinfo {year} {1964})}\BibitemShut {NoStop}%
\bibitem [{\citenamefont {Hutchings}\ \emph {et~al.}(1992)\citenamefont
  {Hutchings}, \citenamefont {Sheik-Bahae}, \citenamefont {Hagan},\ and\
  \citenamefont {Stryland}}]{hutchings2}%
  \BibitemOpen
  \bibfield  {author} {\bibinfo {author} {\bibfnamefont {D.}~\bibnamefont
  {Hutchings}}, \bibinfo {author} {\bibfnamefont {M.}~\bibnamefont
  {Sheik-Bahae}}, \bibinfo {author} {\bibfnamefont {D.}~\bibnamefont {Hagan}},
  \ and\ \bibinfo {author} {\bibfnamefont {E.~V.}\ \bibnamefont {Stryland}},\
  }\href@noop {} {\bibfield  {journal} {\bibinfo  {journal} {Opt. Quantum
  Electron.}\ }\textbf {\bibinfo {volume} {24}},\ \bibinfo {pages} {1--30}
  (\bibinfo {year} {1992})}\BibitemShut {NoStop}%
\bibitem [{\citenamefont {Kogan}(1963)}]{kogan}%
  \BibitemOpen
  \bibfield  {author} {\bibinfo {author} {\bibfnamefont {S.}~\bibnamefont
  {Kogan}},\ }\href@noop {} {\bibfield  {journal} {\bibinfo  {journal} {Sov.
  Phys. JETP}\ }\textbf {\bibinfo {volume} {16}},\ \bibinfo {pages} {217}
  (\bibinfo {year} {1963})}\BibitemShut {NoStop}%
\bibitem [{\citenamefont {Gibbs}(1972)}]{nussenzweig}%
  \BibitemOpen
  \bibfield  {author} {\bibinfo {author} {\bibfnamefont {H.}~\bibnamefont
  {Gibbs}},\ }\href@noop {} {\emph {\bibinfo {title} {{Causality} {and}
  {Dispersion} {Relations}}}}\ (\bibinfo  {publisher} {Academic Press},\
  \bibinfo {address} {New York},\ \bibinfo {year} {1972})\BibitemShut {NoStop}%
\bibitem [{\citenamefont {Price}(1964)}]{price}%
  \BibitemOpen
  \bibfield  {author} {\bibinfo {author} {\bibfnamefont {P.}~\bibnamefont
  {Price}},\ }\href@noop {} {\bibfield  {journal} {\bibinfo  {journal} {Phys.
  Rev.}\ }\textbf {\bibinfo {volume} {130}},\ \bibinfo {pages} {1792} (\bibinfo
  {year} {1964})}\BibitemShut {NoStop}%
\bibitem [{\citenamefont {Jr.}\ and\ \citenamefont {Jr.}(1975)}]{ridener}%
  \BibitemOpen
  \bibfield  {author} {\bibinfo {author} {\bibfnamefont {F.~R.}\ \bibnamefont
  {Jr.}}\ and\ \bibinfo {author} {\bibfnamefont {R.~G.}\ \bibnamefont {Jr.}},\
  }\href@noop {} {\bibfield  {journal} {\bibinfo  {journal} {Phys. Rev. B}\
  }\textbf {\bibinfo {volume} {11}},\ \bibinfo {pages} {2768} (\bibinfo {year}
  {1975})}\BibitemShut {NoStop}%
\bibitem [{\citenamefont {Sheik-Bahae}, \citenamefont {Hagan},\ and\
  \citenamefont {Styland}(1990)}]{sheikbahae}%
  \BibitemOpen
  \bibfield  {author} {\bibinfo {author} {\bibfnamefont {M.}~\bibnamefont
  {Sheik-Bahae}}, \bibinfo {author} {\bibfnamefont {D.}~\bibnamefont {Hagan}},
  \ and\ \bibinfo {author} {\bibfnamefont {E.~V.}\ \bibnamefont {Styland}},\
  }\href@noop {} {\bibfield  {journal} {\bibinfo  {journal} {Phys. Rev. Lett.}\
  }\textbf {\bibinfo {volume} {65}},\ \bibinfo {pages} {96--99} (\bibinfo
  {year} {1990})}\BibitemShut {NoStop}%
\bibitem [{\citenamefont {Sheik-Bahae}\ \emph {et~al.}(1991)\citenamefont
  {Sheik-Bahae}, \citenamefont {Hutchings}, \citenamefont {Hagan},\ and\
  \citenamefont {Stryland}}]{sheikbahae2}%
  \BibitemOpen
  \bibfield  {author} {\bibinfo {author} {\bibfnamefont {M.}~\bibnamefont
  {Sheik-Bahae}}, \bibinfo {author} {\bibfnamefont {D.}~\bibnamefont
  {Hutchings}}, \bibinfo {author} {\bibfnamefont {D.}~\bibnamefont {Hagan}}, \
  and\ \bibinfo {author} {\bibfnamefont {E.~V.}\ \bibnamefont {Stryland}},\
  }\href@noop {} {\bibfield  {journal} {\bibinfo  {journal} {IEEE J. Quantum
  Electron.}\ }\textbf {\bibinfo {volume} {27}},\ \bibinfo {pages} {1296--1309}
  (\bibinfo {year} {1991})}\BibitemShut {NoStop}%
\bibitem [{\citenamefont {Toll}(1956)}]{toll}%
  \BibitemOpen
  \bibfield  {author} {\bibinfo {author} {\bibfnamefont {J.}~\bibnamefont
  {Toll}},\ }\href@noop {} {\bibfield  {journal} {\bibinfo  {journal} {Phys.
  Rev.}\ }\textbf {\bibinfo {volume} {104}},\ \bibinfo {pages} {1760} (\bibinfo
  {year} {1956})}\BibitemShut {NoStop}%
\bibitem [{\citenamefont {Gibbs}(1985)}]{gibbs}%
  \BibitemOpen
  \bibfield  {author} {\bibinfo {author} {\bibfnamefont {H.}~\bibnamefont
  {Gibbs}},\ }\href@noop {} {\emph {\bibinfo {title} {{Optical} {Bistability:}
  {Controlling} {Light} {with} {Light}}}}\ (\bibinfo  {publisher} {Academic
  Press},\ \bibinfo {address} {Orlando},\ \bibinfo {year} {1985})\BibitemShut
  {NoStop}%
\bibitem [{\citenamefont {Stegeman}\ and\ \citenamefont
  {Wright}(1990)}]{stegeman}%
  \BibitemOpen
  \bibfield  {author} {\bibinfo {author} {\bibfnamefont {G.}~\bibnamefont
  {Stegeman}}\ and\ \bibinfo {author} {\bibfnamefont {E.}~\bibnamefont
  {Wright}},\ }\href@noop {} {\bibfield  {journal} {\bibinfo  {journal} {Opt.
  Quantum Electron.}\ }\textbf {\bibinfo {volume} {22}},\ \bibinfo {pages}
  {95--122} (\bibinfo {year} {1990})}\BibitemShut {NoStop}%
\bibitem [{\citenamefont {Reintjes}\ and\ \citenamefont
  {McGroddy}(1973)}]{reintjes}%
  \BibitemOpen
  \bibfield  {author} {\bibinfo {author} {\bibfnamefont {J.}~\bibnamefont
  {Reintjes}}\ and\ \bibinfo {author} {\bibfnamefont {J.}~\bibnamefont
  {McGroddy}},\ }\href@noop {} {\bibfield  {journal} {\bibinfo  {journal}
  {Phys. Rev. Lett.}\ }\textbf {\bibinfo {volume} {30}},\ \bibinfo {pages}
  {901--903} (\bibinfo {year} {1973})}\BibitemShut {NoStop}%
\bibitem [{\citenamefont {Bechtel}\ and\ \citenamefont
  {Smith}(1976)}]{bechtel}%
  \BibitemOpen
  \bibfield  {author} {\bibinfo {author} {\bibfnamefont {J.}~\bibnamefont
  {Bechtel}}\ and\ \bibinfo {author} {\bibfnamefont {W.}~\bibnamefont
  {Smith}},\ }\href@noop {} {\bibfield  {journal} {\bibinfo  {journal} {Phys.
  Rev. B}\ }\textbf {\bibinfo {volume} {13}},\ \bibinfo {pages} {3515--3522}
  (\bibinfo {year} {1976})}\BibitemShut {NoStop}%
\bibitem [{\citenamefont {Bosacchi}, \citenamefont {Bessey},\ and\
  \citenamefont {Jain}(1978)}]{bosacchi}%
  \BibitemOpen
  \bibfield  {author} {\bibinfo {author} {\bibfnamefont {B.}~\bibnamefont
  {Bosacchi}}, \bibinfo {author} {\bibfnamefont {J.}~\bibnamefont {Bessey}}, \
  and\ \bibinfo {author} {\bibfnamefont {F.}~\bibnamefont {Jain}},\ }\href@noop
  {} {\bibfield  {journal} {\bibinfo  {journal} {J. Appl. Phys.}\ }\textbf
  {\bibinfo {volume} {49}},\ \bibinfo {pages} {4609--4611} (\bibinfo {year}
  {1978})}\BibitemShut {NoStop}%
\bibitem [{\citenamefont {Stryland}\ \emph {et~al.}(1994)\citenamefont
  {Stryland}, \citenamefont {Sheik-Bahae}, \citenamefont {Xia}, \citenamefont
  {Wamsley}, \citenamefont {Wang}, \citenamefont {Said},\ and\ \citenamefont
  {Hagan}}]{vanstryland}%
  \BibitemOpen
  \bibfield  {author} {\bibinfo {author} {\bibfnamefont {E.~V.}\ \bibnamefont
  {Stryland}}, \bibinfo {author} {\bibfnamefont {M.}~\bibnamefont
  {Sheik-Bahae}}, \bibinfo {author} {\bibfnamefont {T.}~\bibnamefont {Xia}},
  \bibinfo {author} {\bibfnamefont {C.}~\bibnamefont {Wamsley}}, \bibinfo
  {author} {\bibfnamefont {Z.}~\bibnamefont {Wang}}, \bibinfo {author}
  {\bibfnamefont {A.}~\bibnamefont {Said}}, \ and\ \bibinfo {author}
  {\bibfnamefont {D.}~\bibnamefont {Hagan}},\ }\href@noop {} {\bibfield
  {journal} {\bibinfo  {journal} {Int. J. Nonlinear Opt. Phys.}\ }\textbf
  {\bibinfo {volume} {3}},\ \bibinfo {pages} {489--500} (\bibinfo {year}
  {1994})}\BibitemShut {NoStop}%
\bibitem [{\citenamefont {Sheik-Bahae}\ and\ \citenamefont
  {Stryland}(1999)}]{sheikbahae3}%
  \BibitemOpen
  \bibfield  {author} {\bibinfo {author} {\bibfnamefont {M.}~\bibnamefont
  {Sheik-Bahae}}\ and\ \bibinfo {author} {\bibfnamefont {E.~V.}\ \bibnamefont
  {Stryland}},\ }\href@noop {} {\emph {\bibinfo {title} {{Semiconductors} {and}
  {Semimetals}}}},\ edited by\ \bibinfo {editor} {\bibfnamefont
  {R.}~\bibnamefont {Willardson}}\ and\ \bibinfo {editor} {\bibfnamefont
  {E.}~\bibnamefont {Weber}},\ Vol.~\bibinfo {volume} {58}\ (\bibinfo
  {publisher} {Academic Press},\ \bibinfo {address} {San Diego},\ \bibinfo
  {year} {1999})\ pp.\ \bibinfo {pages} {257--318}\BibitemShut {NoStop}%
\bibitem [{\citenamefont {Reitze}\ \emph {et~al.}(1990)\citenamefont {Reitze},
  \citenamefont {Zhang}, \citenamefont {Wood},\ and\ \citenamefont
  {Downer}}]{reitze}%
  \BibitemOpen
  \bibfield  {author} {\bibinfo {author} {\bibfnamefont {D.}~\bibnamefont
  {Reitze}}, \bibinfo {author} {\bibfnamefont {T.}~\bibnamefont {Zhang}},
  \bibinfo {author} {\bibfnamefont {W.}~\bibnamefont {Wood}}, \ and\ \bibinfo
  {author} {\bibfnamefont {M.}~\bibnamefont {Downer}},\ }\href@noop {}
  {\bibfield  {journal} {\bibinfo  {journal} {J. Opt. Soc. Am. B}\ }\textbf
  {\bibinfo {volume} {7}},\ \bibinfo {pages} {84--89} (\bibinfo {year}
  {1990})}\BibitemShut {NoStop}%
\bibitem [{\citenamefont {Hurlbut}\ \emph {et~al.}(2007)\citenamefont
  {Hurlbut}, \citenamefont {Lee}, \citenamefont {Vodopyanov}, \citenamefont
  {Kuo},\ and\ \citenamefont {Fejer}}]{hurlbut}%
  \BibitemOpen
  \bibfield  {author} {\bibinfo {author} {\bibfnamefont {W.}~\bibnamefont
  {Hurlbut}}, \bibinfo {author} {\bibfnamefont {Y.}~\bibnamefont {Lee}},
  \bibinfo {author} {\bibfnamefont {K.}~\bibnamefont {Vodopyanov}}, \bibinfo
  {author} {\bibfnamefont {P.}~\bibnamefont {Kuo}}, \ and\ \bibinfo {author}
  {\bibfnamefont {M.}~\bibnamefont {Fejer}},\ }\href@noop {} {\bibfield
  {journal} {\bibinfo  {journal} {Opt. Lett.}\ }\textbf {\bibinfo {volume}
  {32}},\ \bibinfo {pages} {668--670} (\bibinfo {year} {2007})}\BibitemShut
  {NoStop}%
\bibitem [{\citenamefont {Fishman}\ \emph {et~al.}(2011)\citenamefont
  {Fishman}, \citenamefont {Cirloganu}, \citenamefont {Webster}, \citenamefont
  {Padilha}, \citenamefont {Monroe}, \citenamefont {Hagan},\ and\ \citenamefont
  {Stryland}}]{fishman}%
  \BibitemOpen
  \bibfield  {author} {\bibinfo {author} {\bibfnamefont {D.}~\bibnamefont
  {Fishman}}, \bibinfo {author} {\bibfnamefont {C.}~\bibnamefont {Cirloganu}},
  \bibinfo {author} {\bibfnamefont {S.}~\bibnamefont {Webster}}, \bibinfo
  {author} {\bibfnamefont {L.}~\bibnamefont {Padilha}}, \bibinfo {author}
  {\bibfnamefont {M.}~\bibnamefont {Monroe}}, \bibinfo {author} {\bibfnamefont
  {D.}~\bibnamefont {Hagan}}, \ and\ \bibinfo {author} {\bibfnamefont {E.~V.}\
  \bibnamefont {Stryland}},\ }\href@noop {} {\bibfield  {journal} {\bibinfo
  {journal} {Nat. Photonics}\ }\textbf {\bibinfo {volume} {5}},\ \bibinfo
  {pages} {561--565} (\bibinfo {year} {2011})}\BibitemShut {NoStop}%
\bibitem [{\citenamefont {Shank}, \citenamefont {Ippen},\ and\ \citenamefont
  {Shapiro}(1978)}]{shank}%
  \BibitemOpen
  \bibinfo {editor} {\bibfnamefont {C.}~\bibnamefont {Shank}}, \bibinfo
  {editor} {\bibfnamefont {E.}~\bibnamefont {Ippen}}, \ and\ \bibinfo {editor}
  {\bibfnamefont {S.}~\bibnamefont {Shapiro}},\ eds.,\ \href@noop {} {\emph
  {\bibinfo {title} {{Picosecond} {Phenomena}}}}\ (\bibinfo  {publisher}
  {Springer-Verlag},\ \bibinfo {address} {Hilton Head},\ \bibinfo {year}
  {1978})\BibitemShut {NoStop}%
\bibitem [{\citenamefont {Benis}\ \emph {et~al.}(2020)\citenamefont {Benis},
  \citenamefont {Cirloganu}, \citenamefont {Cox}, \citenamefont {Ensley},
  \citenamefont {Hu}, \citenamefont {Nootz}, \citenamefont {Olszak},
  \citenamefont {Padilha}, \citenamefont {Peceli}, \citenamefont {Reichert},
  \citenamefont {Webster}, \citenamefont {Woodall}, \citenamefont {Hagan},\
  and\ \citenamefont {Stryland}}]{benis}%
  \BibitemOpen
  \bibfield  {author} {\bibinfo {author} {\bibfnamefont {S.}~\bibnamefont
  {Benis}}, \bibinfo {author} {\bibfnamefont {C.}~\bibnamefont {Cirloganu}},
  \bibinfo {author} {\bibfnamefont {N.}~\bibnamefont {Cox}}, \bibinfo {author}
  {\bibfnamefont {T.}~\bibnamefont {Ensley}}, \bibinfo {author} {\bibfnamefont
  {H.}~\bibnamefont {Hu}}, \bibinfo {author} {\bibfnamefont {G.}~\bibnamefont
  {Nootz}}, \bibinfo {author} {\bibfnamefont {P.}~\bibnamefont {Olszak}},
  \bibinfo {author} {\bibfnamefont {L.}~\bibnamefont {Padilha}}, \bibinfo
  {author} {\bibfnamefont {D.}~\bibnamefont {Peceli}}, \bibinfo {author}
  {\bibfnamefont {M.}~\bibnamefont {Reichert}}, \bibinfo {author}
  {\bibfnamefont {S.}~\bibnamefont {Webster}}, \bibinfo {author} {\bibfnamefont
  {M.}~\bibnamefont {Woodall}}, \bibinfo {author} {\bibfnamefont
  {D.}~\bibnamefont {Hagan}}, \ and\ \bibinfo {author} {\bibfnamefont {E.~V.}\
  \bibnamefont {Stryland}},\ }\href@noop {} {\bibfield  {journal} {\bibinfo
  {journal} {Optica}\ }\textbf {\bibinfo {volume} {7}},\ \bibinfo {pages}
  {888--899} (\bibinfo {year} {2020})}\BibitemShut {NoStop}%
\bibitem [{\citenamefont {Cirloganu}(2010)}]{cirloganu}%
  \BibitemOpen
  \bibfield  {author} {\bibinfo {author} {\bibfnamefont {C.}~\bibnamefont
  {Cirloganu}},\ }\emph {\bibinfo {title} {Experimental And Theoretical
  Approaches To Characterization Of Electronic Nonlinearities In Direct--gap
  Semiconductors}},\ \href@noop {} {Ph.D. thesis},\ \bibinfo  {school}
  {University of Central Florida} (\bibinfo {year} {2010})\BibitemShut
  {NoStop}%
\bibitem [{\citenamefont {Boyd}(2008)}]{boyd}%
  \BibitemOpen
  \bibfield  {author} {\bibinfo {author} {\bibfnamefont {R.}~\bibnamefont
  {Boyd}},\ }\href@noop {} {\emph {\bibinfo {title} {{Nonlinear} {Optics}}}},\
  \bibinfo {edition} {3rd}\ ed.\ (\bibinfo  {publisher} {Academic Press -
  Elsevier, Inc.},\ \bibinfo {address} {Burlington},\ \bibinfo {year}
  {2008})\BibitemShut {NoStop}%
\bibitem [{\citenamefont {Popov}, \citenamefont {Svirko},\ and\ \citenamefont
  {Zheludev}(1995)}]{popov}%
  \BibitemOpen
  \bibfield  {author} {\bibinfo {author} {\bibfnamefont {S.}~\bibnamefont
  {Popov}}, \bibinfo {author} {\bibfnamefont {Y.}~\bibnamefont {Svirko}}, \
  and\ \bibinfo {author} {\bibfnamefont {N.}~\bibnamefont {Zheludev}},\
  }\href@noop {} {\emph {\bibinfo {title} {{Susceptibility} {Tensors} {for}
  {Nonlinear} {Optics}}}},\ Optics and Optoelectronics Series\ (\bibinfo
  {publisher} {Institute of Physics Publishing},\ \bibinfo {address}
  {Bristol},\ \bibinfo {year} {1995})\BibitemShut {NoStop}%
\bibitem [{\citenamefont {Kleinman}(1962)}]{kleinman1}%
  \BibitemOpen
  \bibfield  {author} {\bibinfo {author} {\bibfnamefont {D.}~\bibnamefont
  {Kleinman}},\ }\href@noop {} {\bibfield  {journal} {\bibinfo  {journal}
  {Phys. Rev.}\ }\textbf {\bibinfo {volume} {126}},\ \bibinfo {pages} {1977}
  (\bibinfo {year} {1962})}\BibitemShut {NoStop}%
\bibitem [{\citenamefont {Murayama}\ and\ \citenamefont
  {Nakayama}(1995)}]{murayama}%
  \BibitemOpen
  \bibfield  {author} {\bibinfo {author} {\bibfnamefont {M.}~\bibnamefont
  {Murayama}}\ and\ \bibinfo {author} {\bibfnamefont {T.}~\bibnamefont
  {Nakayama}},\ }\href@noop {} {\bibfield  {journal} {\bibinfo  {journal}
  {Phys. Rev. B}\ }\textbf {\bibinfo {volume} {52}},\ \bibinfo {pages}
  {4986--4997} (\bibinfo {year} {1995})}\BibitemShut {NoStop}%
\bibitem [{\citenamefont {Noffsinger}\ \emph {et~al.}(2012)\citenamefont
  {Noffsinger}, \citenamefont {Kioupakis}, \citenamefont {de~Welle},
  \citenamefont {Louie},\ and\ \citenamefont {Cohen}}]{noffsinger}%
  \BibitemOpen
  \bibfield  {author} {\bibinfo {author} {\bibfnamefont {J.}~\bibnamefont
  {Noffsinger}}, \bibinfo {author} {\bibfnamefont {E.}~\bibnamefont
  {Kioupakis}}, \bibinfo {author} {\bibfnamefont {C.~V.}\ \bibnamefont
  {de~Welle}}, \bibinfo {author} {\bibfnamefont {S.}~\bibnamefont {Louie}}, \
  and\ \bibinfo {author} {\bibfnamefont {M.}~\bibnamefont {Cohen}},\
  }\href@noop {} {\bibfield  {journal} {\bibinfo  {journal} {Phys. Rev. Lett.}\
  }\textbf {\bibinfo {volume} {108}},\ \bibinfo {pages} {167402--167407}
  (\bibinfo {year} {2012})}\BibitemShut {NoStop}%
\bibitem [{\citenamefont {Jones}\ and\ \citenamefont {Reiss}(1977)}]{jones}%
  \BibitemOpen
  \bibfield  {author} {\bibinfo {author} {\bibfnamefont {H.}~\bibnamefont
  {Jones}}\ and\ \bibinfo {author} {\bibfnamefont {H.}~\bibnamefont {Reiss}},\
  }\href@noop {} {\bibfield  {journal} {\bibinfo  {journal} {Phys. Rev. B}\
  }\textbf {\bibinfo {volume} {16}},\ \bibinfo {pages} {2466--2473} (\bibinfo
  {year} {1977})}\BibitemShut {NoStop}%
\bibitem [{\citenamefont {Wherrett}(1984)}]{wherrett}%
  \BibitemOpen
  \bibfield  {author} {\bibinfo {author} {\bibfnamefont {B.}~\bibnamefont
  {Wherrett}},\ }\href@noop {} {\bibfield  {journal} {\bibinfo  {journal} {J.
  Opt. Soc. Am. B}\ }\textbf {\bibinfo {volume} {1}},\ \bibinfo {pages}
  {67--72} (\bibinfo {year} {1984})}\BibitemShut {NoStop}%
\bibitem [{\citenamefont {Garcia}\ and\ \citenamefont
  {Kalyanaraman}(2006)}]{garcia1}%
  \BibitemOpen
  \bibfield  {author} {\bibinfo {author} {\bibfnamefont {H.}~\bibnamefont
  {Garcia}}\ and\ \bibinfo {author} {\bibfnamefont {R.}~\bibnamefont
  {Kalyanaraman}},\ }\href@noop {} {\bibfield  {journal} {\bibinfo  {journal}
  {J. Phys. B}\ }\textbf {\bibinfo {volume} {39}},\ \bibinfo {pages}
  {2737--2746} (\bibinfo {year} {2006})}\BibitemShut {NoStop}%
\bibitem [{\citenamefont {Garcia}(2006)}]{garcia2}%
  \BibitemOpen
  \bibfield  {author} {\bibinfo {author} {\bibfnamefont {H.}~\bibnamefont
  {Garcia}},\ }\href@noop {} {\bibfield  {journal} {\bibinfo  {journal} {Phys.
  Rev. B}\ }\textbf {\bibinfo {volume} {74}},\ \bibinfo {pages} {035212}
  (\bibinfo {year} {2006})}\BibitemShut {NoStop}%
\bibitem [{\citenamefont {Aspnes}(1972)}]{aspnes1}%
  \BibitemOpen
  \bibfield  {author} {\bibinfo {author} {\bibfnamefont {D.}~\bibnamefont
  {Aspnes}},\ }\href@noop {} {\bibfield  {journal} {\bibinfo  {journal} {Phys.
  Rev. B}\ }\textbf {\bibinfo {volume} {6}},\ \bibinfo {pages} {4648--4659}
  (\bibinfo {year} {1972})}\BibitemShut {NoStop}%
\bibitem [{\citenamefont {Moss}, \citenamefont {Sipe},\ and\ \citenamefont {van
  Driel}(1987)}]{moss}%
  \BibitemOpen
  \bibfield  {author} {\bibinfo {author} {\bibfnamefont {D.}~\bibnamefont
  {Moss}}, \bibinfo {author} {\bibfnamefont {J.}~\bibnamefont {Sipe}}, \ and\
  \bibinfo {author} {\bibfnamefont {H.}~\bibnamefont {van Driel}},\ }\href@noop
  {} {\bibfield  {journal} {\bibinfo  {journal} {Phys. Rev. B}\ }\textbf
  {\bibinfo {volume} {36}},\ \bibinfo {pages} {9708--9721} (\bibinfo {year}
  {1987})}\BibitemShut {NoStop}%
\bibitem [{\citenamefont {Zallen}\ and\ \citenamefont {Paul}(1964)}]{zallen}%
  \BibitemOpen
  \bibfield  {author} {\bibinfo {author} {\bibfnamefont {R.}~\bibnamefont
  {Zallen}}\ and\ \bibinfo {author} {\bibfnamefont {W.}~\bibnamefont {Paul}},\
  }\href@noop {} {\bibfield  {journal} {\bibinfo  {journal} {Phys. Rev.}\
  }\textbf {\bibinfo {volume} {134}},\ \bibinfo {pages} {A1628--A1641}
  (\bibinfo {year} {1964})}\BibitemShut {NoStop}%
\bibitem [{\citenamefont {Diels}\ and\ \citenamefont {Rudolph}(1996)}]{diels}%
  \BibitemOpen
  \bibfield  {author} {\bibinfo {author} {\bibfnamefont {J.}~\bibnamefont
  {Diels}}\ and\ \bibinfo {author} {\bibfnamefont {W.}~\bibnamefont
  {Rudolph}},\ }\href@noop {} {\emph {\bibinfo {title} {{Ultrafaser} {Laser}
  {Pulse} {Phenomena:} {Fundamentals,} {Techniques,} {and} {Applications} {on}
  {a} {Femtosecond} {Time} {Scale}}}}\ (\bibinfo  {publisher} {Academic
  Press},\ \bibinfo {address} {San Diego},\ \bibinfo {year} {1996})\BibitemShut
  {NoStop}%
\bibitem [{\citenamefont {Skauli}\ \emph {et~al.}(2003)\citenamefont {Skauli},
  \citenamefont {Kuo}, \citenamefont {Vodopyanov}, \citenamefont {Pinguet},
  \citenamefont {Levi}, \citenamefont {Eyres}, \citenamefont {Harris},
  \citenamefont {Fejer}, \citenamefont {Gerard}, \citenamefont {Becouarn},\
  and\ \citenamefont {Lallier}}]{skauli}%
  \BibitemOpen
  \bibfield  {author} {\bibinfo {author} {\bibfnamefont {T.}~\bibnamefont
  {Skauli}}, \bibinfo {author} {\bibfnamefont {P.}~\bibnamefont {Kuo}},
  \bibinfo {author} {\bibfnamefont {K.}~\bibnamefont {Vodopyanov}}, \bibinfo
  {author} {\bibfnamefont {T.}~\bibnamefont {Pinguet}}, \bibinfo {author}
  {\bibfnamefont {O.}~\bibnamefont {Levi}}, \bibinfo {author} {\bibfnamefont
  {L.}~\bibnamefont {Eyres}}, \bibinfo {author} {\bibfnamefont
  {J.}~\bibnamefont {Harris}}, \bibinfo {author} {\bibfnamefont
  {M.}~\bibnamefont {Fejer}}, \bibinfo {author} {\bibfnamefont
  {B.}~\bibnamefont {Gerard}}, \bibinfo {author} {\bibfnamefont
  {L.}~\bibnamefont {Becouarn}}, \ and\ \bibinfo {author} {\bibfnamefont
  {E.}~\bibnamefont {Lallier}},\ }\href@noop {} {\bibfield  {journal} {\bibinfo
   {journal} {J. Appl. Phys.}\ }\textbf {\bibinfo {volume} {94}},\ \bibinfo
  {pages} {6447--6455} (\bibinfo {year} {2003})}\BibitemShut {NoStop}%
\bibitem [{\citenamefont {Li}(1980)}]{li}%
  \BibitemOpen
  \bibfield  {author} {\bibinfo {author} {\bibfnamefont {H.}~\bibnamefont
  {Li}},\ }\href@noop {} {\bibfield  {journal} {\bibinfo  {journal} {J. Phys.
  Chem. Ref. Data}\ }\textbf {\bibinfo {volume} {9}},\ \bibinfo {pages}
  {561--658} (\bibinfo {year} {1980})}\BibitemShut {NoStop}%
\bibitem [{\citenamefont {Smirl}\ \emph {et~al.}(1988)\citenamefont {Smirl},
  \citenamefont {Valley}, \citenamefont {Bohnert},\ and\ \citenamefont
  {T.F.~Boggess}}]{smirl}%
  \BibitemOpen
  \bibfield  {author} {\bibinfo {author} {\bibfnamefont {A.}~\bibnamefont
  {Smirl}}, \bibinfo {author} {\bibfnamefont {G.}~\bibnamefont {Valley}},
  \bibinfo {author} {\bibfnamefont {K.}~\bibnamefont {Bohnert}}, \ and\
  \bibinfo {author} {\bibfnamefont {J.}~\bibnamefont {T.F.~Boggess}},\
  }\href@noop {} {\bibfield  {journal} {\bibinfo  {journal} {IEEE J. Quantum
  Electron.}\ }\textbf {\bibinfo {volume} {24}},\ \bibinfo {pages} {289--303}
  (\bibinfo {year} {1988})}\BibitemShut {NoStop}%
\bibitem [{\citenamefont {Valley}\ and\ \citenamefont {Smirl}(1988)}]{valley}%
  \BibitemOpen
  \bibfield  {author} {\bibinfo {author} {\bibfnamefont {G.}~\bibnamefont
  {Valley}}\ and\ \bibinfo {author} {\bibfnamefont {A.}~\bibnamefont {Smirl}},\
  }\href@noop {} {\bibfield  {journal} {\bibinfo  {journal} {IEEE J. Quantum
  Electron.}\ }\textbf {\bibinfo {volume} {24}},\ \bibinfo {pages} {304--310}
  (\bibinfo {year} {1988})}\BibitemShut {NoStop}%
\bibitem [{\citenamefont {Hutchings}\ and\ \citenamefont
  {Stryland}(1992)}]{hutchings}%
  \BibitemOpen
  \bibfield  {author} {\bibinfo {author} {\bibfnamefont {D.}~\bibnamefont
  {Hutchings}}\ and\ \bibinfo {author} {\bibfnamefont {E.~V.}\ \bibnamefont
  {Stryland}},\ }\href@noop {} {\bibfield  {journal} {\bibinfo  {journal} {J.
  Opt. Soc. Am. B}\ }\textbf {\bibinfo {volume} {9}},\ \bibinfo {pages}
  {2065--2074} (\bibinfo {year} {1992})}\BibitemShut {NoStop}%
\bibitem [{\citenamefont {Bolger}\ \emph {et~al.}(1993)\citenamefont {Bolger},
  \citenamefont {Kar}, \citenamefont {Wherrett}, \citenamefont {DeSalvo},
  \citenamefont {Hutchings},\ and\ \citenamefont {Hagan}}]{bolger}%
  \BibitemOpen
  \bibfield  {author} {\bibinfo {author} {\bibfnamefont {J.}~\bibnamefont
  {Bolger}}, \bibinfo {author} {\bibfnamefont {A.}~\bibnamefont {Kar}},
  \bibinfo {author} {\bibfnamefont {B.}~\bibnamefont {Wherrett}}, \bibinfo
  {author} {\bibfnamefont {R.}~\bibnamefont {DeSalvo}}, \bibinfo {author}
  {\bibfnamefont {D.}~\bibnamefont {Hutchings}}, \ and\ \bibinfo {author}
  {\bibfnamefont {D.}~\bibnamefont {Hagan}},\ }\href@noop {} {\bibfield
  {journal} {\bibinfo  {journal} {Optics Commun.}\ }\textbf {\bibinfo {volume}
  {97}},\ \bibinfo {pages} {203} (\bibinfo {year} {1993})}\BibitemShut
  {NoStop}%
\bibitem [{\citenamefont {Hannes}\ \emph {et~al.}(2019)\citenamefont {Hannes},
  \citenamefont {Krau{\ss}-Kodytek}, \citenamefont {Ruppert}, \citenamefont
  {Betz},\ and\ \citenamefont {Meier}}]{hannes}%
  \BibitemOpen
  \bibfield  {author} {\bibinfo {author} {\bibfnamefont {W.-R.}\ \bibnamefont
  {Hannes}}, \bibinfo {author} {\bibfnamefont {L.}~\bibnamefont
  {Krau{\ss}-Kodytek}}, \bibinfo {author} {\bibfnamefont {C.}~\bibnamefont
  {Ruppert}}, \bibinfo {author} {\bibfnamefont {M.}~\bibnamefont {Betz}}, \
  and\ \bibinfo {author} {\bibfnamefont {T.}~\bibnamefont {Meier}},\ }in\
  \href@noop {} {\emph {\bibinfo {booktitle} {{Ultrafast} {Phenomena} and
  {Nanophotonics} {XXIII}}}},\ Vol.\ \bibinfo {volume} {10916},\ \bibinfo
  {editor} {edited by\ \bibinfo {editor} {\bibfnamefont {M.}~\bibnamefont
  {Betz}}\ and\ \bibinfo {editor} {\bibfnamefont {A.}~\bibnamefont
  {Elezzabi}}},\ \bibinfo {organization} {International Society for Optics and
  Photonics}\ (\bibinfo  {publisher} {SPIE},\ \bibinfo {year} {2019})\ pp.\
  \bibinfo {pages} {52--62}\BibitemShut {NoStop}%
\bibitem [{\citenamefont {Lebedev}\ \emph {et~al.}(2005)\citenamefont
  {Lebedev}, \citenamefont {Misochko}, \citenamefont {Dekorsky},\ and\
  \citenamefont {Georgiev}}]{lebedev}%
  \BibitemOpen
  \bibfield  {author} {\bibinfo {author} {\bibfnamefont {M.}~\bibnamefont
  {Lebedev}}, \bibinfo {author} {\bibfnamefont {O.}~\bibnamefont {Misochko}},
  \bibinfo {author} {\bibfnamefont {T.}~\bibnamefont {Dekorsky}}, \ and\
  \bibinfo {author} {\bibfnamefont {N.}~\bibnamefont {Georgiev}},\ }\href@noop
  {} {\bibfield  {journal} {\bibinfo  {journal} {J. Exp. Theor. Phys.}\
  }\textbf {\bibinfo {volume} {100}},\ \bibinfo {pages} {272--282} (\bibinfo
  {year} {2005})}\BibitemShut {NoStop}%
\bibitem [{\citenamefont {Palfrey}\ and\ \citenamefont
  {Heinz}(1985)}]{palfrey}%
  \BibitemOpen
  \bibfield  {author} {\bibinfo {author} {\bibfnamefont {S.}~\bibnamefont
  {Palfrey}}\ and\ \bibinfo {author} {\bibfnamefont {T.}~\bibnamefont
  {Heinz}},\ }\href@noop {} {\bibfield  {journal} {\bibinfo  {journal} {J. Opt.
  Soc. Am. B}\ }\textbf {\bibinfo {volume} {2}},\ \bibinfo {pages} {674--679}
  (\bibinfo {year} {1985})}\BibitemShut {NoStop}%
\bibitem [{\citenamefont {Penzkofer}\ and\ \citenamefont
  {Bugayev}(1989)}]{penzkofer}%
  \BibitemOpen
  \bibfield  {author} {\bibinfo {author} {\bibfnamefont {A.}~\bibnamefont
  {Penzkofer}}\ and\ \bibinfo {author} {\bibfnamefont {A.}~\bibnamefont
  {Bugayev}},\ }\href@noop {} {\bibfield  {journal} {\bibinfo  {journal} {Opt.
  Quantum Electron.}\ }\textbf {\bibinfo {volume} {21}},\ \bibinfo {pages}
  {283--306} (\bibinfo {year} {1989})}\BibitemShut {NoStop}%
\bibitem [{\citenamefont {Saissy}\ \emph {et~al.}(1978)\citenamefont {Saissy},
  \citenamefont {Azema}, \citenamefont {Botineau},\ and\ \citenamefont
  {Gires}}]{saissy}%
  \BibitemOpen
  \bibfield  {author} {\bibinfo {author} {\bibfnamefont {A.}~\bibnamefont
  {Saissy}}, \bibinfo {author} {\bibfnamefont {A.}~\bibnamefont {Azema}},
  \bibinfo {author} {\bibfnamefont {J.}~\bibnamefont {Botineau}}, \ and\
  \bibinfo {author} {\bibfnamefont {F.}~\bibnamefont {Gires}},\ }\href@noop {}
  {\bibfield  {journal} {\bibinfo  {journal} {Appl. Phys.}\ }\textbf {\bibinfo
  {volume} {15}},\ \bibinfo {pages} {99--102} (\bibinfo {year}
  {1978})}\BibitemShut {NoStop}%
\bibitem [{\citenamefont {Jayaraman}\ and\ \citenamefont
  {Lee}(1972)}]{jayaraman}%
  \BibitemOpen
  \bibfield  {author} {\bibinfo {author} {\bibfnamefont {S.}~\bibnamefont
  {Jayaraman}}\ and\ \bibinfo {author} {\bibfnamefont {C.}~\bibnamefont
  {Lee}},\ }\href@noop {} {\bibfield  {journal} {\bibinfo  {journal} {Appl.
  Phys. Lett.}\ }\textbf {\bibinfo {volume} {20}},\ \bibinfo {pages} {392--395}
  (\bibinfo {year} {1972})}\BibitemShut {NoStop}%
\bibitem [{\citenamefont {Kleinman}, \citenamefont {Miller},\ and\
  \citenamefont {Nordland}(1973)}]{kleinman2}%
  \BibitemOpen
  \bibfield  {author} {\bibinfo {author} {\bibfnamefont {D.}~\bibnamefont
  {Kleinman}}, \bibinfo {author} {\bibfnamefont {R.}~\bibnamefont {Miller}}, \
  and\ \bibinfo {author} {\bibfnamefont {W.}~\bibnamefont {Nordland}},\
  }\href@noop {} {\bibfield  {journal} {\bibinfo  {journal} {Appl. Phys.
  Lett.}\ }\textbf {\bibinfo {volume} {23}},\ \bibinfo {pages} {243--244}
  (\bibinfo {year} {1973})}\BibitemShut {NoStop}%
\bibitem [{\citenamefont {Bepko}(1975)}]{bepko}%
  \BibitemOpen
  \bibfield  {author} {\bibinfo {author} {\bibfnamefont {S.}~\bibnamefont
  {Bepko}},\ }\href@noop {} {\bibfield  {journal} {\bibinfo  {journal} {Phys.
  Rev. B}\ }\textbf {\bibinfo {volume} {12}},\ \bibinfo {pages} {669--672}
  (\bibinfo {year} {1975})}\BibitemShut {NoStop}%
\bibitem [{\citenamefont {Lee}\ and\ \citenamefont {Fan}(1974)}]{lee}%
  \BibitemOpen
  \bibfield  {author} {\bibinfo {author} {\bibfnamefont {C.}~\bibnamefont
  {Lee}}\ and\ \bibinfo {author} {\bibfnamefont {H.}~\bibnamefont {Fan}},\
  }\href@noop {} {\bibfield  {journal} {\bibinfo  {journal} {Phys. Rev. B}\
  }\textbf {\bibinfo {volume} {9}},\ \bibinfo {pages} {3502--3516} (\bibinfo
  {year} {1974})}\BibitemShut {NoStop}%
\bibitem [{\citenamefont {Oksman}\ \emph {et~al.}(1972)\citenamefont {Oksman},
  \citenamefont {Semenov}, \citenamefont {Smirnov},\ and\ \citenamefont
  {Smirnov}}]{oksman}%
  \BibitemOpen
  \bibfield  {author} {\bibinfo {author} {\bibfnamefont {Y.}~\bibnamefont
  {Oksman}}, \bibinfo {author} {\bibfnamefont {A.}~\bibnamefont {Semenov}},
  \bibinfo {author} {\bibfnamefont {V.}~\bibnamefont {Smirnov}}, \ and\
  \bibinfo {author} {\bibfnamefont {O.}~\bibnamefont {Smirnov}},\ }\href@noop
  {} {\bibfield  {journal} {\bibinfo  {journal} {Sov. Phys. Semicond.}\
  }\textbf {\bibinfo {volume} {6}},\ \bibinfo {pages} {629} (\bibinfo {year}
  {1972})}\BibitemShut {NoStop}%
\bibitem [{\citenamefont {Ralston}\ and\ \citenamefont
  {Chang}(1969)}]{ralston}%
  \BibitemOpen
  \bibfield  {author} {\bibinfo {author} {\bibfnamefont {J.}~\bibnamefont
  {Ralston}}\ and\ \bibinfo {author} {\bibfnamefont {R.}~\bibnamefont
  {Chang}},\ }\href@noop {} {\bibfield  {journal} {\bibinfo  {journal} {Appl.
  Phys. Lett.}\ }\textbf {\bibinfo {volume} {15}},\ \bibinfo {pages} {164--166}
  (\bibinfo {year} {1969})}\BibitemShut {NoStop}%
\bibitem [{\citenamefont {Arsenev}\ \emph {et~al.}(1969)\citenamefont
  {Arsenev}, \citenamefont {Dneprovskii}, \citenamefont {Klyshko},\ and\
  \citenamefont {Penin}}]{arsenev}%
  \BibitemOpen
  \bibfield  {author} {\bibinfo {author} {\bibfnamefont {V.}~\bibnamefont
  {Arsenev}}, \bibinfo {author} {\bibfnamefont {V.}~\bibnamefont
  {Dneprovskii}}, \bibinfo {author} {\bibfnamefont {D.}~\bibnamefont
  {Klyshko}}, \ and\ \bibinfo {author} {\bibfnamefont {A.}~\bibnamefont
  {Penin}},\ }\href@noop {} {\bibfield  {journal} {\bibinfo  {journal} {Sov.
  Phys. JETP}\ }\textbf {\bibinfo {volume} {29}},\ \bibinfo {pages} {413--415}
  (\bibinfo {year} {1969})}\BibitemShut {NoStop}%
\bibitem [{\citenamefont {Basov}\ \emph {et~al.}(1966)\citenamefont {Basov},
  \citenamefont {Grasiuk}, \citenamefont {Zubarev}, \citenamefont {Katulin},\
  and\ \citenamefont {Krokhin}}]{basov}%
  \BibitemOpen
  \bibfield  {author} {\bibinfo {author} {\bibfnamefont {N.}~\bibnamefont
  {Basov}}, \bibinfo {author} {\bibfnamefont {A.}~\bibnamefont {Grasiuk}},
  \bibinfo {author} {\bibfnamefont {I.}~\bibnamefont {Zubarev}}, \bibinfo
  {author} {\bibfnamefont {V.}~\bibnamefont {Katulin}}, \ and\ \bibinfo
  {author} {\bibfnamefont {O.}~\bibnamefont {Krokhin}},\ }\href@noop {}
  {\bibfield  {journal} {\bibinfo  {journal} {Sov. Phys. JETP}\ }\textbf
  {\bibinfo {volume} {23}},\ \bibinfo {pages} {366} (\bibinfo {year}
  {1966})}\BibitemShut {NoStop}%
\bibitem [{\citenamefont {Grasyuk}\ \emph {et~al.}(1973)\citenamefont
  {Grasyuk}, \citenamefont {Zubarev}, \citenamefont {Lobko}, \citenamefont
  {Matveets}, \citenamefont {Mironov},\ and\ \citenamefont
  {Shatberashvili}}]{grasyuk}%
  \BibitemOpen
  \bibfield  {author} {\bibinfo {author} {\bibfnamefont {A.}~\bibnamefont
  {Grasyuk}}, \bibinfo {author} {\bibfnamefont {I.}~\bibnamefont {Zubarev}},
  \bibinfo {author} {\bibfnamefont {V.}~\bibnamefont {Lobko}}, \bibinfo
  {author} {\bibfnamefont {Y.}~\bibnamefont {Matveets}}, \bibinfo {author}
  {\bibfnamefont {A.}~\bibnamefont {Mironov}}, \ and\ \bibinfo {author}
  {\bibfnamefont {O.}~\bibnamefont {Shatberashvili}},\ }\href@noop {}
  {\bibfield  {journal} {\bibinfo  {journal} {JETP Lett.}\ }\textbf {\bibinfo
  {volume} {17}},\ \bibinfo {pages} {416--418} (\bibinfo {year}
  {1973})}\BibitemShut {NoStop}%
\bibitem [{\citenamefont {Zubarev}, \citenamefont {Mironov},\ and\
  \citenamefont {Mikhailov}(1977)}]{zubarev}%
  \BibitemOpen
  \bibfield  {author} {\bibinfo {author} {\bibfnamefont {I.}~\bibnamefont
  {Zubarev}}, \bibinfo {author} {\bibfnamefont {A.}~\bibnamefont {Mironov}}, \
  and\ \bibinfo {author} {\bibfnamefont {S.}~\bibnamefont {Mikhailov}},\
  }\href@noop {} {\bibfield  {journal} {\bibinfo  {journal} {Sov. Phys.
  Semicond.}\ }\textbf {\bibinfo {volume} {11}},\ \bibinfo {pages} {239}
  (\bibinfo {year} {1977})}\BibitemShut {NoStop}%
\bibitem [{\citenamefont {DeSalvo}\ \emph {et~al.}(1993)\citenamefont
  {DeSalvo}, \citenamefont {Sheik-Bahae}, \citenamefont {Said}, \citenamefont
  {Hagan},\ and\ \citenamefont {Stryland}}]{desalvo}%
  \BibitemOpen
  \bibfield  {author} {\bibinfo {author} {\bibfnamefont {R.}~\bibnamefont
  {DeSalvo}}, \bibinfo {author} {\bibfnamefont {M.}~\bibnamefont
  {Sheik-Bahae}}, \bibinfo {author} {\bibfnamefont {A.}~\bibnamefont {Said}},
  \bibinfo {author} {\bibfnamefont {D.}~\bibnamefont {Hagan}}, \ and\ \bibinfo
  {author} {\bibfnamefont {E.~V.}\ \bibnamefont {Stryland}},\ }\href@noop {}
  {\bibfield  {journal} {\bibinfo  {journal} {Opt. Lett.}\ }\textbf {\bibinfo
  {volume} {18}},\ \bibinfo {pages} {194--196} (\bibinfo {year}
  {1993})}\BibitemShut {NoStop}%
\bibitem [{\citenamefont {Said}\ \emph {et~al.}(1992)\citenamefont {Said},
  \citenamefont {Sheik-Bahae}, \citenamefont {Hagan}, \citenamefont {Wei},
  \citenamefont {Wang}, \citenamefont {Young},\ and\ \citenamefont
  {Stryland}}]{said}%
  \BibitemOpen
  \bibfield  {author} {\bibinfo {author} {\bibfnamefont {A.}~\bibnamefont
  {Said}}, \bibinfo {author} {\bibfnamefont {M.}~\bibnamefont {Sheik-Bahae}},
  \bibinfo {author} {\bibfnamefont {D.}~\bibnamefont {Hagan}}, \bibinfo
  {author} {\bibfnamefont {T.}~\bibnamefont {Wei}}, \bibinfo {author}
  {\bibfnamefont {J.}~\bibnamefont {Wang}}, \bibinfo {author} {\bibfnamefont
  {J.}~\bibnamefont {Young}}, \ and\ \bibinfo {author} {\bibfnamefont {E.~V.}\
  \bibnamefont {Stryland}},\ }\href@noop {} {\bibfield  {journal} {\bibinfo
  {journal} {J. Opt. Soc. Am. B}\ }\textbf {\bibinfo {volume} {9}},\ \bibinfo
  {pages} {405--414} (\bibinfo {year} {1992})}\BibitemShut {NoStop}%
\bibitem [{\citenamefont {Zhi-hui}\ \emph {et~al.}(2015)\citenamefont
  {Zhi-hui}, \citenamefont {Si}, \citenamefont {Jun},\ and\ \citenamefont
  {Bing}}]{zhihui}%
  \BibitemOpen
  \bibfield  {author} {\bibinfo {author} {\bibfnamefont {C.}~\bibnamefont
  {Zhi-hui}}, \bibinfo {author} {\bibfnamefont {X.}~\bibnamefont {Si}},
  \bibinfo {author} {\bibfnamefont {H.}~\bibnamefont {Jun}}, \ and\ \bibinfo
  {author} {\bibfnamefont {G.}~\bibnamefont {Bing}},\ }\href@noop {} {\bibfield
   {journal} {\bibinfo  {journal} {Chinese J. Lumin.}\ }\textbf {\bibinfo
  {volume} {36}},\ \bibinfo {pages} {969--975} (\bibinfo {year}
  {2015})}\BibitemShut {NoStop}%
\bibitem [{\citenamefont {Lautenschlager}\ \emph
  {et~al.}(1987{\natexlab{a}})\citenamefont {Lautenschlager}, \citenamefont
  {Garriga}, \citenamefont {Logothetidis},\ and\ \citenamefont
  {Cardona}}]{lautenschlager1}%
  \BibitemOpen
  \bibfield  {author} {\bibinfo {author} {\bibfnamefont {P.}~\bibnamefont
  {Lautenschlager}}, \bibinfo {author} {\bibfnamefont {M.}~\bibnamefont
  {Garriga}}, \bibinfo {author} {\bibfnamefont {S.}~\bibnamefont
  {Logothetidis}}, \ and\ \bibinfo {author} {\bibfnamefont {M.}~\bibnamefont
  {Cardona}},\ }\href@noop {} {\bibfield  {journal} {\bibinfo  {journal} {Phys.
  Rev. B}\ }\textbf {\bibinfo {volume} {35}},\ \bibinfo {pages} {9174--9189}
  (\bibinfo {year} {1987}{\natexlab{a}})}\BibitemShut {NoStop}%
\bibitem [{\citenamefont {Hoffman}\ and\ \citenamefont {Mott}(2012)}]{hoffman}%
  \BibitemOpen
  \bibfield  {author} {\bibinfo {author} {\bibfnamefont {R.}~\bibnamefont
  {Hoffman}}\ and\ \bibinfo {author} {\bibfnamefont {A.}~\bibnamefont {Mott}},\
  }\href@noop {} {\bibfield  {journal} {\bibinfo  {journal} {Army Research
  Laboratory}\ }\textbf {\bibinfo {volume} {ARL-TR-6157}} (\bibinfo {year}
  {2012})}\BibitemShut {NoStop}%
\bibitem [{\citenamefont {Dinu}, \citenamefont {Quochi},\ and\ \citenamefont
  {Garcia}(2003)}]{dinu}%
  \BibitemOpen
  \bibfield  {author} {\bibinfo {author} {\bibfnamefont {M.}~\bibnamefont
  {Dinu}}, \bibinfo {author} {\bibfnamefont {F.}~\bibnamefont {Quochi}}, \ and\
  \bibinfo {author} {\bibfnamefont {H.}~\bibnamefont {Garcia}},\ }\href@noop {}
  {\bibfield  {journal} {\bibinfo  {journal} {Appl. Phys. Lett.}\ }\textbf
  {\bibinfo {volume} {82}},\ \bibinfo {pages} {2954--2956} (\bibinfo {year}
  {2003})}\BibitemShut {NoStop}%
\bibitem [{\citenamefont {Lautenschlager}\ \emph
  {et~al.}(1987{\natexlab{b}})\citenamefont {Lautenschlager}, \citenamefont
  {Garriga}, \citenamefont {Vina},\ and\ \citenamefont
  {Cardona}}]{lautenschlager2}%
  \BibitemOpen
  \bibfield  {author} {\bibinfo {author} {\bibfnamefont {P.}~\bibnamefont
  {Lautenschlager}}, \bibinfo {author} {\bibfnamefont {M.}~\bibnamefont
  {Garriga}}, \bibinfo {author} {\bibfnamefont {L.}~\bibnamefont {Vina}}, \
  and\ \bibinfo {author} {\bibfnamefont {M.}~\bibnamefont {Cardona}},\
  }\href@noop {} {\bibfield  {journal} {\bibinfo  {journal} {Phys. Rev. B}\
  }\textbf {\bibinfo {volume} {36}},\ \bibinfo {pages} {4821--4830} (\bibinfo
  {year} {1987}{\natexlab{b}})}\BibitemShut {NoStop}%
\bibitem [{\citenamefont {Stryland}\ \emph {et~al.}(1985)\citenamefont
  {Stryland}, \citenamefont {Woodall}, \citenamefont {Vanherzeele},\ and\
  \citenamefont {Soileau}}]{vanstryland2}%
  \BibitemOpen
  \bibfield  {author} {\bibinfo {author} {\bibfnamefont {E.~V.}\ \bibnamefont
  {Stryland}}, \bibinfo {author} {\bibfnamefont {M.}~\bibnamefont {Woodall}},
  \bibinfo {author} {\bibfnamefont {H.}~\bibnamefont {Vanherzeele}}, \ and\
  \bibinfo {author} {\bibfnamefont {M.}~\bibnamefont {Soileau}},\ }\href@noop
  {} {\bibfield  {journal} {\bibinfo  {journal} {Opt. Lett.}\ }\textbf
  {\bibinfo {volume} {10}},\ \bibinfo {pages} {490} (\bibinfo {year}
  {1985})}\BibitemShut {NoStop}%
\bibitem [{\citenamefont {Bechstedt}, \citenamefont {Adolph},\ and\
  \citenamefont {Schmidt}(1999)}]{bechstedt}%
  \BibitemOpen
  \bibfield  {author} {\bibinfo {author} {\bibfnamefont {F.}~\bibnamefont
  {Bechstedt}}, \bibinfo {author} {\bibfnamefont {B.}~\bibnamefont {Adolph}}, \
  and\ \bibinfo {author} {\bibfnamefont {W.}~\bibnamefont {Schmidt}},\
  }\href@noop {} {\bibfield  {journal} {\bibinfo  {journal} {Braz. J. Phys.}\
  }\textbf {\bibinfo {volume} {29}},\ \bibinfo {pages} {643--651} (\bibinfo
  {year} {1999})}\BibitemShut {NoStop}%
\bibitem [{\citenamefont {Attaccalite}\ and\ \citenamefont
  {Gr\"{u}ning}(2013)}]{attaccalite}%
  \BibitemOpen
  \bibfield  {author} {\bibinfo {author} {\bibfnamefont {C.}~\bibnamefont
  {Attaccalite}}\ and\ \bibinfo {author} {\bibfnamefont {M.}~\bibnamefont
  {Gr\"{u}ning}},\ }\href@noop {} {\bibfield  {journal} {\bibinfo  {journal}
  {Phys. Rev. B}\ }\textbf {\bibinfo {volume} {88}},\ \bibinfo {pages} {235113}
  (\bibinfo {year} {2013})}\BibitemShut {NoStop}%
\bibitem [{\citenamefont {Gr\"{u}ning}\ and\ \citenamefont
  {Attaccalite}(2020)}]{gruening}%
  \BibitemOpen
  \bibfield  {author} {\bibinfo {author} {\bibfnamefont {M.}~\bibnamefont
  {Gr\"{u}ning}}\ and\ \bibinfo {author} {\bibfnamefont {C.}~\bibnamefont
  {Attaccalite}},\ }\href
  {https://meetings.aps.org/Meeting/MAR20/Session/F39.1} {\enquote {\bibinfo
  {title} {Nonlinear optics from first-principles real-time approaches},}\ }
  (\bibinfo {year} {2020}),\ \bibinfo {note} {{APS} {March}
  {Meeting}}\BibitemShut {NoStop}%
\bibitem [{\citenamefont {Anderson}\ \emph {et~al.}(2015)\citenamefont
  {Anderson}, \citenamefont {Tancogne-Dejean}, \citenamefont {Mendoza},\ and\
  \citenamefont {Véniard}}]{anderson}%
  \BibitemOpen
  \bibfield  {author} {\bibinfo {author} {\bibfnamefont {S.}~\bibnamefont
  {Anderson}}, \bibinfo {author} {\bibfnamefont {N.}~\bibnamefont
  {Tancogne-Dejean}}, \bibinfo {author} {\bibfnamefont {B.}~\bibnamefont
  {Mendoza}}, \ and\ \bibinfo {author} {\bibfnamefont {V.}~\bibnamefont
  {Véniard}},\ }\href@noop {} {\bibfield  {journal} {\bibinfo  {journal}
  {Phys. Rev. B}\ }\textbf {\bibinfo {volume} {91}},\ \bibinfo {pages} {075302}
  (\bibinfo {year} {2015})}\BibitemShut {NoStop}%
\bibitem [{\citenamefont {Anderson}\ \emph {et~al.}(2016)\citenamefont
  {Anderson}, \citenamefont {Tancogne-Dejean}, \citenamefont {Mendoza},\ and\
  \citenamefont {Véniard}}]{anderson2}%
  \BibitemOpen
  \bibfield  {author} {\bibinfo {author} {\bibfnamefont {S.}~\bibnamefont
  {Anderson}}, \bibinfo {author} {\bibfnamefont {N.}~\bibnamefont
  {Tancogne-Dejean}}, \bibinfo {author} {\bibfnamefont {B.}~\bibnamefont
  {Mendoza}}, \ and\ \bibinfo {author} {\bibfnamefont {V.}~\bibnamefont
  {Véniard}},\ }\href@noop {} {\bibfield  {journal} {\bibinfo  {journal}
  {Phys. Rev. B}\ }\textbf {\bibinfo {volume} {93}},\ \bibinfo {pages} {235304}
  (\bibinfo {year} {2016})}\BibitemShut {NoStop}%
\bibitem [{\citenamefont {Anderson}(2016)}]{anderson3}%
  \BibitemOpen
  \bibfield  {author} {\bibinfo {author} {\bibfnamefont {S.~M.}\ \bibnamefont
  {Anderson}},\ }\emph {\bibinfo {title} {{Theoretical} {Optical}
  {Second-Harmonic} {Calculations} for {Surfaces}}},\ \href@noop {} {Ph.D.
  thesis},\ \bibinfo  {school} {Centro de Investigaciones en \'{O}ptica, A.C.}
  (\bibinfo {year} {2016})\BibitemShut {NoStop}%
\bibitem [{\citenamefont {Anderson}\ and\ \citenamefont
  {Mendoza}(2016)}]{anderson4}%
  \BibitemOpen
  \bibfield  {author} {\bibinfo {author} {\bibfnamefont {S.}~\bibnamefont
  {Anderson}}\ and\ \bibinfo {author} {\bibfnamefont {B.}~\bibnamefont
  {Mendoza}},\ }\href@noop {} {\bibfield  {journal} {\bibinfo  {journal} {Phys.
  Rev. B}\ }\textbf {\bibinfo {volume} {94}},\ \bibinfo {pages} {115314}
  (\bibinfo {year} {2016})}\BibitemShut {NoStop}%
\end{thebibliography}%


\begin{thebibliography}{11}%
\makeatletter
\providecommand \@ifxundefined [1]{%
 \@ifx{#1\undefined}
}%
\providecommand \@ifnum [1]{%
 \ifnum #1\expandafter \@firstoftwo
 \else \expandafter \@secondoftwo
 \fi
}%
\providecommand \@ifx [1]{%
 \ifx #1\expandafter \@firstoftwo
 \else \expandafter \@secondoftwo
 \fi
}%
\providecommand \natexlab [1]{#1}%
\providecommand \enquote  [1]{``#1''}%
\providecommand \bibnamefont  [1]{#1}%
\providecommand \bibfnamefont [1]{#1}%
\providecommand \citenamefont [1]{#1}%
\providecommand \href@noop [0]{\@secondoftwo}%
\providecommand \href [0]{\begingroup \@sanitize@url \@href}%
\providecommand \@href[1]{\@@startlink{#1}\@@href}%
\providecommand \@@href[1]{\endgroup#1\@@endlink}%
\providecommand \@sanitize@url [0]{\catcode `\\12\catcode `\$12\catcode
  `\&12\catcode `\#12\catcode `\^12\catcode `\_12\catcode `\%12\relax}%
\providecommand \@@startlink[1]{}%
\providecommand \@@endlink[0]{}%
\providecommand \url  [0]{\begingroup\@sanitize@url \@url }%
\providecommand \@url [1]{\endgroup\@href {#1}{\urlprefix }}%
\providecommand \urlprefix  [0]{URL }%
\providecommand \Eprint [0]{\href }%
\providecommand \doibase [0]{http://dx.doi.org/}%
\providecommand \selectlanguage [0]{\@gobble}%
\providecommand \bibinfo  [0]{\@secondoftwo}%
\providecommand \bibfield  [0]{\@secondoftwo}%
\providecommand \translation [1]{[#1]}%
\providecommand \BibitemOpen [0]{}%
\providecommand \bibitemStop [0]{}%
\providecommand \bibitemNoStop [0]{.\EOS\space}%
\providecommand \EOS [0]{\spacefactor3000\relax}%
\providecommand \BibitemShut  [1]{\csname bibitem#1\endcsname}%
\let\auto@bib@innerbib\@empty
\bibitem [{\citenamefont {Boyd}(2008)}]{boyd}%
  \BibitemOpen
  \bibfield  {author} {\bibinfo {author} {\bibfnamefont {R.}~\bibnamefont
  {Boyd}},\ }\href@noop {} {\emph {\bibinfo {title} {{Nonlinear} {Optics}}}},\
  \bibinfo {edition} {3rd}\ ed.\ (\bibinfo  {publisher} {Academic Press -
  Elsevier, Inc.},\ \bibinfo {address} {Burlington},\ \bibinfo {year}
  {2008})\BibitemShut {NoStop}%
\bibitem [{\citenamefont {Dvorak}\ \emph {et~al.}(1994)\citenamefont {Dvorak},
  \citenamefont {Schroeder}, \citenamefont {Andersen}, \citenamefont {Smirl},\
  and\ \citenamefont {Wherrett}}]{dvorak}%
  \BibitemOpen
  \bibfield  {author} {\bibinfo {author} {\bibfnamefont {M.}~\bibnamefont
  {Dvorak}}, \bibinfo {author} {\bibfnamefont {W.}~\bibnamefont {Schroeder}},
  \bibinfo {author} {\bibfnamefont {D.}~\bibnamefont {Andersen}}, \bibinfo
  {author} {\bibfnamefont {A.}~\bibnamefont {Smirl}}, \ and\ \bibinfo {author}
  {\bibfnamefont {B.}~\bibnamefont {Wherrett}},\ }\href@noop {} {\bibfield
  {journal} {\bibinfo  {journal} {IEEE J. Quantum Electron.}\ }\textbf
  {\bibinfo {volume} {30}},\ \bibinfo {pages} {256} (\bibinfo {year}
  {1994})}\BibitemShut {NoStop}%
\bibitem [{\citenamefont {DeSalvo}\ \emph {et~al.}(1993)\citenamefont
  {DeSalvo}, \citenamefont {Sheik-Bahae}, \citenamefont {Said}, \citenamefont
  {Hagan},\ and\ \citenamefont {Stryland}}]{desalvo}%
  \BibitemOpen
  \bibfield  {author} {\bibinfo {author} {\bibfnamefont {R.}~\bibnamefont
  {DeSalvo}}, \bibinfo {author} {\bibfnamefont {M.}~\bibnamefont
  {Sheik-Bahae}}, \bibinfo {author} {\bibfnamefont {A.}~\bibnamefont {Said}},
  \bibinfo {author} {\bibfnamefont {D.}~\bibnamefont {Hagan}}, \ and\ \bibinfo
  {author} {\bibfnamefont {E.~V.}\ \bibnamefont {Stryland}},\ }\href@noop {}
  {\bibfield  {journal} {\bibinfo  {journal} {Opt. Lett.}\ }\textbf {\bibinfo
  {volume} {18}},\ \bibinfo {pages} {194} (\bibinfo {year} {1993})}\BibitemShut
  {NoStop}%
\bibitem [{\citenamefont {Murayama}\ and\ \citenamefont
  {Nakayama}(1995)}]{murayama}%
  \BibitemOpen
  \bibfield  {author} {\bibinfo {author} {\bibfnamefont {M.}~\bibnamefont
  {Murayama}}\ and\ \bibinfo {author} {\bibfnamefont {T.}~\bibnamefont
  {Nakayama}},\ }\href@noop {} {\bibfield  {journal} {\bibinfo  {journal}
  {Phys. Rev. B}\ }\textbf {\bibinfo {volume} {52}},\ \bibinfo {pages} {4986}
  (\bibinfo {year} {1995})}\BibitemShut {NoStop}%
\bibitem [{\citenamefont {Bepko}(1975)}]{bepko}%
  \BibitemOpen
  \bibfield  {author} {\bibinfo {author} {\bibfnamefont {S.}~\bibnamefont
  {Bepko}},\ }\href@noop {} {\bibfield  {journal} {\bibinfo  {journal} {Phys.
  Rev. B}\ }\textbf {\bibinfo {volume} {12}},\ \bibinfo {pages} {669} (\bibinfo
  {year} {1975})}\BibitemShut {NoStop}%
\bibitem [{\citenamefont {Bechtel}\ and\ \citenamefont
  {Smith}(1976)}]{bechtel}%
  \BibitemOpen
  \bibfield  {author} {\bibinfo {author} {\bibfnamefont {J.}~\bibnamefont
  {Bechtel}}\ and\ \bibinfo {author} {\bibfnamefont {W.}~\bibnamefont
  {Smith}},\ }\href@noop {} {\bibfield  {journal} {\bibinfo  {journal} {Phys.
  Rev. B}\ }\textbf {\bibinfo {volume} {13}},\ \bibinfo {pages} {3515}
  (\bibinfo {year} {1976})}\BibitemShut {NoStop}%
\bibitem [{\citenamefont {Said}\ \emph {et~al.}(1992)\citenamefont {Said},
  \citenamefont {Sheik-Bahae}, \citenamefont {Hagan}, \citenamefont {Wei},
  \citenamefont {Wang}, \citenamefont {Young},\ and\ \citenamefont
  {Stryland}}]{said}%
  \BibitemOpen
  \bibfield  {author} {\bibinfo {author} {\bibfnamefont {A.}~\bibnamefont
  {Said}}, \bibinfo {author} {\bibfnamefont {M.}~\bibnamefont {Sheik-Bahae}},
  \bibinfo {author} {\bibfnamefont {D.}~\bibnamefont {Hagan}}, \bibinfo
  {author} {\bibfnamefont {T.}~\bibnamefont {Wei}}, \bibinfo {author}
  {\bibfnamefont {J.}~\bibnamefont {Wang}}, \bibinfo {author} {\bibfnamefont
  {J.}~\bibnamefont {Young}}, \ and\ \bibinfo {author} {\bibfnamefont {E.~V.}\
  \bibnamefont {Stryland}},\ }\href@noop {} {\bibfield  {journal} {\bibinfo
  {journal} {J. Opt. Soc. Am. B}\ }\textbf {\bibinfo {volume} {9}},\ \bibinfo
  {pages} {405} (\bibinfo {year} {1992})}\BibitemShut {NoStop}%
\bibitem [{\citenamefont {Bristow}\ \emph {et~al.}(2007)\citenamefont
  {Bristow}, \citenamefont {Rotenberg},\ and\ \citenamefont {van
  Driel}}]{bristow}%
  \BibitemOpen
  \bibfield  {author} {\bibinfo {author} {\bibfnamefont {A.}~\bibnamefont
  {Bristow}}, \bibinfo {author} {\bibfnamefont {N.}~\bibnamefont {Rotenberg}},
  \ and\ \bibinfo {author} {\bibfnamefont {H.}~\bibnamefont {van Driel}},\
  }\href@noop {} {\bibfield  {journal} {\bibinfo  {journal} {Appl. Phys.
  Lett.}\ }\textbf {\bibinfo {volume} {90}},\ \bibinfo {pages} {191104}
  (\bibinfo {year} {2007})}\BibitemShut {NoStop}%
\bibitem [{woo(2020)}]{woollam}%
  \BibitemOpen
  \href@noop {} {\enquote {\bibinfo {title} {Ellipsometry tutorial},}\
  }\bibinfo {howpublished}
  {\url{https://www.jawoollam.com/resources/ellipsometry-tutorial}} (\bibinfo
  {year} {2020})\BibitemShut {NoStop}%
\bibitem [{\citenamefont {Skauli}\ \emph {et~al.}(2003)\citenamefont {Skauli},
  \citenamefont {Kuo}, \citenamefont {Vodopyanov}, \citenamefont {Pinguet},
  \citenamefont {Levi}, \citenamefont {Eyres}, \citenamefont {Harris},
  \citenamefont {Fejer}, \citenamefont {Gerard}, \citenamefont {Becouarn},\
  and\ \citenamefont {Lallier}}]{skauli}%
  \BibitemOpen
  \bibfield  {author} {\bibinfo {author} {\bibfnamefont {T.}~\bibnamefont
  {Skauli}}, \bibinfo {author} {\bibfnamefont {P.}~\bibnamefont {Kuo}},
  \bibinfo {author} {\bibfnamefont {K.}~\bibnamefont {Vodopyanov}}, \bibinfo
  {author} {\bibfnamefont {T.}~\bibnamefont {Pinguet}}, \bibinfo {author}
  {\bibfnamefont {O.}~\bibnamefont {Levi}}, \bibinfo {author} {\bibfnamefont
  {L.}~\bibnamefont {Eyres}}, \bibinfo {author} {\bibfnamefont
  {J.}~\bibnamefont {Harris}}, \bibinfo {author} {\bibfnamefont
  {M.}~\bibnamefont {Fejer}}, \bibinfo {author} {\bibfnamefont
  {B.}~\bibnamefont {Gerard}}, \bibinfo {author} {\bibfnamefont
  {L.}~\bibnamefont {Becouarn}}, \ and\ \bibinfo {author} {\bibfnamefont
  {E.}~\bibnamefont {Lallier}},\ }\href@noop {} {\bibfield  {journal} {\bibinfo
   {journal} {J. Appl. Phys.}\ }\textbf {\bibinfo {volume} {94}},\ \bibinfo
  {pages} {6447} (\bibinfo {year} {2003})}\BibitemShut {NoStop}%
\bibitem [{\citenamefont {Li}(1993)}]{li}%
  \BibitemOpen
  \bibfield  {author} {\bibinfo {author} {\bibfnamefont {H.}~\bibnamefont
  {Li}},\ }\href@noop {} {\bibfield  {journal} {\bibinfo  {journal} {J. Phys.
  Chem. Ref. Data}\ }\textbf {\bibinfo {volume} {9}},\ \bibinfo {pages} {561}
  (\bibinfo {year} {1993})}\BibitemShut {NoStop}%
\end{thebibliography}%

\end{document}


\title{\Large \bfseries Supplementary Material:\\
\large Im\{$\mathbf{\chi^{(3)}}$\} spectra of $\mathbf{110}$-cut GaAs, GaP, and Si near the \\
two-photon absorption band edge}

\author{Brandon J. Furey}
\email{furey@utexas.edu.}
\affiliation{Physics Department, University of Texas at Austin, 2515 Speedway, C1600, Austin, TX, USA 78712}
\author{Rodrigo M. Barba-Barba}
\affiliation{Centro de Investigaciones en \'{O}ptica, A.C., Loma del Bosque 115, Colonia Lomas del Campestre, Le\'{o}n, Gto., M\'{e}xico 37150}
\author{Ramon Carriles}
\email{ramon@cio.mx.}
\affiliation{Centro de Investigaciones en \'{O}ptica, A.C., Loma del Bosque 115, Colonia Lomas del Campestre, Le\'{o}n, Gto., M\'{e}xico 37150}
\author{Alan Bernal}
\affiliation{Centro de Investigaciones en \'{O}ptica, A.C., Loma del Bosque 115, Colonia Lomas del Campestre, Le\'{o}n, Gto., M\'{e}xico 37150}
\author{Bernardo S. Mendoza}
\affiliation{Centro de Investigaciones en \'{O}ptica, A.C., Loma del Bosque 115, Colonia Lomas del Campestre, Le\'{o}n, Gto., M\'{e}xico 37150}
\author{Michael C. Downer}
\email{downer@physics.utexas.edu.}
\affiliation{Physics Department, University of Texas at Austin, 2515 Speedway, C1600, Austin, TX, USA 78712}

\maketitle

\makeatletter
\def\l@subsubsection#1#2{}
\makeatother
{\hypersetup{linkcolor=black}\tableofcontents}

\clearpage

\section{\label{sec:waveequation}Third-order nonlinear optical fundamental frequency response}

Here we present a complete derivation of the relation between the imaginary part of the third-order nonlinear susceptibility tensor and the 2PA coefficient as well as the real part of the third-order nonlinear susceptibility tensor and the nonlinear refractive index in the SI system of units. For nearly collinear beams co-propagating along the laboratory $Z$-axis, the probe and pump fields in laboratory $XYZ$ coordinates are
\begin{align}
\vec{E}_1&=\big[E^X_1(\omega,\vec{k}_1) + c.c.\big] \hat{X}= A_1(Z) e^{i(k_1 Z - \omega t)} \hat{X} + c.c.,\\
\begin{split}
\vec{E}_2&=\big[E^X_2(\omega,\vec{k}_2) + c.c.\big] \hat{X} + \big[E^Y_2(\omega,\vec{k}_2) + c.c.\big] \hat{Y}\\
&= \big[A_2^X (Z) \hat{X} + A_2^Y(Z) \hat{Y}\big] e^{i(k_2 Z - \omega t + \gamma)} + c.c.,
\end{split}
\end{align}
where $A_i = |A_i|e^{i \phi_i}$ are the complex field amplitudes, $A_2^X(Z) = A_2(Z) \cos \zeta$, $A_2^Y(Z) = A_2(Z) \sin \zeta$, $\zeta$ is the angle between linearly-polarized probe and pump fields, $\phi_i$ are the phases of the complex field amplitudes, and $\gamma$ is an arbitrary phase shift between the probe and pump fields (assumed to not vary in time over the duration of the pulse such that the two pulses are mutually coherent). The field experienced by the nonlinear medium is just the sum field
\begin{equation}
\vec{E} = \vec{E}_1 + \vec{E}_2,
\end{equation}
and the intensity of the fields as defined is\cite{boyd}
\begin{equation}
\label{eqintensity}
I_i(Z) = 2\epsilon_0 n c |A_i(Z)|^2,
\end{equation}
where $\epsilon_0$ is the vacuum permittivity, $n$ the refractive index, and $c$ the speed of light. 

If the beams are co-polarized, i.e. $\zeta = 0$, then the third-order polarization density $\vec{P}^{(3)}(Z,t)$ is
\begin{equation}
\begin{split}
P_X^{(3)}(Z,t) &=  \epsilon_0 \chi_{XXXX}^{(3)} E^X E^X E^X + c.c.\\
&= \epsilon_0 \chi_{XXXX}^{(3)} \Big( A_1^3 e^{3i(k_1 Z - \omega t)} + 3 A_1^2 A_2 e^{i\big[(2k_1 + k_2)Z - 3 \omega t + \gamma \big]} + 3 A_1 A_2^2 e^{i\big[(k_1 + 2 k_2) Z - 3 \omega t + 2 \gamma \big]}\\
&\ \ \ \ \ + A_2^3 e^{3i(k_2 Z - \omega t + \gamma)} + \underline{3 A_1 |A_1|^2 e^{i (k_1 Z - \omega t)}} + \underline{6 A_1 |A_2|^2 e^{i (k_1 Z - \omega t)}} + 3 A_1^2 A_2^* e^{i \big[(2k_1 - k_2)Z - \omega t - \gamma \big]}\\
&\ \ \ \ \ + 6 |A_1|^2 A_2 e^{i(k_2 Z - \omega t + \gamma)} + 3 A_2 |A_2|^2 e^{i(k_2 Z - \omega t + \gamma)} + 3 A_1 (A_2^*)^2 e^{i\big[(k_1 - 2 k_2)Z + \omega t - 2\gamma \big]}\Big) + c.c.,
\end{split}
\end{equation}
where the underlined terms (and their complex conjugates) are the Fourier components at $\pm \omega, \pm \vec{k}_1$, applicable for self-action and two-beam effects in the probe field including degenerate 2PA and nonlinear refraction. We have neglected dispersion in $\chi^{(3)}_{XXXX}$, since we are only interested in the response at $\pm \omega$. Thus the Fourier component $\vec{P}^{(3)}(\pm \omega, \pm \vec{k}_1)$ is
\begin{equation}
P_X^{(3)}(\pm \omega,\pm \vec{k}_1) = 3 \epsilon_0 \chi^{(3)}_{XXXX}(\omega) \big[ A_1(Z) |A_1(Z)|^2 + 2 A_1(Z) |A_2(Z)|^2\big] e^{i(k_1 Z - \omega t)} + c.c.
\end{equation}
Here $\chi^{(3)}_{abcd}(\omega; \omega, \omega, -\omega) = \chi^{(3)}_{abcd}(\omega) \in \mathbb{C}$ is the degenerate third-order nonlinear optical susceptibility tensor describing a polarization density response at the same frequency as the incident frequency.

The nonlinear wave equation is\cite{boyd}
\begin{equation}
\nabla^2 \vec{E} - \frac{\epsilon_{rel}}{c^2}\frac{\partial^2}{\partial t^2} \vec{E} = \frac{1}{\epsilon_0 c^2}\frac{\partial^2}{\partial t^2} \vec{P}^{NL},
\end{equation}
where $\epsilon_{rel} = \epsilon/\epsilon_0$, and $\vec{P}^{NL}$ is the nonlinear polarization density driving the optical response. In the plane wave basis defined in this case, $\nabla^2 \rightarrow \frac{\mathrm{d}^2}{\mathrm{d}Z^2}$ and thus we have
\begin{multline}
\frac{\mathrm{d}^2}{\mathrm{d}Z^2} \Big( A_1 e^{i(k_1 Z - \omega t)} \Big) - \frac{\epsilon_{rel}}{c^2} \frac{\partial^2}{\partial t^2} \Big( A_1 e^{i(k_1 Z - \omega t)}\Big) + c.c.\\
= \frac{1}{\epsilon_0 c^2}\frac{\partial^2}{\partial t^2} \bigg[ 3 \epsilon_0 \chi_{XXXX}^{(3)}(\omega) \Big(A_1 |A_1|^2 + 2 A_1 |A_2|^2\Big) e^{i(k_1 Z - \omega t)} + c.c.\bigg].
\end{multline}
Applying the derivatives, we obtain
\begin{multline}
\frac{\mathrm{d}^2A_1}{\mathrm{d}Z^2} e^{i(k_1 Z - \omega t)} + 2 i k_1 \frac{\mathrm{d}A_1}{\mathrm{d}Z} e^{i(k_1 Z - \omega t)} - k_1^2 A_1 e^{i(k_1 Z - \omega t)} + \frac{\epsilon_{rel} \omega^2}{c^2} A_1 e^{i(k_1 Z - \omega t)} + c.c.\\
= \frac{- 3 \omega^2}{c^2} \chi^{(3)}_{XXXX}(\omega) \Big( A_1 |A_1|^2 + 2 A_1 |A_2|^2 \Big) e^{i(k_1 Z - \omega t)} + c.c.,
\end{multline}
where we can neglect the $\frac{\mathrm{d}^2A_1}{\mathrm{d}Z^2} e^{i(k_1 Z - \omega t)}$ term due to the slowly-varying amplitude approximation and the third and fourth terms on the left-hand side cancel since $\epsilon_{rel} \omega ^2 / c^2 = k_1^2$. Then we can explicitly write out the complex conjugate terms, and we have
\begin{multline}
2 i k_1 \frac{\mathrm{d}A_1}{\mathrm{d}Z}e^{i(k_1 Z - \omega t)} - 2 i k_1 \frac{\mathrm{d}A_1^*}{\mathrm{d}Z}e^{-i(k_1 Z - \omega t)}\\
= \frac{-3 \omega^2}{c^2} \chi^{(3)}_{XXXX}(\omega) \Big(A_1 |A_1|^2 + 2 A_1 |A_2|^2\Big)e^{i(k_1 Z - \omega t)} - \frac{3 \omega^2}{c^2} \big[\chi^{(3)}_{XXXX}(\omega)\big]^* \Big(A_1^*|A_1|^2 + 2 A_1^* |A_2|^2\Big)e^{-i(k_1 Z - \omega t)}.
\end{multline}
Comparing like terms in oscillatory behavior, we obtain two differential equations for the amplitudes of the $e^{i(k_1 Z - \omega t)}$ and $e^{-i(k_1 Z - \omega t)}$ terms which must simultaneously hold. We have for the $e^{i(k_1 Z - \omega t)}$ terms
\begin{equation}
2 i k_1 \frac{\mathrm{d}A_1}{\mathrm{d}Z} = \frac{-3 \omega^2}{c^2} \chi^{(3)}_{XXXX}(\omega) \Big(A_1 |A_1|^2 + 2 A_1 |A_2|^2\Big),
\end{equation}
and for the $e^{-i(k_1 Z - \omega t)}$ terms
\begin{equation}
-2 i k_1 \frac{\mathrm{d}A_1^*}{\mathrm{d}Z} = \frac{-3 \omega^2}{c^2} \big[\chi^{(3)}_{XXXX}(\omega)\big]^* \Big(A_1^* |A_1|^2 + 2 A_1^* |A_2|^2\Big),
\end{equation}
which simplifies to
\begin{equation}
\label{eqdiff1}
\frac{\mathrm{d}A_1}{\mathrm{d}Z} = \frac{3 i \omega}{2 n(\omega) c} \chi^{(3)}_{XXXX}(\omega) \Big(A_1 |A_1|^2 + 2 A_1 |A_2|^2\Big),
\end{equation}
and
\begin{equation}
\label{eqdiff2}
\frac{\mathrm{d}A_1^*}{\mathrm{d}Z} = \frac{-3 i \omega}{2 n(\omega) c} \big[\chi^{(3)}_{XXXX}(\omega)\big]^* \Big(A_1^* |A_1|^2 + 2 A_1^* |A_2|^2\Big),
\end{equation}
where we have used $k_1 = n(\omega)\ \omega / c$. 

\subsection{\label{sec:chibeta}Derivation of the relation between $\mathrm{Im}\{\chi^{(3)}\}$ and $\beta$}

The change in intensity with propagation distance due to 2PA can be calculated by making use of Eq. \ref{eqintensity} to write
\begin{equation}
\begin{split}
\frac{\mathrm{d}I_1}{\mathrm{d}Z} &=2 n(\omega) \epsilon_0 c \bigg(A_1 \frac{\mathrm{d}A_1^*}{\mathrm{d}Z} + A_1^* \frac{\mathrm{d}A_1}{\mathrm{d}Z}\bigg)\\
&= 3 \epsilon_0 \omega \Big(|A_1|^4 + 2 |A_1|^2 |A_2|^2\Big) \bigg(i \chi^{(3)}_{XXXX}(\omega) -  i \big[\chi^{(3)}_{XXXX}(\omega)\big]^* \bigg),
\end{split}
\end{equation}
where
\begin{equation}
\begin{split}
i \chi^{(3)}_{XXXX}(\omega) -  i \big[\chi^{(3)}_{XXXX}(\omega)\big]^* &= i \big[\mathrm{Re}\{\chi^{(3)}_{XXXX}(\omega)\} +  i \mathrm{Im}\{\chi^{(3)}_{XXXX}(\omega)\}\big]\\
&\ \ \ \ \ -  i \big[\mathrm{Re}\{\chi^{(3)}_{XXXX}(\omega)\} - i \mathrm{Im}\{\chi^{(3)}_{XXXX}(\omega)\}\big]\\
&= - 2 \mathrm{Im}\{\chi^{(3)}_{XXXX}(\omega)\}.
\end{split}
\end{equation}

Thus we have
\begin{equation}
\begin{split}
\frac{\mathrm{d}I_1}{\mathrm{d}Z} &= - 6 \epsilon_0 \omega \mathrm{Im}\{\chi^{(3)}_{XXXX}(\omega)\} \Big(|A_1|^4 + 2 |A_1|^2 |A_2|^2\Big)\\
&= \frac{-3 \omega}{2 \epsilon_0 \big[n(\omega)\big]^2 c^2} \mathrm{Im}\{\chi^{(3)}_{XXXX}(\omega)\} \Big( I_1^2 + 2 I_1 I_2 \Big).
\end{split}
\end{equation}
Equating this differential equation with the relevant terms of Eq. 1 in the article, we find that
\begin{equation}
\beta_{12}^{\parallel}(\omega) = \frac{3 \omega}{\epsilon_0 \big[ n(\omega)\big]^2 c^2} \mathrm{Im}\{\chi^{(3)}_{XXXX}(\omega)\} = 2 \beta_{11}^{\parallel}(\omega).
\end{equation}

If the beams are cross-polarized, i.e. $\zeta = \pi/2$, the relevant Fourier component $\vec{P}^{(3)}(\pm \omega, \pm \vec{k}_1)$ is
\begin{equation}
P_X^{(3)}(\pm \omega,\pm \vec{k}_1) = 3 \epsilon_0 \big[\chi^{(3)}_{XXXX}(\omega) A_1(Z)^3 + 2 \chi^{(3)}_{XXYY} A_2(Z)^2 A_1(Z)\big] e^{i(k_1 Z - \omega t)} + c.c.,
\end{equation}
for which a similar derivation leads to the relation between the $\mathrm{Im}\{\chi^{(3)}_{XXYY}(\omega)\}$ component and the $\beta_{12}^{\perp}(\omega)$, 
\begin{equation}
\beta_{12}^{\perp}(\omega) = \frac{3 \omega}{\epsilon_0 \big[ n(\omega)\big]^2 c^2} \mathrm{Im}\{\chi^{(3)}_{XXYY}(\omega)\}.
\end{equation}

\subsection{\label{sec:chin2}Derivation of the relation between $\mathrm{Re}\{\chi^{(3)}\}$ and $n_2$}

The change in phase of the probe amplitude with propagation distance due to the nonlinear refractive index can be calculated by making use of Eq. \ref{eqdiff1} to write
\begin{equation}
\begin{split}
\frac{\mathrm{d}A_1}{\mathrm{d}Z} &= \frac{\mathrm{d}\big(|A_1| e^{i \phi_1}\big)}{\mathrm{d}Z}\\
&= \bigg( \frac{\mathrm{d}|A_1|}{\mathrm{d}Z}\bigg)e^{i \phi_1} + |A_1| \frac{\mathrm{d} e^{i \phi_1}}{\mathrm{d}Z},
\end{split}
\end{equation}
where
\begin{equation}
\frac{\mathrm{d}|A_1|}{\mathrm{d}Z} = \frac{1}{2 |A_1|} \bigg(A_1 \frac{\mathrm{d}A_1^*}{\mathrm{d}Z} + A_1^* \frac{\mathrm{d}A_1}{\mathrm{d}Z}\bigg).
\end{equation}
Thus we can write
\begin{equation}
\begin{split}
\frac{\mathrm{d}e^{i \phi_1}}{\mathrm{d}Z} &= \bigg( \frac{\mathrm{d}A_1}{\mathrm{d}Z}\bigg) \bigg(\frac{1}{|A_1|}\bigg) - \bigg(\frac{\mathrm{d}|A_1|}{\mathrm{d}Z}\bigg)\bigg(\frac{e^{i \phi_1}}{|A_1|}\bigg)\\
&= \frac{3 i \omega}{2 n(\omega) c} e^{i \phi_1} \big( |A_1|^2 + 2 |A_2|^2 \big) \big( \chi^{(3)}_{XXXX}(\omega) - i \mathrm{Im}\{\chi^{(3)}_{XXXX}(\omega)\} \big)\\
&= \frac{3 i \omega}{2 n(\omega) c} \mathrm{Re}\{\chi^{(3)}_{XXXX}(\omega)\} e^{i \phi_1} \big( |A_1|^2 + 2 |A_2|^2 \big).
\end{split}
\end{equation}

Noting that
\begin{equation}
\frac{\mathrm{d}e^{i \phi_1}}{\mathrm{d}Z} = i e^{i \phi_1} \frac{\mathrm{d} \phi_1}{\mathrm{d}Z},
\end{equation}
we have 
\begin{equation}
\begin{split}
\frac{\mathrm{d}\phi_1}{\mathrm{d}Z} &= \frac{3 \omega}{2 n(\omega) c} \mathrm{Re}\{\chi^{(3)}_{XXXX}(\omega)\}\big( |A_1|^2 + 2 |A_2|^2 \big)\\
&= \frac{3 \omega}{4 \epsilon_0 \big[n(\omega)\big]^2 c^2} \mathrm{Re}\{\chi^{(3)}_{XXXX}(\omega)\}\big( I_1 + 2 I_2 \big).
\end{split}
\end{equation}

For the remainder of this subsection, we identify the linear refractive index as $n_0(\omega)$ and the nonlinear refractive index as $n_2^{\parallel,\perp}(\omega)$. The phase change can be expressed in terms of the nonlinear refractive index by
\begin{equation}
\frac{\mathrm{d}\phi_1}{\mathrm{d}Z} = \frac{\omega}{c} \Big[ n_{2\ (11)}^{\parallel}(\omega)I_1 + n_{2\ (12)}^{\parallel}(\omega) I_2 \Big],
\end{equation}
where $n_{2\ (11)}^{\parallel}(\omega)$ is the prope-probe nonlinear refractive index and $n_{2\ (12)}^{\parallel}(\omega)$ is the pump-probe nonlinear refractive index in the co-polarized geometry. Thus the pump-probe nonlinear refractive index is\cite{boyd}
\begin{equation}
n_{2\ (12)}^{\parallel}(\omega) = \frac{3}{2 \epsilon_0 \big[n_0(\omega)\big]^2 c} \mathrm{Re}\{\chi^{(3)}_{XXXX}(\omega)\} = 2 n_{2\ (11)}^{\parallel}(\omega).
\end{equation}

If the beams are cross-polarized, i.e. $\zeta = \pi/2$, the relevant Fourier component $\vec{P}^{(3)}(\pm \omega, \pm \vec{k}_1)$ is
\begin{equation}
P_X^{(3)}(\pm \omega,\pm \vec{k}_1) = 3 \epsilon_0 \big[\chi^{(3)}_{XXXX}(\omega) A_1(Z)^3 + 2 \chi^{(3)}_{XXYY} A_2(Z)^2 A_1(Z)\big] e^{i(k_1 Z - \omega t)} + c.c.,
\end{equation}
for which a similar derivation leads to the relation between the $\mathrm{Re}\{\chi^{(3)}_{XXYY}(\omega)\}$ component and the $n_{2\ (12)}^{\perp}(\omega)$, 
\begin{equation}
n_{2\ (12)}^{\perp}(\omega) = \frac{3}{2 \epsilon_0 \big[n_0(\omega)\big]^2 c} \mathrm{Re}\{\chi^{(3)}_{XXYY}(\omega)\}.
\end{equation}

In this work we assume that $|\mathrm{d}\phi_{1,2}/\mathrm{d}Z| \ll \pi$, i.e. $n_{2\ (11)}^{\parallel,\perp}(\omega)I_1 + n_{2\ (12)}^{\parallel,\perp}(\omega)I_2 \rightarrow 0$ and $n_{2\ (12)}^{\parallel,\perp}(\omega)I_1 + n_{2\ (22)}^{\parallel,\perp}(\omega)I_2 \rightarrow 0$, which is valid when $\hbar \omega > E_g/2$ such that it is expected that $\mathrm{Im}\{\chi^{(3)}(\omega)\} \sim \mathrm{Re}\{\chi^{(3)}(\omega)\}$, and the incident intensity is small such that pump depletion by 2PA and self-focusing by the optical Kerr effect are both small effects. In this case the measurement of 2PA and nonlinear refractive index are essentially decoupled. However, if 2PA is being measured in a regime where the optical Kerr effect is strong, this will have to be accounted for as it modifies the intensity of the pulse during propagation through the nonlinear medium.

\subsection{\label{sec:chisign}Constraints on $\mathrm{Im}\{\chi^{(3)}(\omega)\}$, $\sigma$, and $\eta$}

It is useful to understand the constraints on the third-order nonlinear susceptibility tensor and the anisotropy parameters for setting bounds on fit parameters as well as understanding the physics behind the third-order interactions. As long as there is no population inversion in the system, and certainly if the system is in the ground state, we must have 
\begin{align}
\mathrm{sgn}\Big[\beta_{12}^{\parallel}(\omega)\Big] = 1,\\
\mathrm{sgn}\Big[\beta_{12}^{\perp}(\omega)\Big] = 1,
\end{align}
for $\beta_{12}^{\parallel,\perp}(\omega)$ to represent pump-probe 2PA and not stimulated two-photon emission (TPE). We then have
\begin{align}
\mathrm{sgn}\Big[\mathrm{Im}\{\chi^{(3)}_{XXXX}(\omega)\}\Big] &= \mathrm{sgn}\Big[\beta_{12}^{\parallel}(\omega)\Big],\\
\mathrm{sgn}\Big[\mathrm{Im}\{\chi^{(3)}_{XXYY}(\omega)\}\Big] &= \mathrm{sgn}\Big[\beta_{12}^{\perp}(\omega)\Big],
\end{align}
by Eqs. 19 -- 20 in the article. We immediately note that if the rotation matrix is just $R = \hat{I}$, the identity matrix, i.e. the incident electric fields are polarized along the crystal axes, we have the constraints
\begin{align}
\mathrm{sgn}\Big[\mathrm{Im}\{\chi^{(3)}_{xxxx}(\omega)\}\Big] &= 1,\\
\mathrm{sgn}\Big[\mathrm{Im}\{\chi^{(3)}_{xxyy}(\omega)\}\Big] &= 1.
\end{align}

We can estimate the sign of the third tensor component by inspecting the form of $\mathrm{Im}\{\chi^{(3)}_{xyyx}(\omega)\}$ calculated using the perturbative classical anharmonic oscillator extension of the Lorentz model. The third-order response can be obtained assuming the potential energy function $U(\tilde{x})$ of the electrons in the crystal has the form\cite{boyd}
\begin{equation}
U(\tilde{x}) = \frac{m \omega_0^2}{2} \tilde{x}^2 - \frac{m b}{4}\tilde{x}^4,
\end{equation}
where $\tilde{x}$ is the electron position from equilibrium, $m$ is the mass of the electron, $\omega_0$ is the resonance frequency, and $b$ characterizes the strength of the nonlinearity. This model gives a degenerate third-order nonlinear optical susceptibility tensor applicable for degenerate 2PA\cite{boyd}
\begin{equation}
\chi^{(3)}_{ijkl}(\omega; \omega, \omega, - \omega) = \frac{b m \epsilon_0^3}{3 N^3 e^4}\Bigg\{\Big[ \chi^{(1)}(\omega)\Big]^3 \Big[\chi^{(1)}(\omega)\Big]^*\Bigg\} (\delta_{ij} \delta_{kl} + \delta_{ik}\delta_{jl} + \delta_{il}\delta_{jk}),
\end{equation}
where $N > 0$ is the number density of atoms, $e$ is the electric charge (and $e^4 > 0$), $\chi^{(1)}(\omega) \in \mathbb{C}$ is the linear optical susceptibility, and $\delta_{ij}$ is the Kronecker delta. We also note that $b >0$, $m>0$, $\epsilon_0 > 0$, and that $\delta_{ij} \delta_{kl} + \delta_{ik}\delta_{jl} + \delta_{il}\delta_{jk} \geq 0$. In particular, for $\mathrm{Im}\{\chi^{(3)}_{xyyx}(\omega)\}$, we have $\delta_{ij} \delta_{kl} + \delta_{ik}\delta_{jl} + \delta_{il}\delta_{jk} = 1 > 0$. 

The linear optical susceptibility (in the scalar limit) is related to the complex refractive index by
\begin{align}
1 + \mathrm{Re}\{\chi^{(1)}(\omega)\} &= n^2 - k^2,\\
\mathrm{Im}\{\chi^{(1)}(\omega)\} &= 2 n k,
\end{align}
where $n$ is the real part of the linear refractive index and $k$ is the imaginary part. We note that $n > 1$ for most materials, and certainly $n > 0$ except for metamaterials. In addition $k \geq 0$ for materials in the ground state, or at least with no population inversion. Thus the signs of the real and imaginary parts of the linear optical susceptibilities are
\begin{align}
\mathrm{sgn}\Big(\mathrm{Re}\{\chi^{(1)}(\omega)\}\Big) &= \mathrm{sgn}\Big[ (n^2 - 1) - k^2\Big] = (-1,0,1),\\
\mathrm{sgn}\Big(\mathrm{Im}\{\chi^{(1)}(\omega)\}\Big) &= \mathrm{sgn}(nk) = 1.
\end{align}
Thus the real part of the linear optical susceptibility is in principle unconstrained while the imaginary part is strictly positive. 

Now we can compute the signs of the real and imaginary parts of this third-order nonlinear optical susceptibility tensor component
\begin{align}
\begin{split}
\mathrm{sgn}\Big[ \mathrm{Re}\{\chi^{(3)}_{xyyx}(\omega)\}\Big] &= \mathrm{sgn}\bigg\{ \Big[\mathrm{Re}\{\chi^{(1)}(\omega)\}\Big]^4 - \Big[\mathrm{Im}\{\chi^{(1)}(\omega)\}\Big]^4\bigg\}\\
&= \mathrm{sgn}\bigg\{ \Big[ (n^2 - 1) - k^2\Big]^4 - (2 n k)^4 \bigg\} = (-1,0,1),
\end{split}\\
\mathrm{sgn}\Big[ \mathrm{Im}\{\chi^{(3)}_{xyyx}(\omega)\}\Big] &= \mathrm{sgn}\Big[ \mathrm{Re}\{\chi^{(1)}(\omega)\} \Big] = \mathrm{sgn}\Big[ (n^2 - 1) - k^2\Big] = (-1,0,1),
\end{align}
and thus both the real and imaginary parts of this third-order nonlinear optical susceptibility tensor component may in general be unconstrained. 

In this work, however, GaAs, GaP, and Si have $n \approx 3$ and $k \approx 0$ over the entire spectral range studied. Thus we expect
\begin{align}
\mathrm{sgn}\Big[ \mathrm{Re}\{\chi^{(3)}_{xyyx}(\omega)\}\Big] &\approx \mathrm{sgn}\bigg\{ \Big[ (3^2 - 1) - 0^2\Big]^4 - \Big[(2(3)(0)\Big]^4\bigg\} = \mathrm{sgn}(8^4) = 1,\\
\mathrm{sgn}\Big[ \mathrm{Im}\{\chi^{(3)}_{xyyx}(\omega)\}\Big] &\approx \mathrm{sgn}\Big[(3^2 - 1) - 0^2 \Big] = \mathrm{sgn}(8) = 1,
\end{align}
and thus we expect $\mathrm{Im}\{\chi^{(3)}_{xyyx}(\omega)\} \geq 0$. 

We next obtain the constraints on the anisotropy parameters. We immediately observe that 
\begin{equation}
\eta(\omega) = \frac{\mathrm{Im}\{\chi^{(3)}_{xyyx}(\omega)\}}{\mathrm{Im}\{\chi^{(3)}_{xxxx}(\omega)\}} \geq 0.
\end{equation}

The anisotropy parameter $\sigma$ is constrained by the upper bound $\sigma < 1.5$ by the condition that\cite{dvorak}
\begin{equation}
\beta_{12}^{\parallel} \sim \Big( 1 - \sigma + \sigma \sum_i |e_i|^4\Big) \geq 0,
\end{equation}
where $e_i$ are the pump direction cosines, defined by $e_i = \hat{X} \cdot \hat{i}$ which can be represented in spherical coordinates by 
\begin{equation}
\begin{split}
e_1 &= \sin \theta \cos \phi\\
e_2 &= \sin \theta \sin \phi\\
e_3 &= \cos \theta,
\end{split}
\end{equation}
for which we can calculate the range
\begin{equation}
\mathcal{R}\Big\{ \sum_i |e_i|^4 \Big\} = [1/3,1],
\end{equation}
and thus $\sigma \leq 3/2$. This same bound can be obtained from the 110-direction used in this work by considering Eq. 23 in the article, whereby
\begin{equation}
\Big[3 \cos(4\theta)-4\cos(2\theta)-7\Big]\sigma(\omega) + 16 \geq 0.
\end{equation}

We can obtain an even tighter bound by considering the cross-polarized 2PA coefficient. We have the condition\cite{dvorak}
\begin{equation}
\beta_{12}^{\perp} \sim \Big( \frac{1 - \sigma - \eta}{2} + \sigma \sum_i |e_i|^2 |p_i|^2 \Big) \geq 0
\end{equation}
where $p_i$ are the probe direction cosines, defined by $p_i = \hat{Y} \cdot{i}$ and $p_i e^i = 0$, which can be represented in spherical coordinates by
\begin{equation}
\begin{split}
p_1 &= \sin (\theta+\pi/2) \cos \phi\\
p_2 &= \sin (\theta+\pi/2) \sin \phi\\
p_3 &= \cos (\theta+\pi/2).
\end{split}
\end{equation}
We can then calculate the range
\begin{equation}
\mathcal{R}\Big\{  \sum_i |e_i|^2 |p_i|^2 \Big\} = [0,1/2],
\end{equation}
and thus $\sigma \leq 1 - \eta$ and $\eta \leq 1$. The first of these bounds can also be obtained from the 110-direction used in this work by referencing Eq. 24 in the article for the cross-polarized 2PA coefficient for which we must have
\begin{equation}
8 \Big[ 1 - \eta(\omega)\Big] - \Big[ 3 \cos(4\theta) + 5\Big] \sigma(\omega) \geq 0,
\end{equation}
for the interaction between the two beams to be 2PA and not stimulated 2PE. This implies the constraint $\sigma \leq 1 - \eta$.

Thus we conclude that in this work, we expect
\begin{align}
\Big\{\mathrm{Im}\{\chi^{(3)}_{xxxx}(\omega)\}, \mathrm{Im}\{\chi^{(3)}_{xxyy}(\omega)\}, \mathrm{Im}\{\chi^{(3)}_{xyyx}(\omega)\}\Big\} &\geq 0,\\
\sigma &\leq 1 - \eta,\\
0 &\leq \eta \leq 1.
\end{align}

\section{\label{sec:comparisons}Relationships and definitions for comparisons of results from this work to literature}

\subsection{\label{sec:betacomparisons}Comparison of co-polarized pump-probe $\beta_{12}^{\parallel}$ to single-beam $\beta$ in literature}

The degenerate co-polarized pump-probe 2PA coefficients reported in this work ($\beta_{12}^{\parallel})$ can be compared to single-beam 2PA coefficients reported in literature ($\beta$) by
\begin{equation}
\label{eqbeta}
\beta = \frac{1}{2} \beta_{12}^{\parallel}
\end{equation}
where $\beta_{12}^{\parallel}$ is the 2-beam 2PA coefficient in co-polarized geometry at angle $-\theta_0$ from the $\overline{1}10$-axis and $\beta$ is the single-beam 2PA coefficient. Often the orientation of the crystal relative to the polarization is not recorded, nor is the anisotropy. 

\subsection{\label{sec:anisocomparisons}Comparison of anisotropy parameters $\sigma$ and $\eta$ to literature}

The anisotropy parameter 
\begin{equation}
\sigma(\omega)= 1 - \Bigg[\frac{2\ \mathrm{Im}\{\chi_{xxyy}^{(3)}(\omega)\} + \mathrm{Im}\{\chi_{xyyx}^{(3)}(\omega)\}}{\mathrm{Im}\{\chi_{xxxx}^{(3)}(\omega)\}}\Bigg]
\end{equation}
is a material property independent of the orientation of the crystal relative to the polarization and thus allows direct comparison to literature results. $\sigma$ is reported explicitly for GaAs in Dvorak, et al.,\cite{dvorak} as $\sigma = -0.76 \pm 0.08$ at $\hbar \omega = 1.31$ eV, for GaAs in DeSalvo, et al.,\cite{desalvo} as $\sigma = -0.74 \pm 0.18$ at $\hbar \omega = 1.17$ eV, and for GaAs and Si in ab initio calculations by Murayama, et al.\cite{murayama} It was calculated from related parameters for GaAs in other works: In Bepko, et al.,\cite{bepko} $\sigma = -A = -0.225 \pm 0.011$ at $\hbar \omega = 1.17$ eV for the 110-direction; and in Bechtel, et al.\cite{bechtel} a magnitude of anisotropy was reported as $\max\ \delta \beta_{12}(\theta)/\beta_{12} = 20\%$ and in Said, et al.\cite{said} $\max\ \delta \beta_{12}(\theta)/\beta_{12} \leq 10\%$ for the 110-direction for which $\sigma = -\frac{3}{2}\ \max\ \delta \beta_{12}(\theta)/\beta_{12}$, which gives $\sigma = -0.3$ and $|\sigma| \leq 0.15$, respectively, both at $\hbar \omega = 1.17$ eV. An upper bound on $|\sigma|$ in Si was calculated for a result reported by Bristow, et al.\cite{bristow} that anisotropy larger than $\max\ \delta \beta_{12}(\theta)/\beta_{12} = 5\%$ was not detectable in the 001-direction, for which $\sigma = -\max\ \delta \beta_{12}(\theta)/\beta_{12} = 0.00 \pm 0.05$ over the range $\hbar \omega = 0.56$ -- $0.89$ eV. 

The values of $\eta$ from Dvorak, et al.\cite{dvorak} in GaAs and from ab initio calculations by Murayama, et al.\cite{murayama} can be calculated directly from the reported relative ratio of $\mathrm{Im}\{\chi^{(3)}_{xxxx}\}$ to $\mathrm{Im}\{\chi^{(3)}_{xyyx}\}$ by
\begin{equation}
\eta(\omega) = \frac{\mathrm{Im}\{\chi_{xyyx}^{(3)}(\omega)\}}{\mathrm{Im}\{\chi_{xxxx}^{(3)}(\omega)\}},
\end{equation}
for which $\eta = 0.6 \pm 0.2$ at $\hbar \omega = 1.31$ eV.

\subsection{\label{sec:chicomparisons}Comparison of $\mathrm{Im}\{\chi^{(3)}(\omega)\}$ to literature}

The definition of $\mathrm{Im}\{\chi^{(3)}(\omega)\}$ used by Dvorak, et al.\cite{dvorak} and their reported values of the tensor components for GaAs at $\hbar \omega = 1.31$ eV differ by the treatment we use here by a factor of 3. They define
\begin{equation}
\beta_{pe} = \frac{\omega}{\epsilon_0 \big[n(\omega)\big]^2 c^2} \mathrm{Im}\{\chi^{(3)}_{eff}\},
\end{equation}
where $pe$ refers to probe-excite (probe-pump) 2PA and $\mathrm{Im}\{\chi^{(3)}_{eff}\}$ is the effective imaginary part of the third-order nonlinear optical susceptibility tensor, which is equivalent to $\mathrm{Im}\{\chi^{(3)}_{XXXX}\}$ for co-polarized beams and $\mathrm{Im}\{\chi^{(3)}_{XXYY}\}$ for cross-polarized beams in this work. Thus, in comparing the values of the tensor components in Dvorak, et al. to this work, we must divide Dvorak, et al.'s results by a factor of 3. In addition, both Dvorak, et al. and the ab initio calculations by Murayama, et al.\cite{murayama} define the degenerate third-order nonlinear optical susceptibility tensor for the fundamental frequency as
\begin{equation}
\chi^{(3)}_{abcd}(\omega; - \omega, \omega, \omega),
\end{equation}
and thus by intrinsic permutation symmetry (simultaneous interchange of indices and corresponding frequency arguments),\cite{boyd} we can compare the results in Dvorak, et al. to our work by making the transformation
\begin{equation}
\mathrm{Im}\{\chi^{(3)}_{abcd}(\omega)\} = \frac{1}{3} \mathrm{Im}\{X^{(3)}_{adcb}(\omega)\},
\end{equation}
and we can compare the ab initio calculations in Murayama, et al. to our work by making the transformation
\begin{equation}
\mathrm{Im}\{\chi^{(3)}_{abcd}(\omega)\} = \mathrm{Im}\{X^{(3)}_{adcb}(\omega)\},
\end{equation}
where $\mathrm{Im}\{\chi^{(3)}_{abcd}(\omega)\}$ is the transformed tensor directly comparable to results from this work, and $\mathrm{Im}\{X^{(3)}_{adcb}(\omega)\}$ are the susceptibility tensor components reported in Dvorak, et al. and Murayama, et al., respectively. We can write this transformation explicitly as
\begin{equation}
\begin{split}
\mathrm{Im}\{\chi^{(3)}_{xxxx}(\omega)\} &= A\ \mathrm{Im}\{X^{(3)}_{xxxx}(\omega)\}\\
\mathrm{Im}\{\chi^{(3)}_{xxyy}(\omega)\} &= A\ \mathrm{Im}\{X^{(3)}_{xyyx}(\omega)\}\\
\mathrm{Im}\{\chi^{(3)}_{xyyx}(\omega)\} &= A\ \mathrm{Im}\{X^{(3)}_{xxyy}(\omega)\},
\end{split}
\end{equation}
where $A = 1/3$ for the results from Dvorak, et al. and $A = 1$ for the ab initio calculations from Murayama, et al. In addition, the orientation of the lab $X$-axis defined in Dvorak, et al. differs trivially from this work by a rotation about $Z = 110$ of $\theta_0 = \pm \pi/2$. 
 
 \clearpage

\section{\label{sec:ellipsometry}Linear optical measurements}

Linear optical properties can be inferred from the complex reflectance ratio $\rho$, 
\begin{equation}
\label{eqcomplexreflectance}
\rho = \frac{r_p}{r_s} = \tan \Psi\ e^{i \Delta},
\end{equation}
which can be measured by variable-angle spectroscopic ellipsometry (VASE). Here $(r_p, r_s)$ are the ($p$-polarized, $s$-polarized) complex Fresnel amplitude reflection coefficients, $\Psi$ characterizes the modulus of the complex reflectance ratio, and $\Delta$ characterizes the phase. 

\begin{figure}[ht]
\begin{centering}
  \includegraphics[scale=0.85]{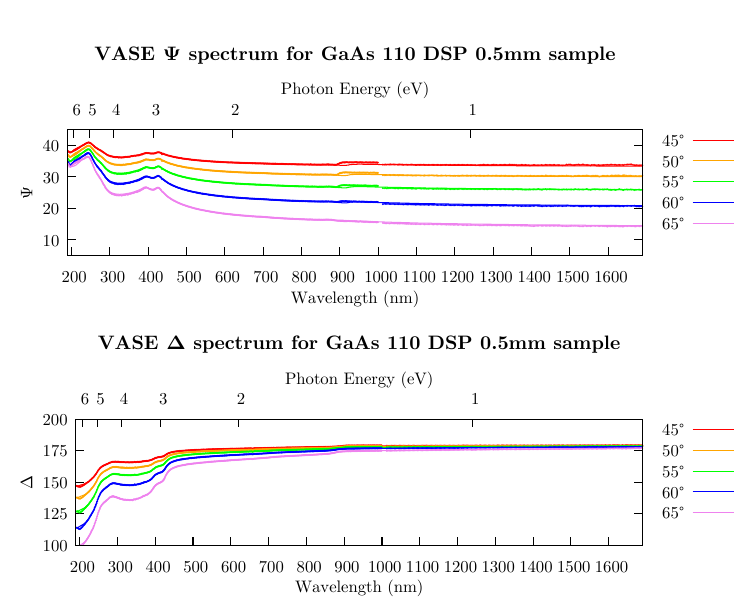}
  \caption{\label{fig:ellipsometrygaas}VASE spectra of $\Psi$ (top) and $\Delta$ (bottom) measured for GaAs (points) and corresponding global fits to determine the complex refractive index (curves).}
  \end{centering}
\end{figure}

\begin{figure}[ht]
\begin{centering}
  \includegraphics[scale=0.85]{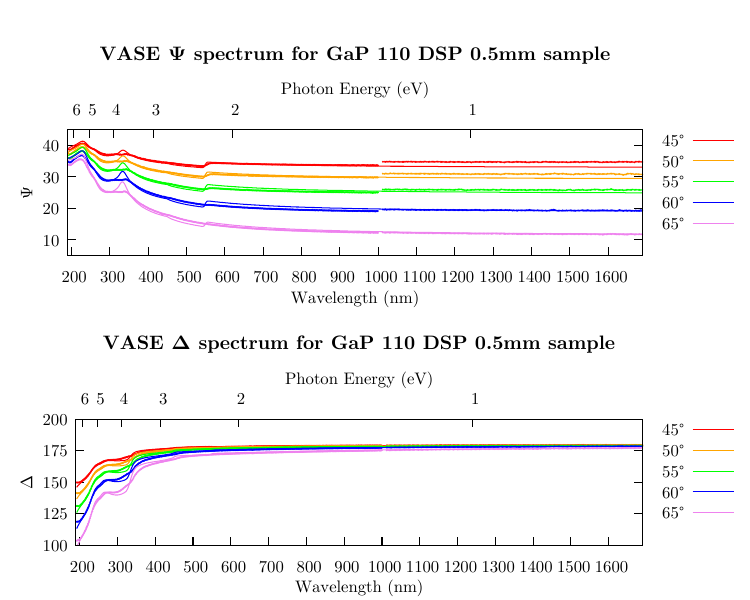}
  \caption{\label{fig:ellipsometrygap}VASE spectra of $\Psi$ (top) and $\Delta$ (bottom) measured for GaP (points) and corresponding global fits to determine the complex refractive index (curves).}
  \end{centering}
\end{figure}

\begin{figure}[ht]
\begin{centering}
  \includegraphics[scale=0.85]{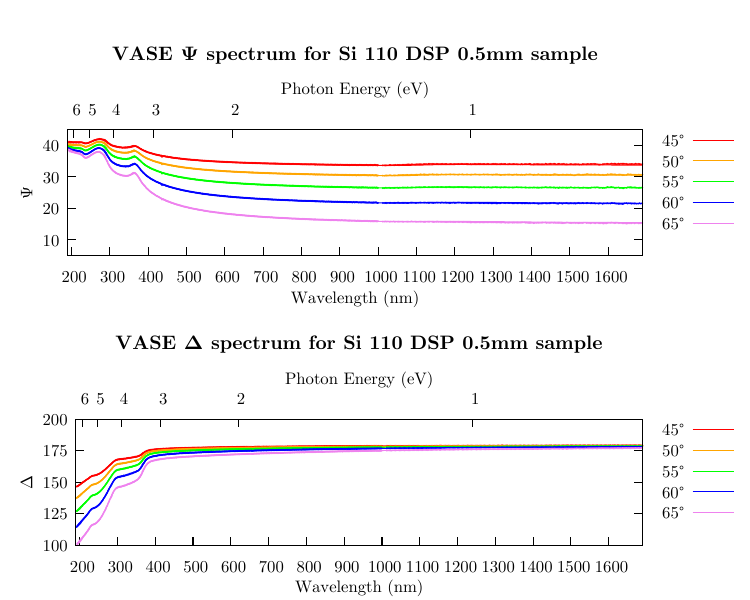}
  \caption{\label{fig:ellipsometrysi}VASE spectra of $\Psi$ (top) and $\Delta$ (bottom) measured for Si (points) and corresponding global fits to determine the complex refractive index (curves).}
  \end{centering}
\end{figure}

\begin{figure}[ht]
\begin{centering}
  \includegraphics[scale=1.0]{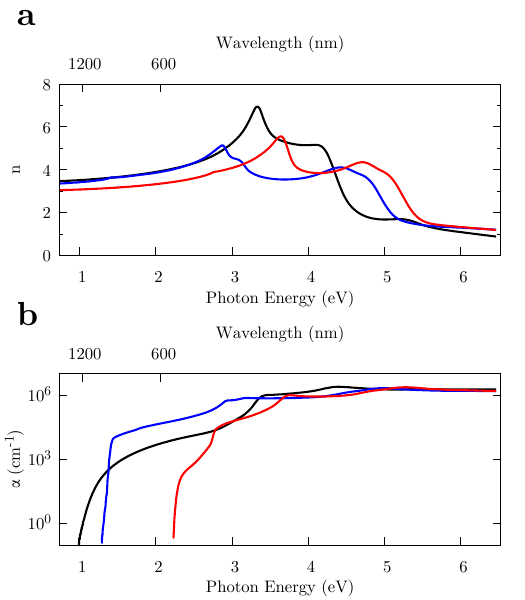}
  \caption{\label{fig:ellipsometry}Linear optical properties of samples from resulting fits to data measured using VASE. a) Linear refractive index $n$ spectrum of GaAs (blue), GaP (red), and Si (black). b) Linear absorption coefficient $\alpha$ spectrum of GaAs (blue), GaP (red), and Si (black).}
  \end{centering}
\end{figure}

The samples studied in this work were measured at incidence angles of $45^{\circ}$, $50^{\circ}$, $55^{\circ}$, $60^{\circ}$, and $65^{\circ}$ over the spectral range 190 -- 1690 nm using a J.A. Woollam M2000 spectroscopic ellipsometer. The spectra of $\Psi$ and $\Delta$ are shown in Figs. \ref{fig:ellipsometrygaas} -- \ref{fig:ellipsometrysi}. A global fit using J.A. Woollam CompleteEase software\cite{woollam} obtained the complex refractive index $\tilde{n} = n + i k$ of each sample. Fig. \ref{fig:ellipsometry} shows the real part of the refractive index $n$ and the linear absorption coefficient $\alpha = 4 \pi k / \lambda$ for each sample. The mean squared errors of fit were 5.22, 17.1, and 2.76 for GaAs, GaP, and Si, respectively. The values of $n$ are used in calculations of the Fresnel power reflection coefficients and $\mathrm{Im}\big\{ \chi_{abcd}^{(3)}(\omega)\big\}$, and 2PA measurements were performed where $\alpha \approx 0$. Values of $n$ and $\alpha$ for $\lambda > 1690$ nm were obtained from literature.\cite{skauli,li}

\FloatBarrier

\section{\label{sec:beamchar}Spatiotemporal profile characterization}

\subsection{\label{sec:beam1}Beam radii, Rayleigh ranges, and intersection angles}

\begin{figure}[ht]
\begin{centering}
  \includegraphics[scale=1.0]{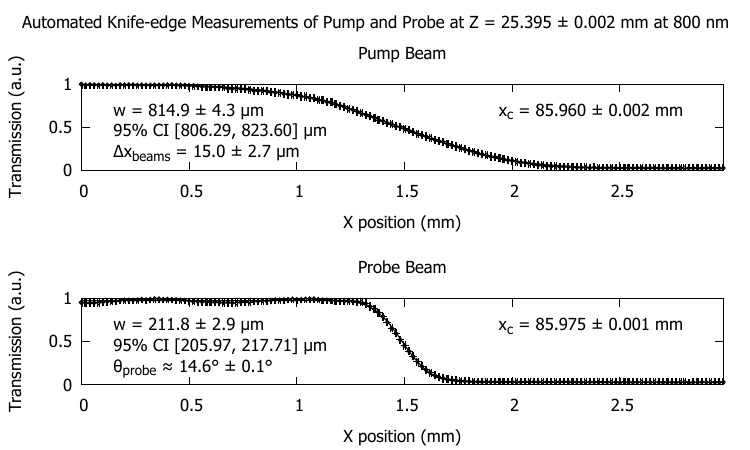}
  \caption{\label{fig:knifeedge}Knife-edge transects of pump (top) and probe (bottom) beams at overlap position at $\lambda = 800$ nm.}
  \end{centering}
\end{figure}

\begin{figure}[ht]
\begin{centering}
  \includegraphics[scale=1.0]{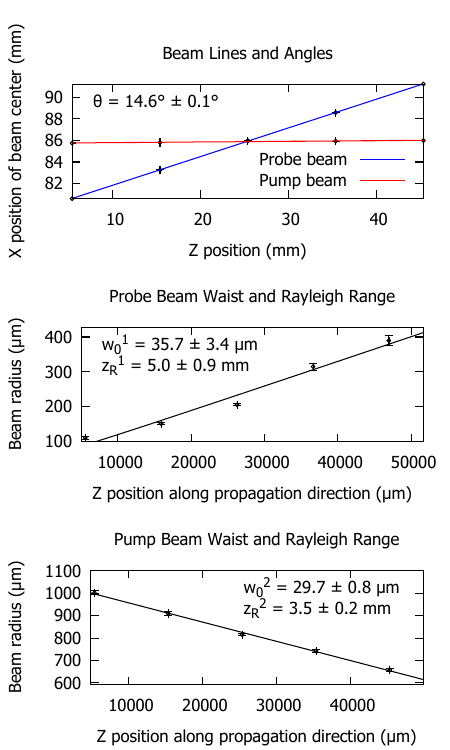}
  \caption{\label{fig:rayleighangles}Characterization of pump and probe beams at $\lambda = 800$ nm. Top) Beam center positions as a function of $Z$ distance to calculate the intersection angle between the two beams. Middle) Probe beam radius as a function of $Z$ fit to Gaussian beam to calculate Rayleigh range $Z_R$. Bottom) Pump beam radius as a function of $Z$ fit to Gaussian beam to calculate $Z_R$.}
  \end{centering}
\end{figure}

The Light Conversion TOPAS-C optical parametric amplifier with the Coherent Libra HE USP titanium-doped sapphire regenerative amplifier nominally provide a Gaussian transverse mode and Gaussian temporal profile. Additionally, we used a spatial filter to ensure good transverse beam quality, and the beam radii were measured using an automated knife-edge technique. The transmitted intensity $T$ as a funciton of knife-edge position $X$ is
\begin{equation}
\label{eqknifeedge}
T(\Delta X) = \frac{a}{2} \Bigg\{ 1 - \mathrm{erf}\bigg[\frac{\sqrt{2}\Delta X}{w(Z)}\bigg] \Bigg\} + b,
\end{equation}
where $a$ is an amplitude parameter, $b$ is an intensity offset parameter, $\Delta X = X - X_0$ is the transverse position of the knife-edge, $X_0$ is the beam center position parameter, and $w(Z)$ is the Gaussian beam radius parameter at longitudinal position $Z$. The knife-edge is positioned to measure the beam radii of both the pump and probe at the overlap position and again at postions $\pm 10$ and $\pm 20$ mm from the overlap in order to determine the Rayleigh ranges ($Z_R$) and beam intersection angles ($\Psi$). The intersection angle was used to calculate the Fresnel power reflection coefficient ($R_j$) and used to correct for the projection of the measured probe beam radius due to the knife-edge transecting perpendicularly to the pump beam. An example of this analysis is shown in Figs. \ref{fig:knifeedge} - \ref{fig:rayleighangles}. 

\FloatBarrier

\subsection{\label{sec:pulseduration}Pulse duration}

The pulse duration can be measured using a second-order autocorrelation technique, where the pump and probe beams propagate through a barium borate nonlinear crystal. Second-harmonic generation (SHG) occurs along the bisector of the two beams when the two beams are temporally and spatially overlapped and phase-matching conditions are met. Adjusting the relative time delay between the two pulses $\Delta t = t - t_0$ produces a peak of the form
\begin{equation}
\label{eqshg}
T(\Delta t) = \frac{3 a}{\sinh^2(2.7196\ \Delta t/\tau_a)}\Bigg[ \bigg(\frac{2.7196\ \Delta t}{\tau_a}\bigg) \coth\bigg(\frac{2.7196\ \Delta t}{\tau_a} - 1\bigg) \Bigg] + b,
\end{equation}
where $a$ is an amplitude parameter, $b$ is an intensity offset parameter, $\Delta t = t - t_0$ is the time delay, measured time $t$ from the peak center position parameter $t_0$, and $\tau_a = 1.54 \tau_{FWHM}$ is the autocorrelation width parameter which is related to the full-width at half-maximum (FWHM) pulse duration for a hyperbolic secant squared pulse by a factor of 1.54. In our article, a Gaussian profile was assumed, for which the corresponding Gaussian pulse duration parameter is related to the FWHM pulse duration by $\tau_g = \tau_{FWHM}/\sqrt{2 \ln{2}}$. 

\begin{figure}[ht]
\begin{centering}
  \includegraphics[scale=1.0]{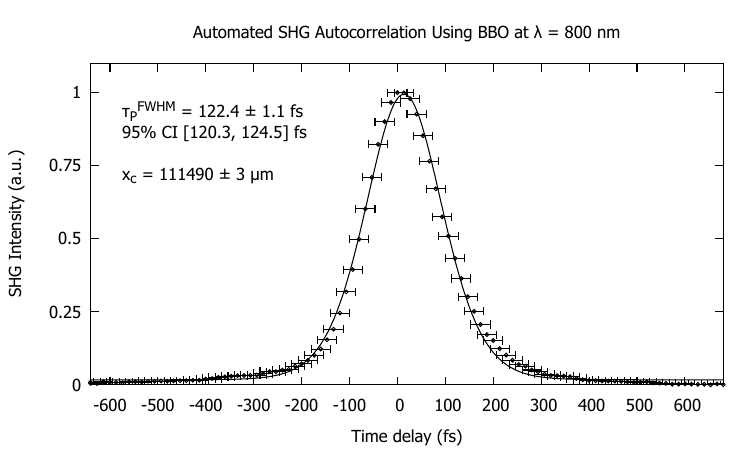}
  \caption{\label{fig:shg}Noncollinear two-beam SHG second-order autocorrelation at $\lambda = 800$ nm using a barium borate nonlinear crystal (black circles) and fit (black curve).}
  \end{centering}
\end{figure}

This method allows verification of the pulse duration independently of the two-photon absorption (2PA) time delay scans, which allows 2PA to be distinguished from competing nonlinear optical effects by the width and shape of the transmission peak. It also enabled the rough temporal overlap position to be determined before proceeding with 2PA time delay scans. An example of this analysis is shown in Fig. \ref{fig:shg}.

\FloatBarrier

\section{\label{sec:systemsi}Systematic correction for $Z$-position of Si in time delay scans}

The sample holder thickness for Si differed by $\delta Z = 412 \pm 13$ $\mu$m from the sample holders for GaAs and GaP. Therefore, a systematic correction was needed to account for the probe beam being slightly off coincidence with respect to the pump beam on the Si sample. Eq. 4 in the article can be modified with the probe beam off-center with respect to the pump beam by an amount $\delta X = \delta Z \tan(\psi)$ where $\psi$ is the angle between the two beams. The correction factor for $\beta_{12}$ is then
\begin{equation}
\label{eqsizcorrect}
\frac{\beta_{12}^{\mathrm{corrected}}}{\beta_{12}^{\mathrm{original}}} = \Bigg[ \frac{\pi}{2\big(\frac{1}{w_1^2} + \frac{1}{w_2^2}\big)}\Bigg] \bigg/ \int_{- \infty}^{+ \infty} \int_{- \infty}^{+ \infty} e^{-2\big[ (X - \delta X)^2 + Y^2\big]/w_1^2} e^{-2(X^2 + Y^2)/w_2^2}\ \mathrm{d}X\ \mathrm{d}Y.
\end{equation}

The result of the correction factors used in this analysis are summarized in Table \ref{tab:sizcorrect}. Note that the correction factors for $\lambda = 1800$ and $\lambda = 2000$ nm are identity since the Si sample $z$-position was independently optimized after performing scans on the other samples at these excitation wavelengths. These values agreed with an empirical method by calculating the ratio between $\Delta T_1/T_1$ in the $z$-scan of the other samples between the peak and the position $\delta Z$ from the peak. This systematic error was small, on the order of 10\%, but is nonetheless documented and corrected for.

\begin{table}[ht]
\centering
 \caption{\label{tab:sizcorrect}2PA coefficient correction factors for $z$-offset in Si sample.}
  \begin{tabular}{c c}
  \hline
  $\lambda$ (nm) & $\beta_{12}^{\mathrm{corrected}}/\beta_{12}^{\mathrm{original}}$\\
  \hline
   1200 & 1.062\\
   1300 & 1.072\\
   1400 & 1.135\\
   1600 & 1.138\\
   1800 & 1.000\\
   2000 & 1.000\\
   \hline
  \end{tabular}
\end{table}

\FloatBarrier

\section{\label{sec:datarchive}Data archive}

Plots of the data from the time delay and rotation scans used in the 2PA analysis of this article are presented here along with tables of the fit parameters. The raw data that support the findings of this study are available from the corresponding author upon reasonable request.

The run number, which identifies the scan of a given sample, is located in the lower right corner of each plot. These can be used to lookup the corresponding parameters in the following tables. The values and uncertainties of measured parameters are either reported as a mean value and propagated symmetric uncertainty of the form $p \pm \delta p$, or as a mean value and confidence interval (CI) of the form $p,\ [p - \delta p^{lo}, p + \delta p ^{up}]$. When multiple scans were performed at a single excitation energy, the mean value of $\beta_{12}(\omega)$ and $\mathrm{Im}\{\chi_{abcd}^{(3)}(\omega)\}$ was reported and the uncertainty was conservatively quantified by the maximum of the set including the standard error of the mean, the propagated uncertainty of the mean, and the propagated uncertainty in the individual measurements.

\FloatBarrier

\subsection{\label{sec:2paparameters}Summary of 2PA parameters}

The co-polarized pump-probe 2PA coefficients reported here can be compared to single-beam values by Eq. \ref{eqbeta}.

\begin{table}[ht]
\centering
 \caption{\label{tab:gaas2pasum}Summary of 2PA coefficients for GaAs.}
  \begin{tabular}{c c c c c c c c c}
  \hline
  $\lambda$& $\hbar \omega$& Pol. & $\beta_{12}^{\parallel}$& CI & $\beta_{12}^{\perp}$& CI\\
  \begin{tiny}(nm)\end{tiny}& \begin{tiny}(eV)\end{tiny}& & \begin{tiny} (cm GW$^{-1}$)\end{tiny}& \begin{tiny} (cm GW$^{-1}$)\end{tiny}& \begin{tiny} (cm GW$^{-1}$)\end{tiny}& \begin{tiny} (cm GW$^{-1}$)\end{tiny}\\
  \hline
  1000 & 1.24 & s & 16 & [12,19] & 3.6 & [2.6,4.4]\\
  1100 & 1.13 & s & 120 & [60,150] & 24 & [13,30]\\
  1200 & 1.03 & s & 54 & [21,101] & 13 & [9,17]\\
  1300 & 0.95 & s & 23 & [20,41] & 6.14 & [5.14,9.64]\\
  1400 & 0.89 & s & 2.6 & [2.1,5.7] & 0.87 & [0.71,1.83]\\
  1600 & 0.77 & s & 2.9 & [2.4,4.8] & 1.1 & [0.9,1.7]\\
  \hline
  \end{tabular}
\end{table}

\begin{table}[ht]
\centering
 \caption{\label{tab:gaasanisosum}Summary of 2PA anisotropy parameters for GaAs.}
  \begin{tabular}{c c c c c c c c c}
  \hline
  $\lambda$& $\hbar \omega$& Pol. & $\sigma$ & $\delta \sigma$ & $\eta$ & $\delta \eta$ & $\theta_0$& $\delta \theta_0$\\
  \begin{tiny}(nm)\end{tiny}& \begin{tiny}(eV)\end{tiny}& & & & & \begin{tiny}(deg)\end{tiny}& \begin{tiny}(deg)\end{tiny}\\
  \hline
  1000 & 1.24 & s & -0.23 & 0.05 & 0.75 & 0.07 & 37.1 & 1.6\\
  1100 & 1.13 & s & -0.40 & 0.05 & 0.62 & 0.09 & 32.5 & 1.4\\
  1200 & 1.03 & s & -0.27 & 0.08 & 0.5 & 1.6 & 31 & 4\\
  1300 & 0.95 & s & -0.29 & 0.05 & 0.61 & 0.10 & 31.5 & 1.7\\
  1400 & 0.89 & s & -0.36 & 0.06 & 0.2 & 0.3 & 38 & 2\\
  1600 & 0.77 & s & -0.13 & 0.08 & 0.74 & 0.17 & 34.0 & 1.7\\
  \hline
  \end{tabular}
\end{table}

\begin{table}[ht]
\centering
 \caption{\label{tab:gap2pasum}Summary of 2PA coefficients for GaP.}
  \begin{tabular}{c c c c c c c c c}
  \hline
 $\lambda$& $\hbar \omega$& Pol. & $\beta_{12}^{\parallel}$& CI & $\beta_{12}^{\perp}$& CI\\
  \begin{tiny}(nm)\end{tiny}& \begin{tiny}(eV)\end{tiny}& & \begin{tiny} (cm GW$^{-1}$)\end{tiny}& \begin{tiny} (cm GW$^{-1}$)\end{tiny}& \begin{tiny} (cm GW$^{-1}$)\end{tiny}& \begin{tiny} (cm GW$^{-1}$)\end{tiny}\\
  \hline
  700 & 1.77 & p & 5.4 & [3.7,11.9] & 0.9 & [0.6,2.1]\\
  800 & 1.55 & p & 3.9 & [3.4,4.6] & 1.03 & [0.93,1.24]\\
  900 & 1.38 & s & 0.561 & [0.466,1.100] & 0.24 & [0.29,0.39]\\
  \hline
  \end{tabular}
\end{table}

\begin{table}[ht]
\centering
 \caption{\label{tab:gapanisosum}Summary of 2PA anisotropy parameters for GaP.}
  \begin{tabular}{c c c c c c c c c}
  \hline
  $\lambda$& $\hbar \omega$& Pol. & $\sigma$ & $\delta \sigma$ & $\eta$ & $\delta \eta$ & $\theta_0$& $\delta \theta_0$\\
  \begin{tiny}(nm)\end{tiny}& \begin{tiny}(eV)\end{tiny}& & & & & \begin{tiny}(deg)\end{tiny}& \begin{tiny}(deg)\end{tiny}\\
  \hline
  700 & 1.77 & p & -0.61 & 0.09 & 0.4 & 0.6 & 133.4 & 1.7\\
  800 & 1.55 & p & -0.88 & 0.10 & 0.31 & 0.18 & 132.5 & 1.0\\
  900 & 1.38 & s & -0.53 & 0.06 & 0.35 & 0.14 & 47.1 & 1.2\\
  \hline
  \end{tabular}
\end{table}

\begin{table}[ht]
\centering
 \caption{\label{tabsi2pasum}Summary of 2PA coefficients for Si.}
  \begin{tabular}{c c c c c c c c c}
  \hline
  $\lambda$& $\hbar \omega$& Pol. & $\beta_{12}^{\parallel}$& CI & $\beta_{12}^{\perp}$& CI\\
  \begin{tiny}(nm)\end{tiny}& \begin{tiny}(eV)\end{tiny}& & \begin{tiny} (cm GW$^{-1}$)\end{tiny}& \begin{tiny} (cm GW$^{-1}$)\end{tiny}& \begin{tiny} (cm GW$^{-1}$)\end{tiny}& \begin{tiny} (cm GW$^{-1}$)\end{tiny}\\
  \hline
  1200 & 1.03 & s & 4.5 & [3.2,5.8] & 2.2 & [1.6,2.9]\\
  1300 & 0.95 & s & 0.41 & [0.33,1.33] & 0.18 & [0.14,0.22]\\
  1400 & 0.89 & s &0.29 & [0.23,0.80] & 0.16 & [0.13,0.34]\\
  1600 & 0.77 & s &0.43 & [0.31,0.51] & 0.20 & [0.15,0.24]\\
  \hline
  \end{tabular}
\end{table}

\begin{table}[ht]
\centering
 \caption{\label{tab:sianisosum}Summary of 2PA anisotropy parameters for Si.}
  \begin{tabular}{c c c c c c c c c}
  \hline
  $\lambda$& $\hbar \omega$& Pol. & $\sigma$ & $\delta \sigma$ & $\eta$ & $\delta \eta$ & $\theta_0$& $\delta \theta_0$\\
  \begin{tiny}(nm)\end{tiny}& \begin{tiny}(eV)\end{tiny}& & & & & \begin{tiny}(deg)\end{tiny}& \begin{tiny}(deg)\end{tiny}\\
  \hline
  1200 & 1.03  & s & -0.29 & 0.08 & 0.3 & 0.6 & 80 & 4\\
  1300 & 0.95 & s & -0.15 & 0.14 & 0.84 & 0.16 & 40 & 40\\
  1400 & 0.89 & s & -0.03 & 0.03 & 0.96 & 0.06 & 0 & 2\\
  1600 & 0.77 & s & -0.55 & 0.09 & 0.4 & 0.3 & 93 & 2\\
  \hline
  \end{tabular}
\end{table}

\FloatBarrier

\subsection{\label{sec:chi3summary}Summary of $\mathrm{Im}\{\chi^{(3)}(\omega)\}$ tensor spectra}

There are three independent, non-vanishing components of the imaginary part of the degenerate third-order nonlinear optical susceptibility tensor describing degenerate 2PA for crystals of $\overline{4}3m$ (GaAs, GaP) and $m3m$ (Si) symmetries. These are $\mathrm{Im}\{\chi_{xxxx}^{(3)}(\omega)\}$, $\mathrm{Im}\{\chi_{xxyy}^{(3)}(\omega)\} = \mathrm{Im}\{\chi_{xyxy}^{(3)}(\omega)\}$, and $\mathrm{Im}\{\chi_{xyyx}^{(3)}(\omega)\}$. 

\begin{table}[ht]
\centering
 \caption{\label{tab:gaaschisum}Summary of $\mathrm{Im}\{\chi^{(3)}_{abcd}(\omega)\}$ for GaAs. The units of $\mathrm{Im}\{ \chi^{(3)}_{abcd}\}$ are indicated as [$\chi^{(3)}$] = (E-21 m$^2$ V$^{-2}$).}
  \begin{tabular}{c c c c c c c c}
  \hline
  $\lambda$& $\hbar \omega$& $\mathrm{Im}\{\chi^{(3)}_{xxxx}\}$ & CI & $\mathrm{Im}\{\chi^{(3)}_{xxyy}\}$ & CI & $\mathrm{Im}\{\chi^{(3)}_{xyyx}\}$ & CI\\
  \begin{tiny}(nm)\end{tiny}& \begin{tiny}(eV)\end{tiny}& \begin{tiny}[$\chi^{(3)}$]\end{tiny}& \begin{tiny}[$\chi^{(3)}$]\end{tiny}& \begin{tiny}[$\chi^{(3)}$]\end{tiny}& \begin{tiny}[$\chi^{(3)}$]\end{tiny}& \begin{tiny}[$\chi^{(3)}$]\end{tiny} & \begin{tiny}[$\chi^{(3)}$]\end{tiny}\\
  \hline
  1000 & 1.24 & 270 & [180,350] & 270 & [180,350] & 200 & [120,290]\\ 
  1100 & 1.13 & 1800 & [800,2400] & 1800 & [800,2400] & 1100 & [400,1700]\\
  1200 & 1.03 & 900 & [0,2100] & 800 & [0,2000] & 400 & [0,2700]\\
  1300 & 0.95 & 420 & [310,750] & 400 & [290,720] & 260 & [150,520]\\
  1400 & 0.89 & 46 & [32,111] & 35 & [20,89] & 7 & [0,42]\\
  1600 & 0.77 & 80 & [50,150] & 80 & [50,150] & 60 & [20,140]\\
  \hline
  \end{tabular}
\end{table}

\begin{table}[ht]
\centering
 \caption{\label{tab:gapchisum}Summary of $\mathrm{Im}\{\chi^{(3)}(\omega)\}$ for GaP. The units of $\mathrm{Im}\{ \chi^{(3)}_{abcd}\}$ are indicated as [$\chi^{(3)}$] = (E-21 m$^2$ V$^{-2}$).}
  \begin{tabular}{c c c c c c c c}
  \hline
  $\lambda$& $\hbar \omega$& $\mathrm{Im}\{\chi^{(3)}_{xxxx}\}$ & CI & $\mathrm{Im}\{\chi^{(3)}_{xxyy}\}$ & CI & $\mathrm{Im}\{\chi^{(3)}_{xyyx}\}$ & CI\\
  \begin{tiny}(nm)\end{tiny}& \begin{tiny}(eV)\end{tiny}& \begin{tiny}[$\chi^{(3)}$]\end{tiny}& \begin{tiny}[$\chi^{(3)}$]\end{tiny}& \begin{tiny}[$\chi^{(3)}$]\end{tiny}& \begin{tiny}[$\chi^{(3)}$]\end{tiny}& \begin{tiny}[$\chi^{(3)}$]\end{tiny} & \begin{tiny}[$\chi^{(3)}$]\end{tiny}\\
  \hline
  700 & 1.77 & 34 & [18,83] & 30 & [14,76] & 11 & [0,52]\\
  800 & 1.55 & 26 & [20,34] & 29 & [23,37] & 8 & [1,16]\\
  900 & 1.38 & 5.4 & [4.1,9.8] & 5.1 & [3.7,9.2] & 1.9 & [0.7,4.3]\\
  \hline
  \end{tabular}
\end{table}

\begin{table}[ht]
\centering
 \caption{\label{tab:sichisum}Summary of $\mathrm{Im}\{\chi^{(3)}(\omega)\}$ for Si. The units of $\mathrm{Im}\{ \chi^{(3)}_{abcd}\}$ are indicated as [$\chi^{(3)}$] = (E-21 m$^2$ V$^{-2}$).}
  \begin{tabular}{c c c c c c c c}
  \hline
 $\lambda$& $\hbar \omega$& $\mathrm{Im}\{\chi^{(3)}_{xxxx}\}$ & CI & $\mathrm{Im}\{\chi^{(3)}_{xxyy}\}$ & CI & $\mathrm{Im}\{\chi^{(3)}_{xyyx}\}$ & CI\\
  \begin{tiny}(nm)\end{tiny}& \begin{tiny}(eV)\end{tiny}& \begin{tiny}[$\chi^{(3)}$]\end{tiny}& \begin{tiny}[$\chi^{(3)}$]\end{tiny}& \begin{tiny}[$\chi^{(3)}$]\end{tiny}& \begin{tiny}[$\chi^{(3)}$]\end{tiny}& \begin{tiny}[$\chi^{(3)}$]\end{tiny} & \begin{tiny}[$\chi^{(3)}$]\end{tiny}\\
  \hline
  1200 & 1.03 & 90 & [50,130] & 70 & [30,110] & 20 & [0,100]\\
  1300 & 0.95 & 10 & [5,36] & 10 & [6,37] & 9 & [3,37]\\
  1400 & 0.89 & 8 & [5, 24] & 8 & [5,25] & 8 & [4,26]\\
  1600 & 0.77 & 11 & [7,14] & 11 & [7,13] & 4 & [0,8]\\
  \hline
  \end{tabular}
\end{table}

\FloatBarrier

\subsection{\label{sec:timedelay}Time delay scans}

The wavelength $\lambda$ was selected by setting the TOPAS-C OPA to the desired setting and using a selection of spectral filters and verified with a Thorlabs CCS-200 fiber spectrometer. The polaraziation (Pol.) is indicated as $s$ or $p$, with $s$ polarization vertical and $p$ polarization horizontal in the lab. The relative angle between the polarizations of the pump and probe beams are given by $\zeta$ in degrees. The co-polarized geometry corresponds to $\zeta = 0^{\circ}$ while the cross-polarized geometry corresponds to $\zeta = 90^{\circ}$. The FWHM pulse duration $\tau_{FWHM}$ is measured using SHG second-order autocorrelation and is given in fs. The intersection angle of the beams (in vacuum) is given by $\Psi$ in degrees as measured using the automated knife-edge technique. The Fresnel power reflection coefficient is calculated using $n(\omega)$ from ellipsometry, or literature if $\lambda > 1690$ nm, and the incident angle of the pump beam onto the sample ($\approx \Psi /2$) for the given polarization. The pump pulse energy $\varepsilon_2$ is calculated from the average power and repetition rate and reported in J. The peak on-axis intensity is indicated by $I_2^0$ and reported in GW cm$^{-2}$. The ratio of the probe beam area to the pump beam area is given by the parameter $w_1^2/w_2^2$ and calculated by the automated knife-edge technique.

The fitting procedure allows for a vertical offset; this can be due to scattered pump light or an electronic offset and is reported with its uncertainty. The 2PA coefficient $\beta_{12}$ is reported with its uncertainty $\delta \beta_{12}$ in cm GW$^{-1}$ for which the single-beam 2PA coefficient can be computed using Eq. \ref{eqbeta}. The value of $b$ and its uncetainty $\delta b$ characterize the FCA and is given in Eqs. 10 -- 11 in the article. The fit was computed up to $N$ orders in the expansion of Eq. 7 in the article such that the variation in $\beta$ was less that 1\% between successive orders.

\subsubsection{\label{sec:tdgaas}GaAs}

\begin{table}[ht]
\centering
\caption{\label{tab:gaas_2pa_td_exp}Time delay scan experimental parameters for GaAs.}
\begin{tabular}{c c c c c c c c c c}
\hline
Run& $\lambda$& Pol. & $\zeta$& $\tau_{FWHM}$& $\Psi$& $R_2$& $\varepsilon_2$& $I_2^0$& $w_1^2/w_2^2$\\
 &\begin{tiny}(nm)\end{tiny} & & \begin{tiny}(deg)\end{tiny}& \begin{tiny}(fs)\end{tiny}& \begin{tiny}(deg)\end{tiny}& & \begin{tiny}(J)\end{tiny}&  \begin{tiny}(GW cm$^{-2}$)\end{tiny}& \\
\hline
1&1000&s&0&185.1&14.7&0.3123&3.49e-07&2.50e-01&1.90e-02\\
2&1000&s&90&185.1&14.7&0.3123&3.49e-07&2.50e-01&1.90e-02\\
3&1100&s&0&182.3&14.4&0.3071&5.54e-08&4.84e-02&2.81e-02\\
4&1100&s&90&182.3&14.4&0.3071&5.54e-08&4.84e-02&2.81e-02\\
5&1200&s&0&96.3&14.6&0.3037&2.06e-07&2.41e-01&2.88e-02\\
6&1200&s&90&96.3&14.6&0.3037&2.06e-07&2.41e-01&2.88e-02\\
7&1200&s&0&96.3&14.6&0.3037&3.53e-07&4.13e-01&2.88e-02\\
8&1200&s&90&96.3&14.6&0.3037&3.53e-07&4.13e-01&2.88e-02\\
9&1300&s&0&104.3&14.7&0.3012&1.46e-07&1.84e-01&4.30e-02\\
10&1300&s&90&104.3&14.7&0.3012&1.46e-07&1.84e-01&4.30e-02\\
11&1400&s&0&97.6&14.7&0.2992&4.95e-07&1.22e+00&6.02e-02\\
12&1400&s&0&97.6&14.7&0.2992&4.95e-07&1.22e+00&6.02e-02\\
13&1400&s&90&97.6&14.7&0.2992&4.95e-07&1.22e+00&6.02e-02\\
14&1600&s&0&123.0&14.7&0.2965&4.93e-07&9.91e-01&6.65e-02\\
15&1600&s&90&123.0&14.7&0.2965&4.93e-07&9.91e-01&6.65e-02\\
\hline
\end{tabular}
\end{table}

\begin{table}[ht]
\centering
\caption{\label{tab:gaas_2pa_td_fit}Time delay scan fit parameters for GaAs.}
\begin{tabular}{c c c c c c c c}
\hline
Run& Offset& $\delta$ offset& $\beta_{12}$& $\delta \beta_{12}$& $b$& $\delta b$& $N$\\
 & & & \begin{tiny}(cm GW$^{-1}$)\end{tiny}& \begin{tiny}(cm GW$^{-1}$)\end{tiny}& & &\\
\hline
1&-2.07e-03&2.90e-04&15.55&3.00&1.11e-12&1.27e-04&3\\
2&-8.20e-04&1.28e-04&3.67&0.73&1.79e-13&1.73e-04&1\\
3&-2.15e-03&1.20e-03&123.47&27.08&4.86e-16&5.98e-04&3\\
4&1.09e-03&7.11e-04&23.59&6.02&5.39e-19&1.03e-03&1\\
5&2.00e-02&4.83e-03&63.45&14.66&2.20e-21&1.18e-03&6\\
6&2.73e-03&3.08e-03&15.79&4.14&1.24e-03&4.58e-03&1\\
7&-7.41e-03&3.33e-03&45.02&10.97&1.43e-04&1.23e-03&4\\
8&-1.88e-03&3.28e-03&10.46&2.90&8.58e-15&4.92e-03&1\\
9&1.66e-03&4.61e-04&23.40&3.82&3.68e-04&2.44e-04&3\\
10&-6.31e-05&1.55e-04&6.14&1.00&1.50e-04&2.31e-04&1\\
11&1.05e-02&4.78e-04&2.62&0.51&3.46e-04&4.33e-04&2\\
12&1.02e-02&5.35e-04&2.62&0.50&4.73e-13&3.75e-04&2\\
13&2.09e-03&1.80e-04&0.87&0.17&3.59e-22&2.74e-04&1\\
14&-2.13e-03&3.57e-04&2.93&0.56&2.56e-04&3.01e-04&2\\
15&-6.21e-04&1.30e-04&1.09&0.21&8.68e-24&2.15e-04&1\\
\hline
\end{tabular}
\end{table}

\FloatBarrier

\begin{figure}[ht]
\begin{centering}
\includegraphics[scale=0.9]{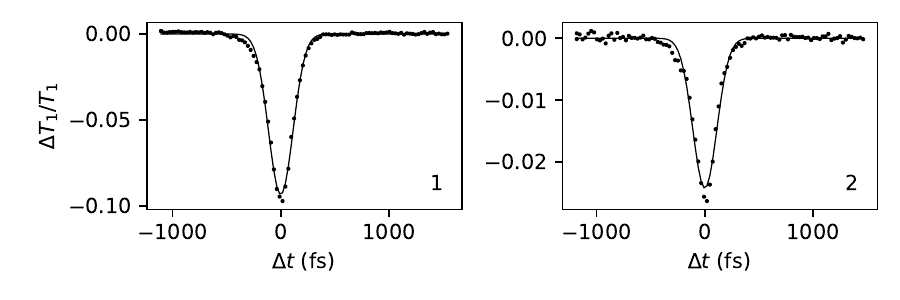}
\caption{\label{fig:td_gaas_1000}Time delay scan data (circles) and fits (curve) for GaAs at $\lambda = 1000$ nm.}
\end{centering}
\end{figure}

\begin{figure}[ht]
\begin{centering}
\includegraphics[scale=0.9]{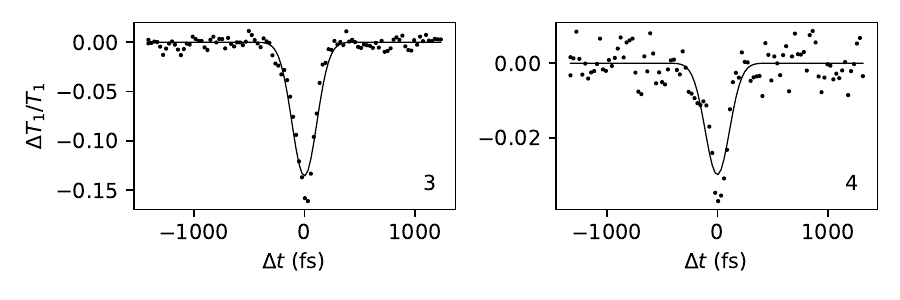}
\caption{\label{fig:td_gaas_1100}Time delay scan data (circles) and fits (curve) for GaAs at $\lambda = 1100$ nm.}
\end{centering}
\end{figure}

\begin{figure}[ht]
\begin{centering}
\includegraphics[scale=0.9]{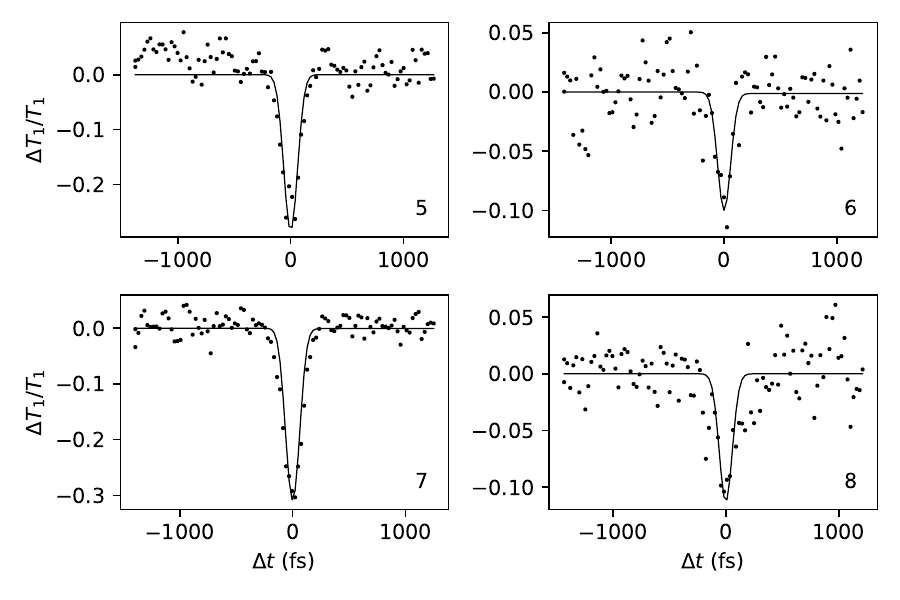}
\caption{\label{fig:td_gaas_1200}Time delay scan data (circles) and fits (curve) for GaAs at $\lambda = 1200$ nm.}
\end{centering}
\end{figure}

\begin{figure}[ht]
\begin{centering}
\includegraphics[scale=0.9]{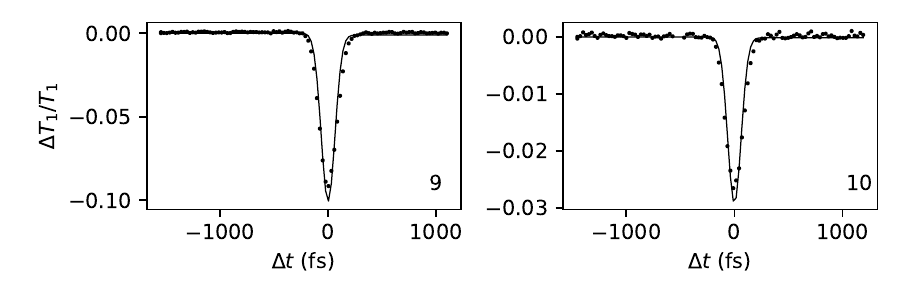}
\caption{\label{fig:td_gaas_1300}Time delay scan data (circles) and fits (curve) for GaAs at $\lambda = 1300$ nm.}
\end{centering}
\end{figure}

\begin{figure}[ht]
\begin{centering}
\includegraphics[scale=0.9]{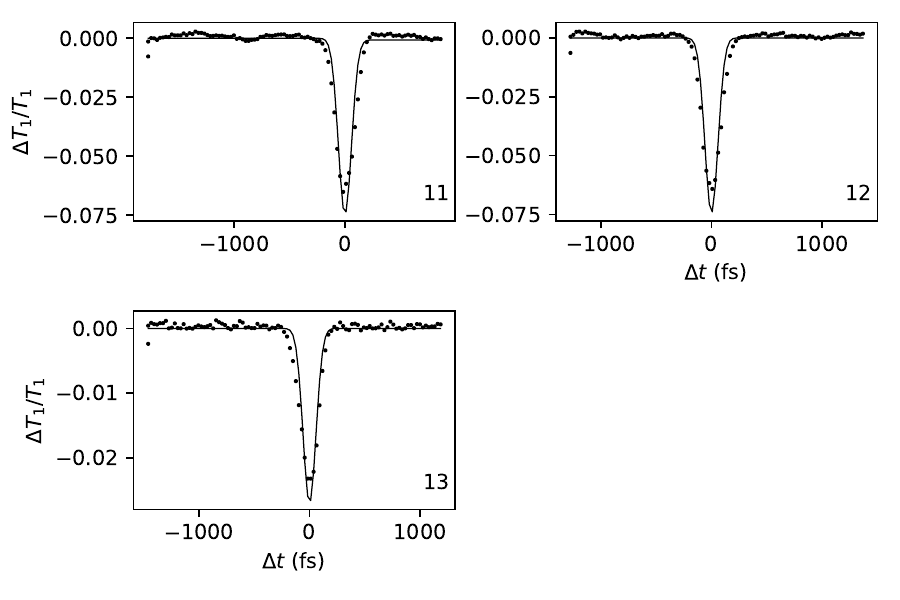}
\caption{\label{fig:td_gaas_1400}Time delay scan data (circles) and fits (curve) for GaAs at $\lambda = 1400$ nm.}
\end{centering}
\end{figure}

\begin{figure}[ht]
\begin{centering}
\includegraphics[scale=0.9]{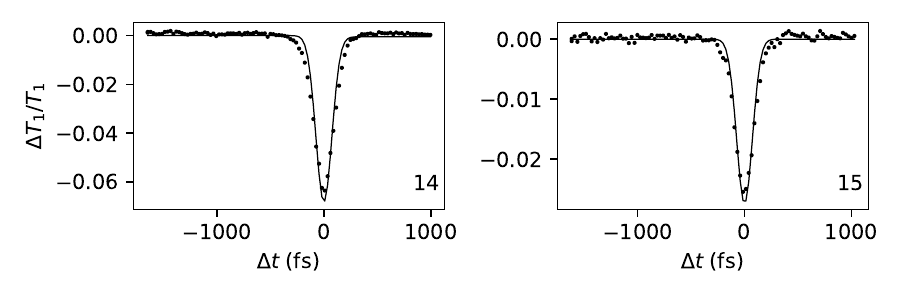}
\caption{\label{fig:td_gaas_1600}Time delay scan data (circles) and fits (curve) for GaAs at $\lambda = 1600$ nm.}
\end{centering}
\end{figure}

\FloatBarrier
\clearpage

\subsubsection{\label{sec:tdgap}GaP}

\begin{table}[ht]
\centering
\caption{\label{tab:gap_2pa_td_exp}Time delay scan experimental parameters for GaP.}
\begin{tabular}{c c c c c c c c c c}
\hline
Run& $\lambda$& Pol. & $\zeta$& $\tau_{FWHM}$& $\Psi$& $R_2$& $\varepsilon_2$& $I_2^0$& $w_1^2/w_2^2$\\
 &\begin{tiny}(nm)\end{tiny} & & \begin{tiny}(deg)\end{tiny}& \begin{tiny}(fs)\end{tiny}& \begin{tiny}(deg)\end{tiny}& & \begin{tiny}(J)\end{tiny}&  \begin{tiny}(GW cm$^{-2}$)\end{tiny}& \\
\hline
1&700&p&0&123.0&14.6&0.2771&4.49e-07&1.03e+00&1.27e-01\\
2&700&p&90&123.0&14.6&0.2771&4.49e-07&1.03e+00&1.27e-01\\
3&700&p&90&123.0&14.6&0.2771&4.49e-07&1.03e+00&1.27e-01\\
4&700&p&0&123.0&14.6&0.2771&2.69e-07&6.15e-01&1.27e-01\\
5&700&p&90&123.0&14.6&0.2771&2.69e-07&6.15e-01&1.27e-01\\
6&700&p&90&123.0&14.6&0.2771&2.69e-07&6.15e-01&1.27e-01\\
7&800&p&0&125.3&14.6&0.2698&8.99e-07&4.72e-01&6.33e-02\\
8&800&p&0&125.3&14.6&0.2698&8.99e-07&4.72e-01&6.33e-02\\
9&800&p&90&125.3&14.6&0.2698&8.99e-07&4.72e-01&6.33e-02\\
10&800&p&0&125.3&14.6&0.2698&6.04e-07&3.17e-01&6.33e-02\\
11&800&p&90&125.3&14.6&0.2698&6.04e-07&3.17e-01&6.33e-02\\
12&800&p&0&125.3&14.6&0.2698&4.49e-07&2.36e-01&6.33e-02\\
13&800&p&90&125.3&14.6&0.2698&4.49e-07&2.36e-01&6.33e-02\\
14&900&s&0&147.8&15.3&0.2709&4.80e-07&5.04e-01&1.58e-01\\
15&900&s&90&147.8&15.3&0.2709&4.80e-07&5.04e-01&1.58e-01\\
\hline
\end{tabular}
\end{table}

\begin{table}[ht]
\centering
\caption{\label{tab:gap_2pa_td_fit}Time delay scan fit parameters for GaP.}
\begin{tabular}{c c c c c c c c}
\hline
Run& Offset& $\delta$ offset& $\beta_{12}$& $\delta \beta_{12}$& $b$& $\delta b$& $N$\\
 & & & \begin{tiny}(cm GW$^{-1}$)\end{tiny}& \begin{tiny}(cm GW$^{-1}$)\end{tiny}& & &\\
\hline
1&-3.66e-03&1.05e-03&4.51&1.06&2.16e-04&5.50e-04&3\\
2&-1.05e-03&3.00e-04&1.23&0.28&1.90e-04&4.70e-04&1\\
3&-1.12e-03&3.52e-04&1.49&0.34&1.16e-04&5.52e-04&1\\
4&-2.38e-03&9.64e-04&6.24&1.68&1.02e-03&6.02e-04&3\\
5&-5.76e-04&9.85e-05&0.44&0.12&5.67e-05&1.52e-04&1\\
6&-5.69e-04&8.99e-05&0.50&0.14&6.01e-04&2.35e-04&1\\
7&1.32e-03&1.39e-04&3.44&0.25&1.22e-04&1.10e-04&2\\
8&4.41e-04&2.80e-04&2.69&0.20&3.09e-22&1.77e-04&2\\
9&-2.65e-04&8.68e-05&0.93&0.07&1.04e-09&1.20e-04&1\\
10&1.47e-03&2.21e-04&4.82&0.32&6.56e-25&1.52e-04&2\\
11&7.47e-04&8.06e-05&1.05&0.08&2.34e-24&1.06e-04&1\\
12&1.94e-03&1.48e-04&4.73&0.34&5.95e-13&1.08e-04&2\\
13&1.05e-03&9.36e-05&1.11&0.10&1.87e-04&1.30e-04&1\\
14&3.54e-04&9.23e-05&0.56&0.09&1.38e-26&1.33e-04&1\\
15&3.96e-04&4.80e-05&0.24&0.04&6.22e-24&6.88e-05&1\\
\hline
\end{tabular}
\end{table}

\FloatBarrier

\begin{figure}[ht]
\begin{centering}
\includegraphics[scale=0.9]{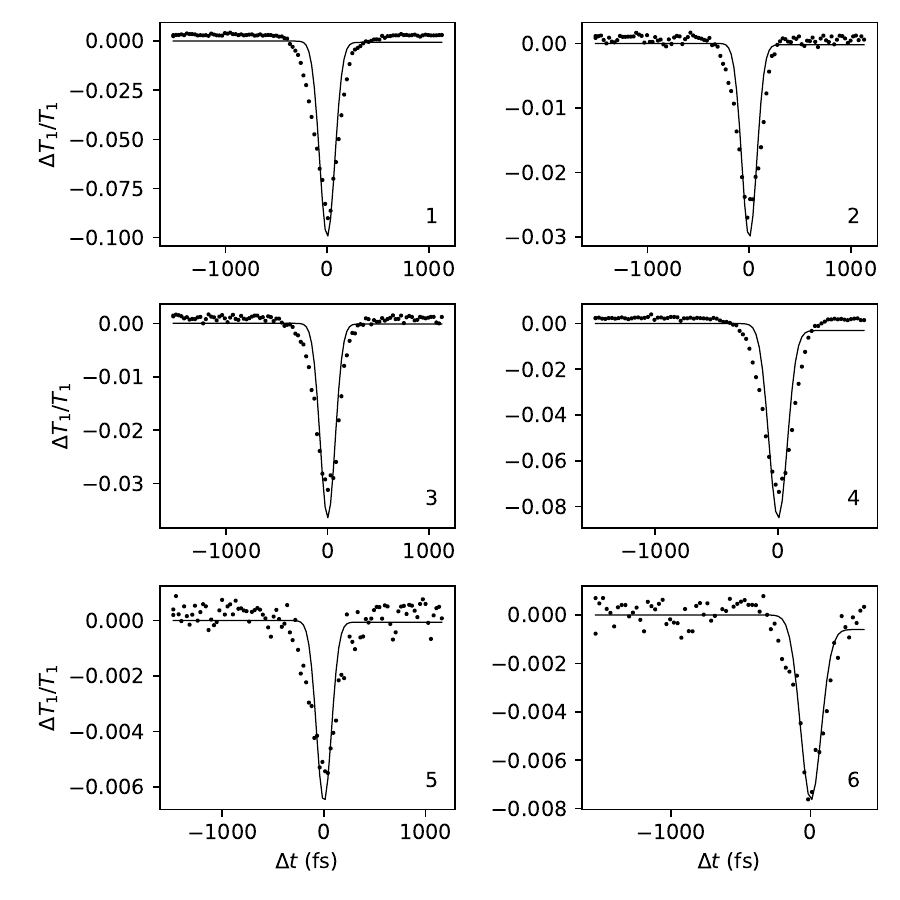}
\caption{\label{fig:td_gap_700}Time delay scan data (circles) and fits (curve) for GaP at $\lambda = 700$ nm.}
\end{centering}
\end{figure}

\begin{figure}[ht]
\begin{centering}
\includegraphics[scale=0.9]{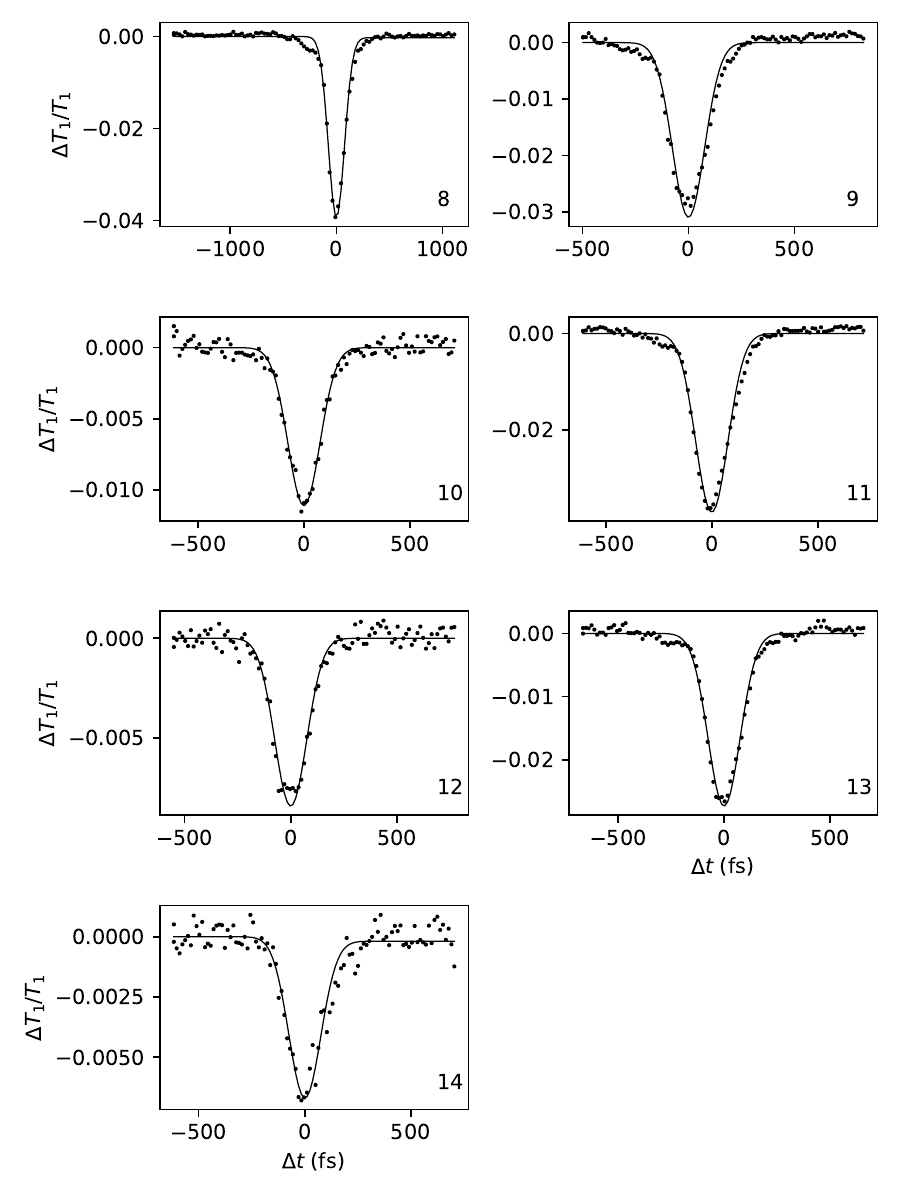}
\caption{\label{fig:td_gap_800}Time delay scan data (circles) and fits (curve) for GaP at $\lambda = 800$ nm.}
\end{centering}
\end{figure}

\begin{figure}[ht]
\begin{centering}
\includegraphics[scale=0.9]{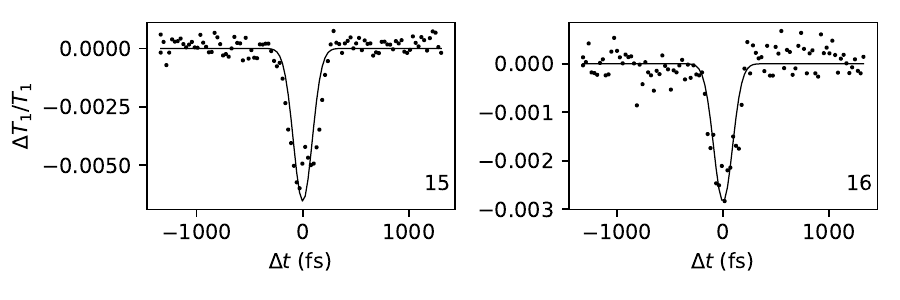}
\caption{\label{fig:td_gap_900}Time delay scan data (circles) and fits (curve) for GaP at $\lambda = 900$ nm.}
\end{centering}
\end{figure}

\FloatBarrier

\subsubsection{\label{sec:tdsi}Si}

Note that the correction factors from Table \ref{tab:sizcorrect} have not been applied to values of $\beta_{12}$ reported in this section.

\begin{table}[ht]
\centering
\caption{\label{tab:si_2pa_td_exp}Time delay scan experimental parameters for Si.}
\begin{tabular}{c c c c c c c c c c}
\hline
Run& $\lambda$& Pol. & $\zeta$& $\tau_{FWHM}$& $\Psi$& $R_2$& $\varepsilon_2$& $I_2^0$& $w_1^2/w_2^2$\\
 &\begin{tiny}(nm)\end{tiny} & & \begin{tiny}(deg)\end{tiny}& \begin{tiny}(fs)\end{tiny}& \begin{tiny}(deg)\end{tiny}& & \begin{tiny}(J)\end{tiny}&  \begin{tiny}(GW cm$^{-2}$)\end{tiny}& \\
\hline
1&1200&s&0&96.3&14.6&0.3142&7.60e-07&8.75e-01&2.88e-02\\
2&1200&s&90&96.3&14.6&0.3142&7.60e-07&8.75e-01&2.88e-02\\
3&1200&s&0&96.3&14.6&0.3142&1.57e-06&1.80e+00&2.88e-02\\
4&1200&s&90&96.3&14.6&0.3142&1.57e-06&1.80e+00&2.88e-02\\
5&1300&s&0&104.3&14.7&0.3124&1.55e-06&1.92e+00&4.30e-02\\
6&1300&s&90&104.3&14.7&0.3124&1.55e-06&1.92e+00&4.30e-02\\
7&1400&s&0&97.6&14.7&0.3110&7.47e-06&1.81e+01&6.02e-02\\
8&1400&s&0&97.6&14.7&0.3110&7.47e-06&1.81e+01&6.02e-02\\
9&1400&s&90&97.6&14.7&0.3110&7.47e-06&1.81e+01&6.02e-02\\
10&1600&s&0&123.0&14.7&0.3089&2.25e-06&4.45e+00&6.65e-02\\
11&1600&s&90&123.0&14.7&0.3089&2.25e-06&4.45e+00&6.65e-02\\
\hline
\end{tabular}
\end{table}

\begin{table}[ht]
\centering
\caption{\label{tab:si_2pa_td_fit}Time delay scan fit parameters for Si.}
\begin{tabular}{c c c c c c c c c c}
\hline
Run& Offset& $\delta$ offset& $\beta_{12}$& $\delta \beta_{12}$& $b$& $\delta b$& $N$\\
 & & & \begin{tiny}(cm GW$^{-1}$)\end{tiny}& \begin{tiny}(cm GW$^{-1}$)\end{tiny}& & &\\
\hline
1&-3.78e-03&1.17e-03&3.05&0.90&4.07e-21&8.47e-04&2\\
2&9.76e-04&1.12e-03&1.93&0.61&6.45e-23&1.59e-03&1\\
3&-4.35e-03&1.44e-03&5.43&1.19&4.23e-04&5.07e-04&4\\
4&-1.26e-03&8.79e-04&2.23&0.49&2.00e-03&1.26e-03&1\\
5&2.67e-04&2.13e-04&0.39&0.08&3.52e-05&1.55e-04&2\\
6&-2.75e-04&1.42e-04&0.17&0.04&1.75e-32&2.06e-04&1\\
7&-2.72e-03&7.22e-04&0.29&0.05&3.76e-04&4.20e-04&3\\
8&-2.66e-03&7.62e-04&0.22&0.04&2.99e-04&3.65e-04&3\\
9&-5.60e-04&4.38e-04&0.14&0.02&2.27e-04&6.32e-04&1\\
10&-9.99e-03&1.21e-04&0.38&0.07&1.63e-07&1.03e-04&2\\
11&-3.11e-04&1.02e-04&0.18&0.03&1.36e-23&1.58e-04&1\\
\hline
\end{tabular}
\end{table}

\FloatBarrier

\begin{figure}[ht]
\begin{centering}
\includegraphics[scale=0.9]{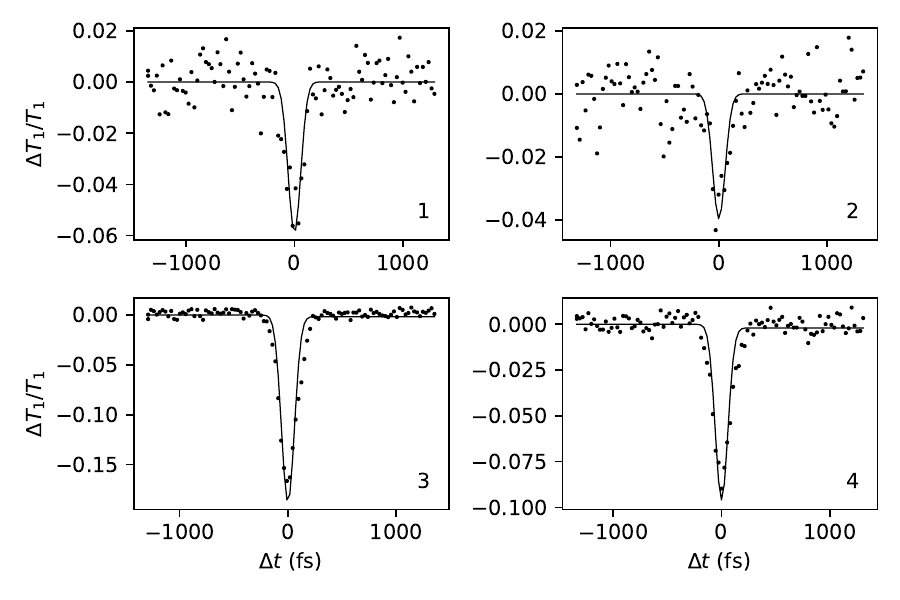}
\caption{\label{fig:td_si_1200}Time delay scan data (circles) and fits (curve) for Si at $\lambda = 1200$ nm.}
\end{centering}
\end{figure}

\begin{figure}[ht]
\begin{centering}
\includegraphics[scale=0.9]{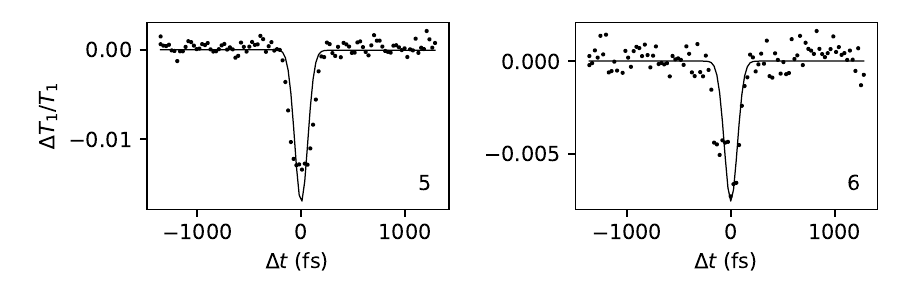}
\caption{\label{fig:td_si_1300}Time delay scan data (circles) and fits (curve) for Si at $\lambda = 1300$ nm.}
\end{centering}
\end{figure}

\begin{figure}[ht]
\begin{centering}
\includegraphics[scale=0.9]{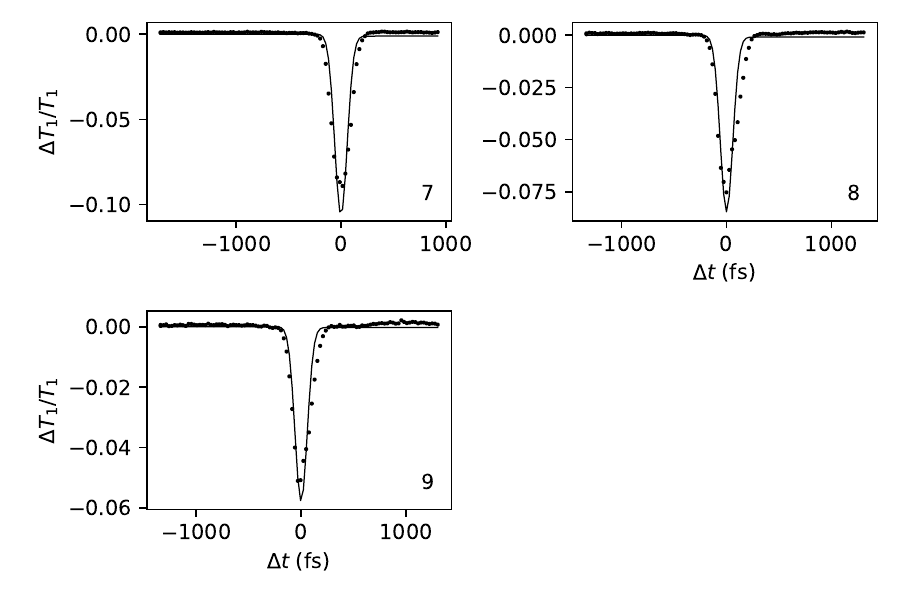}
\caption{\label{fig:td_si_1400}Time delay scan data (circles) and fits (curve) for Si at $\lambda = 1400$ nm.}
\end{centering}
\end{figure}

\begin{figure}[ht]
\begin{centering}
\includegraphics[scale=0.9]{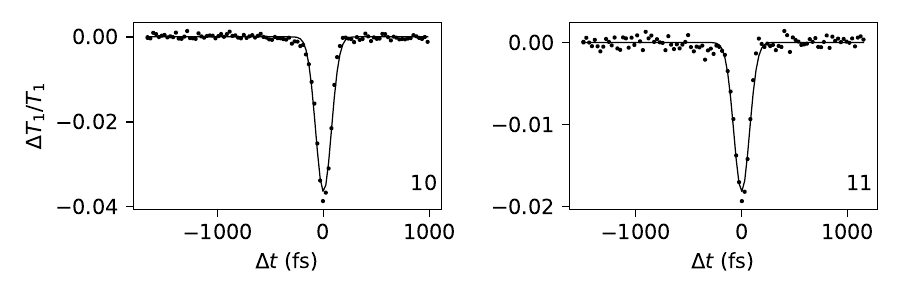}
\caption{\label{fig:td_si_1600}Time delay scan data (circles) and fits (curve) for Si at $\lambda = 1600$ nm.}
\end{centering}
\end{figure}

\FloatBarrier
\clearpage

\subsection{\label{sec:rotation}Rotation scans}

The rotation scan data is presented in a $4 \times 2$ plot format with co-polarized (left column) and cross-polarized (right column) geometries. The title of each set of plots indicates the run number. The top row displays $-\Delta T_1/T_1$ vs. $\theta$ (circles) and fit (curve). The second from top row displays $-\Delta T_1^+/T_1$ vs. $\theta$ (circles) and fit (curve). The second from bottom row displays $-\Delta T_1^-/T_1$ vs. $\theta$. The antisymmetric part is usually much weaker than the symmetric part. The bottom row displays $T_2$ as measured by reference before sample (black) and by pump PD after sample (colored).

The tables present the fit parameters obtained from the global fits of the co- and cross-polarized data. The value of $\theta_0$ and its uncertainty $\delta \theta_0$ describes the angle (in degrees) that the lab $X$-axis is rotated about $Z$ from the crystal $\overline{1}10$ axis. The value of $\sigma$, its uncertainty $\delta \sigma$, and $\eta$, and its uncertainty $\delta \eta$ are described in Eqs. 25 -- 26 in the article.

\subsubsection{\label{sec:rotgaas}GaAs}

\begin{table}[ht]
\centering
\caption{\label{tab:gaas_2pa_rot}Rotation scan experimental and fit parameters for GaAs.}
\begin{tabular}{c c c c c c c c c}
\hline
Run& $\lambda$& Pol.& $\theta_0$& $\delta \theta_0$& $\sigma$& $\delta \sigma$& $\eta$& $\delta \eta$\\
 &\begin{tiny}(nm)\end{tiny}& & \begin{tiny}(deg)\end{tiny}& \begin{tiny}(deg)\end{tiny}& & & & \\
\hline
1&1000&s&37.1&1.6&-0.232&0.046&0.75&0.07\\
2&1100&s&32.1&1.1&-0.432&0.051&0.63&0.07\\
3&1100&s&32.8&1.4&-0.359&0.051&0.61&0.09\\
4&1200&s&35.0&4.2&-0.337&0.085&0.00&1.61\\
5&1200&s&21.0&3.4&-0.385&0.082&0.52&0.50\\
6&1200&s&34.8&1.8&-0.181&0.049&0.69&0.10\\
7&1200&s&32.6&1.6&-0.184&0.051&0.75&0.08\\
8&1300&s&31.9&1.5&-0.321&0.047&0.61&0.10\\
9&1300&s&31.2&1.7&-0.264&0.043&0.62&0.10\\
10&1400&s&36.5&1.9&-0.424&0.061&0.01&0.31\\
11&1400&s&40.5&1.8&-0.295&0.044&0.29&0.18\\
12&1600&s&33.9&1.7&-0.043&0.036&0.91&0.07\\
13&1600&s&34.1&1.6&-0.212&0.040&0.56&0.11\\
\hline
\end{tabular}
\end{table}

\FloatBarrier

\begin{figure}[ht]
\begin{centering}
\includegraphics[scale=0.75]{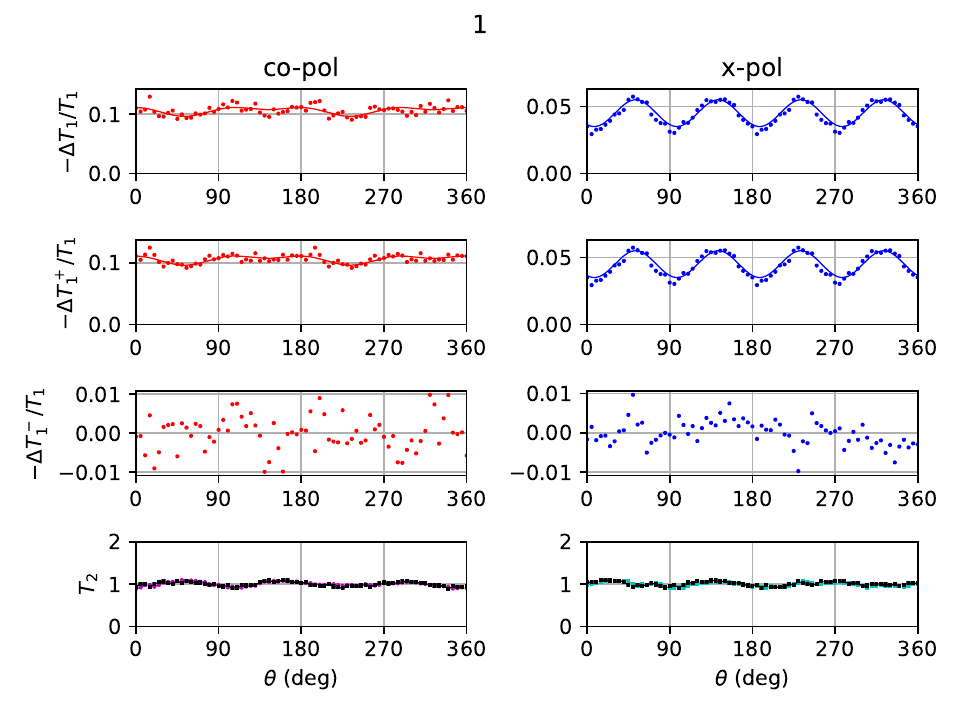}
\caption{\label{fig:rot_gaas_1000_run_1}Rotation scan data for GaAs at $\lambda = 1000$ nm for Run 1.}
\end{centering}
\end{figure}

\begin{figure}[ht]
\begin{centering}
\includegraphics[scale=0.75]{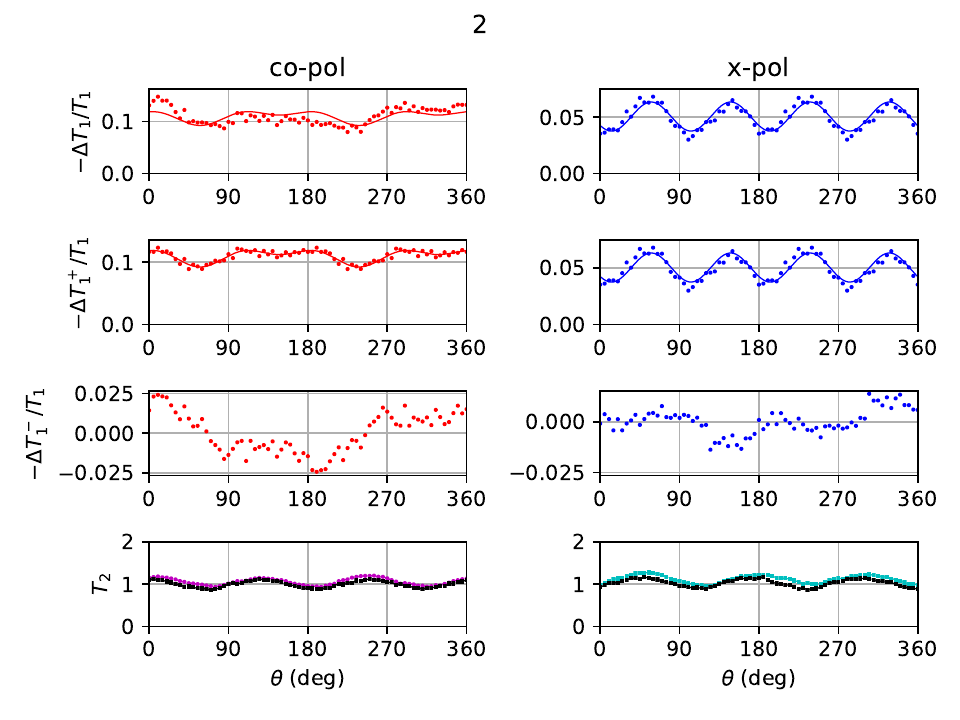}
\caption{\label{fig:rot_gaas_1100_run_2}Rotation scan data for GaAs at $\lambda = 1100$ nm for Run 2.}
\end{centering}
\end{figure}

\begin{figure}[ht]
\begin{centering}
\includegraphics[scale=0.75]{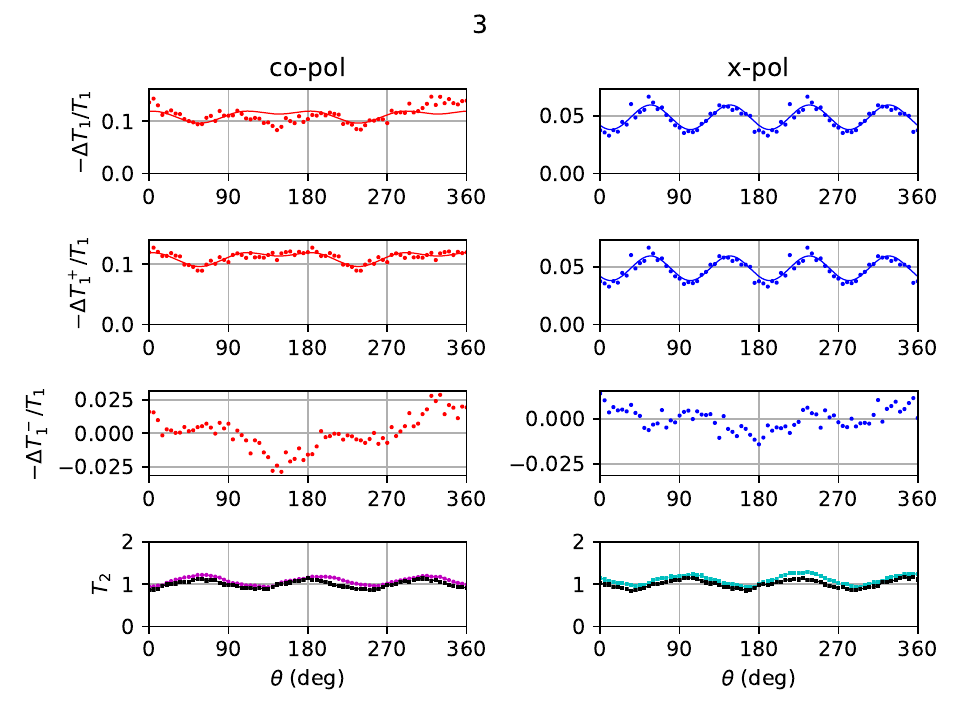}
\caption{\label{fig:rot_gaas_1100_run_3}Rotation scan data for GaAs at $\lambda = 1100$ nm for Run 3.}
\end{centering}
\end{figure}

\begin{figure}[ht]
\begin{centering}
\includegraphics[scale=0.75]{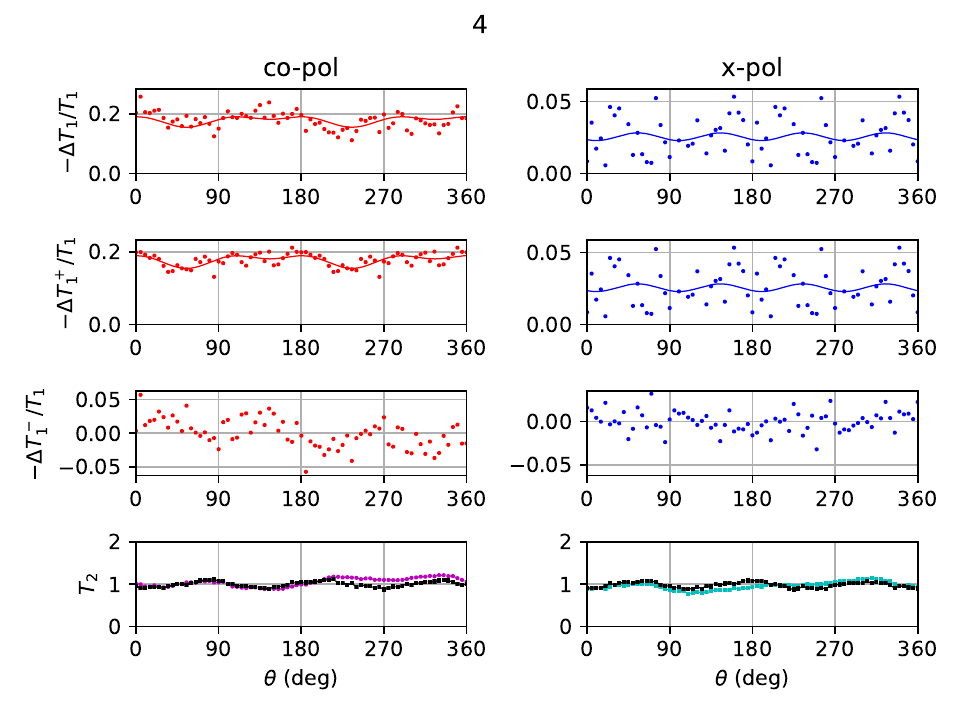}
\caption{\label{fig:rot_gaas_1200_run_4}Rotation scan data for GaAs at $\lambda = 1200$ nm for Run 4.}
\end{centering}
\end{figure}

\begin{figure}[ht]
\begin{centering}
\includegraphics[scale=0.75]{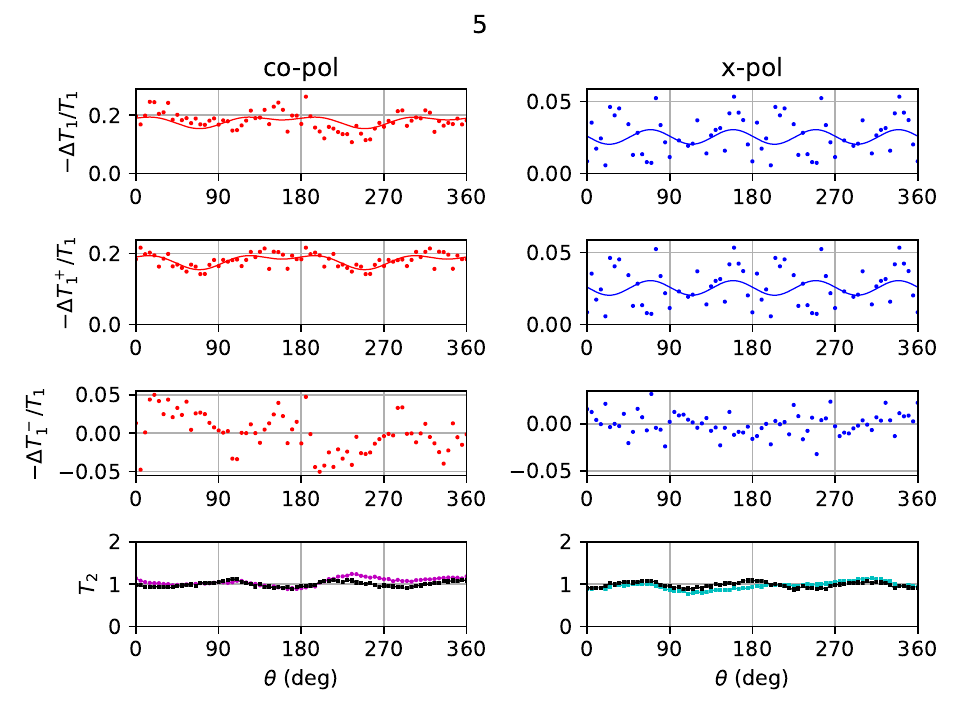}
\caption{\label{fig:rot_gaas_1200_run_5}Rotation scan data for GaAs at $\lambda = 1200$ nm for Run 5.}
\end{centering}
\end{figure}

\begin{figure}[ht]
\begin{centering}
\includegraphics[scale=0.75]{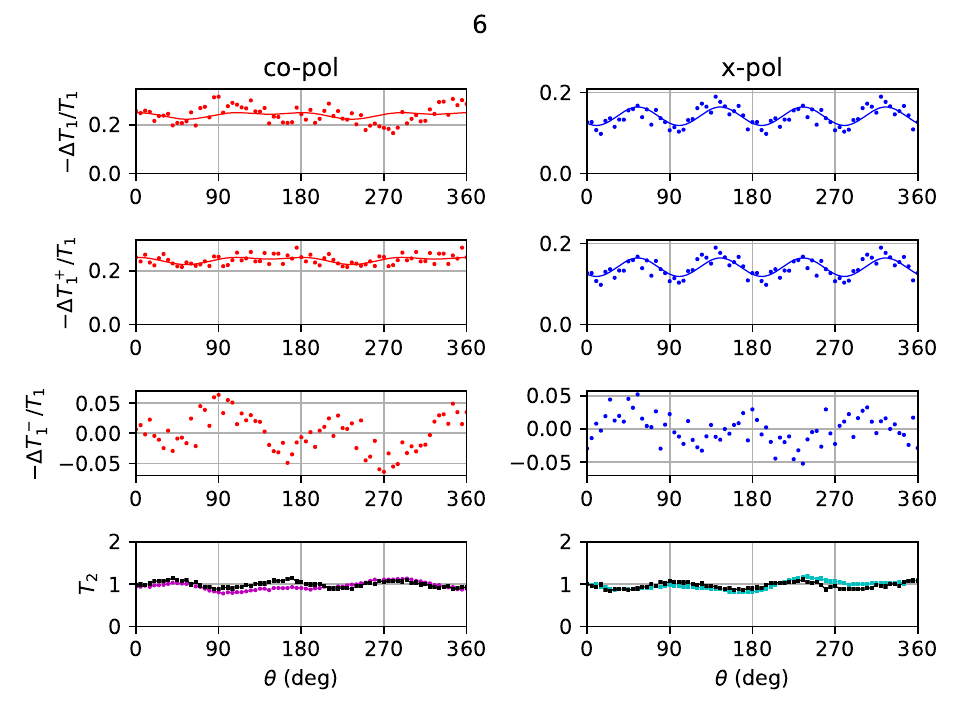}
\caption{\label{fig:rot_gaas_1200_run_6}Rotation scan data for GaAs at $\lambda = 1200$ nm for Run 6.}
\end{centering}
\end{figure}

\begin{figure}[ht]
\begin{centering}
\includegraphics[scale=0.75]{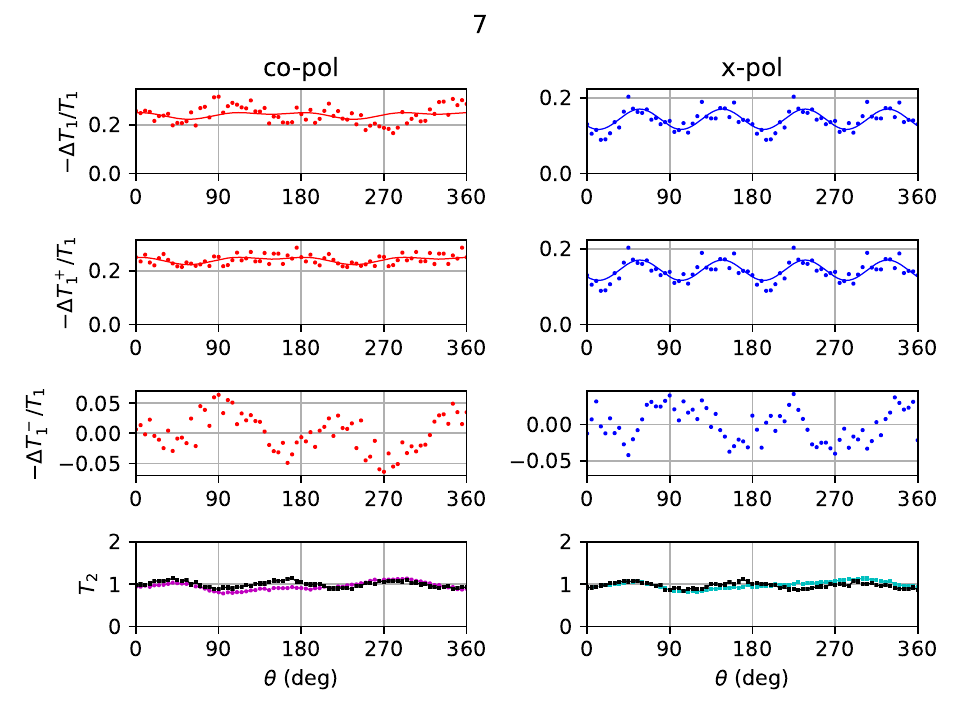}
\caption{\label{fig:rot_gaas_1200_run_7}Rotation scan data for GaAs at $\lambda = 1200$ nm for Run 7.}
\end{centering}
\end{figure}

\begin{figure}[ht]
\begin{centering}
\includegraphics[scale=0.75]{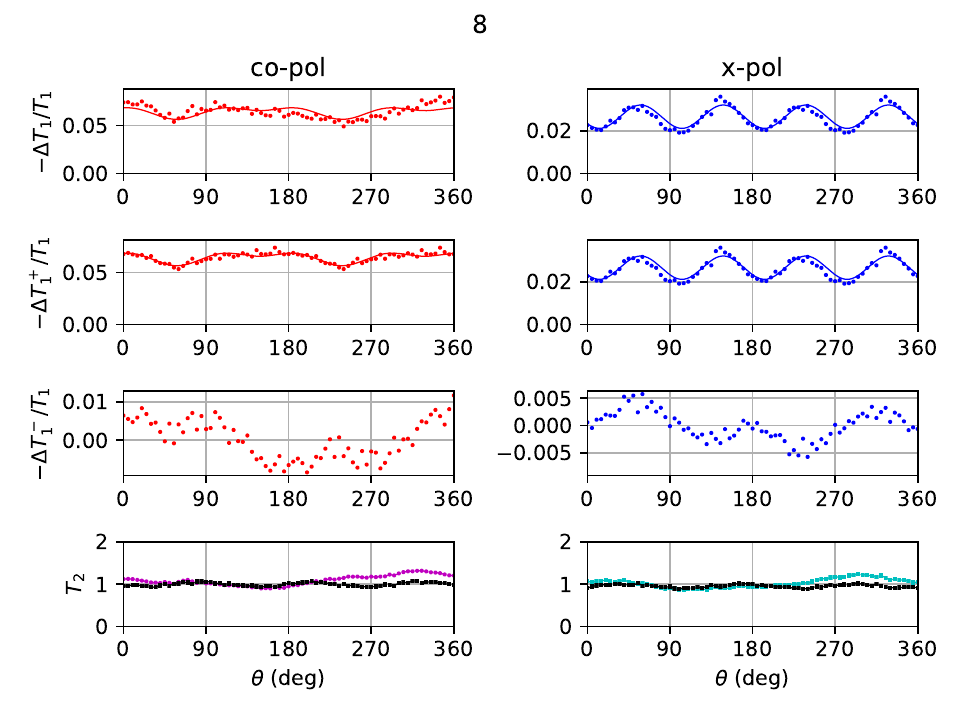}
\caption{\label{fig:rot_gaas_1300_run_8}Rotation scan data for GaAs at $\lambda = 1300$ nm for Run 8.}
\end{centering}
\end{figure}

\begin{figure}[ht]
\begin{centering}
\includegraphics[scale=0.75]{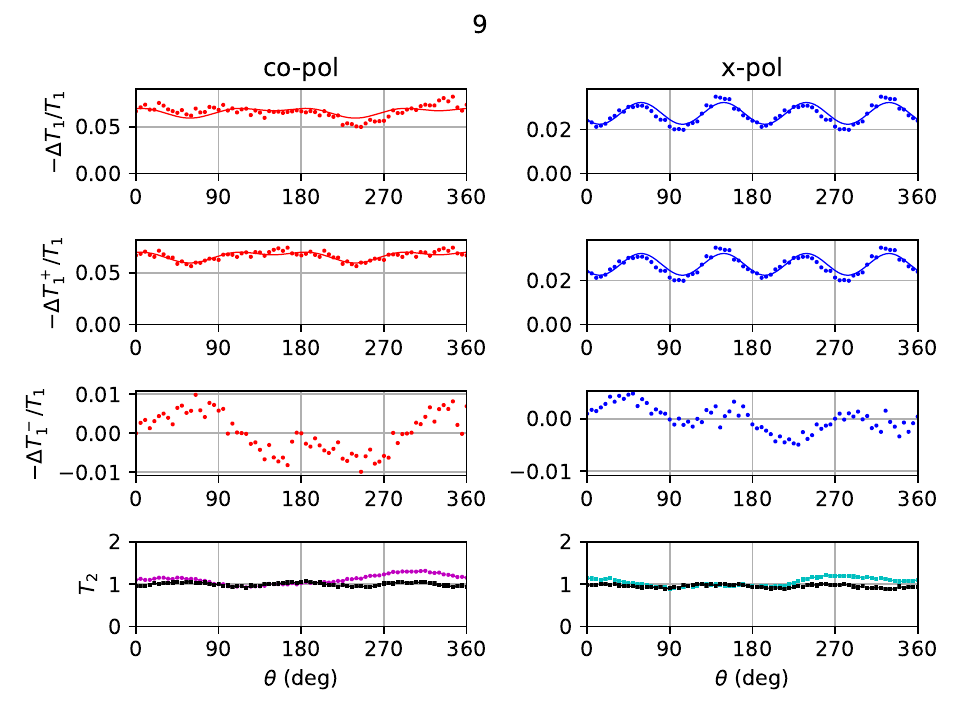}
\caption{\label{fig:rot_gaas_1300_run_9}Rotation scan data for GaAs at $\lambda = 1300$ nm for Run 9.}
\end{centering}
\end{figure}

\begin{figure}[ht]
\begin{centering}
\includegraphics[scale=0.75]{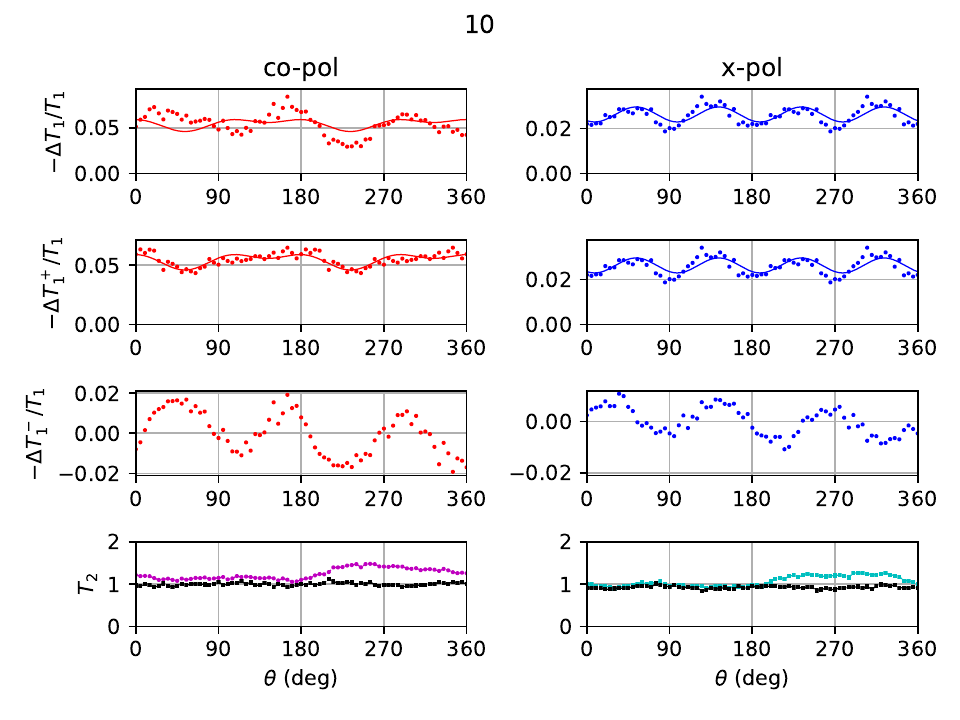}
\caption{\label{fig:rot_gaas_1400_run_10}Rotation scan data for GaAs at $\lambda = 1400$ nm for Run 10.}
\end{centering}
\end{figure}

\begin{figure}[ht]
\begin{centering}
\includegraphics[scale=0.75]{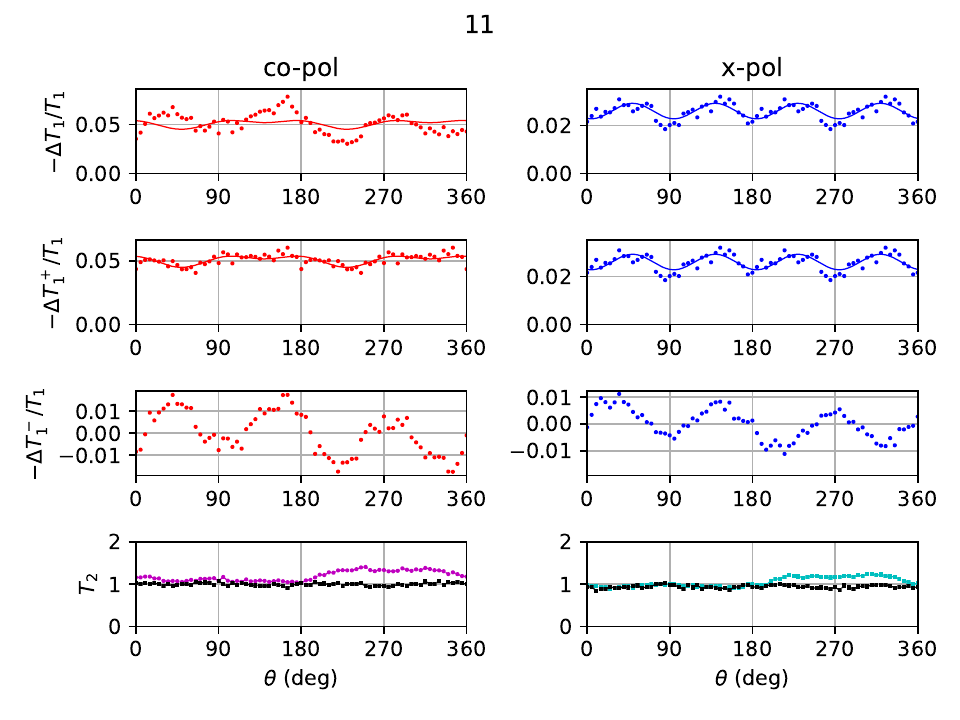}
\caption{\label{fig:rot_gaas_1400_run_11}Rotation scan data for GaAs at $\lambda = 1400$ nm for Run 11.}
\end{centering}
\end{figure}

\begin{figure}[ht]
\begin{centering}
\includegraphics[scale=0.75]{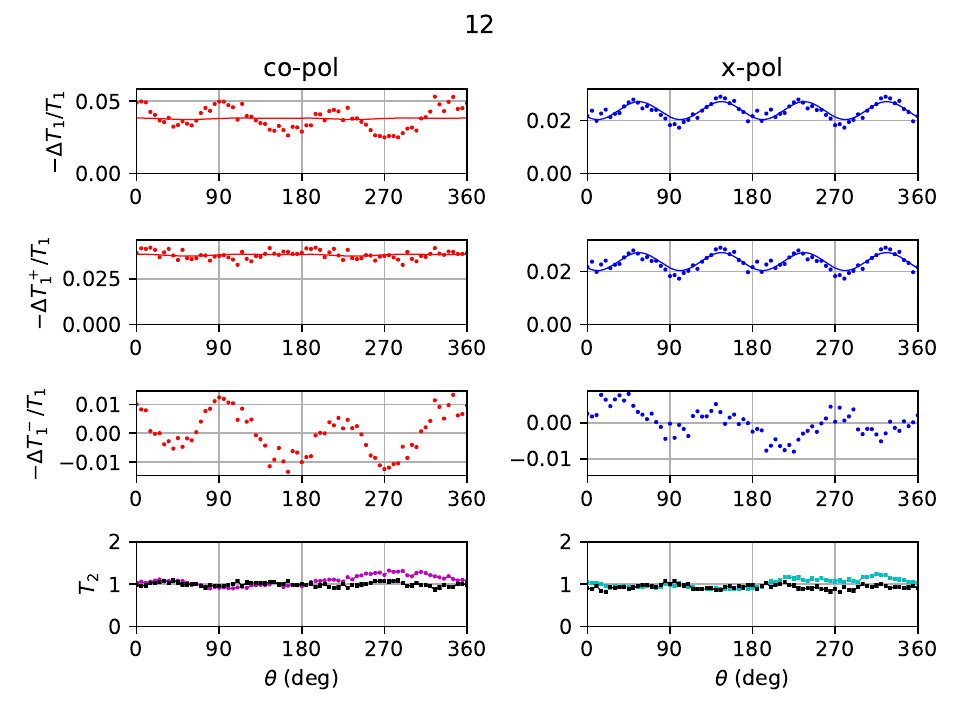}
\caption{\label{fig:rot_gaas_1600_run_12}Rotation scan data for GaAs at $\lambda = 1600$ nm for Run 12.}
\end{centering}
\end{figure}

\begin{figure}[ht]
\begin{centering}
\includegraphics[scale=0.75]{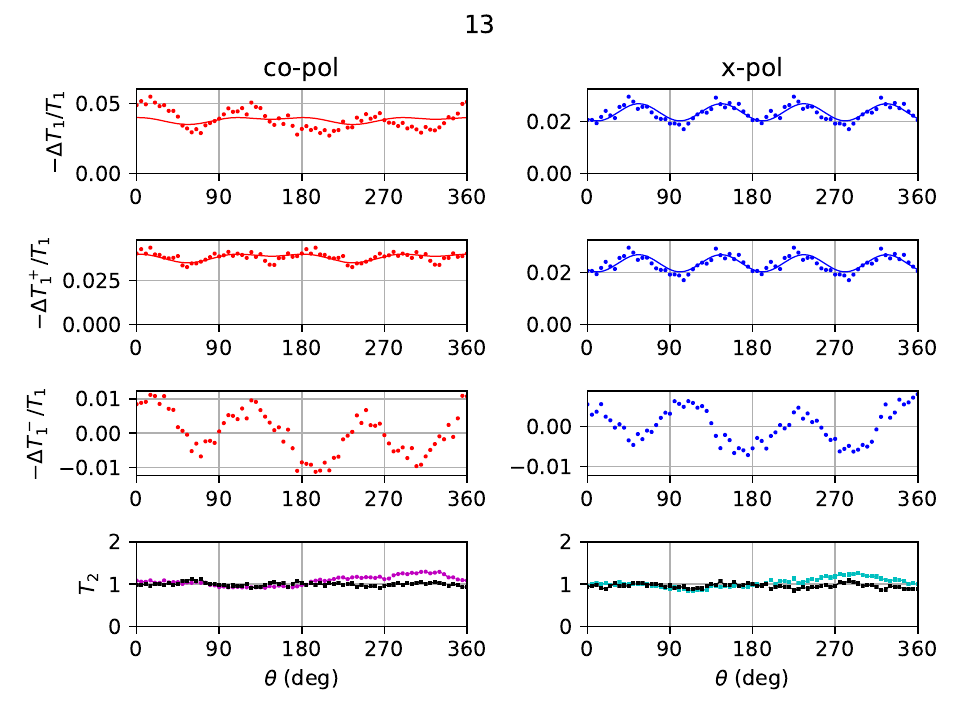}
\caption{\label{fig:rot_gaas_1600_run_13}Rotation scan data for GaAs at $\lambda = 1600$ nm for Run 13.}
\end{centering}
\end{figure}

\FloatBarrier

\subsubsection{\label{sec:rotgap}GaP}

\begin{table}[ht]
\centering
\caption{\label{tab:gap_2pa_rot}Rotation scan experimental and fit parameters for GaP.}
\begin{tabular}{c c c c c c c c c}
\hline
Run& $\lambda$& Pol.& $\theta_0$& $\delta \theta_0$& $\sigma$& $\delta \sigma$& $\eta$& $\delta \eta$\\
 &\begin{tiny}(nm)\end{tiny}& & \begin{tiny}(deg)\end{tiny}& \begin{tiny}(deg)\end{tiny}& & & & \\
\hline
1&700&p&135.2&1.3&-0.479&0.066&0.57&0.10\\
2&700&p&134.5&1.1&-0.604&0.074&0.61&0.09\\
3&700&p&130.4&1.7&-0.732&0.089&0.00&0.57\\
4&800&p&132.3&0.5&-0.935&0.104&0.45&0.07\\
5&800&p&132.2&1.0&-0.860&0.071&0.18&0.18\\
6&800&p&133.1&1.0&-0.856&0.076&0.30&0.16\\
7&900&s&47.1&1.2&-0.530&0.058&0.35&0.14\\
\hline
\end{tabular}
\end{table}

\FloatBarrier

\begin{figure}[ht]
\begin{centering}
\includegraphics[scale=0.75]{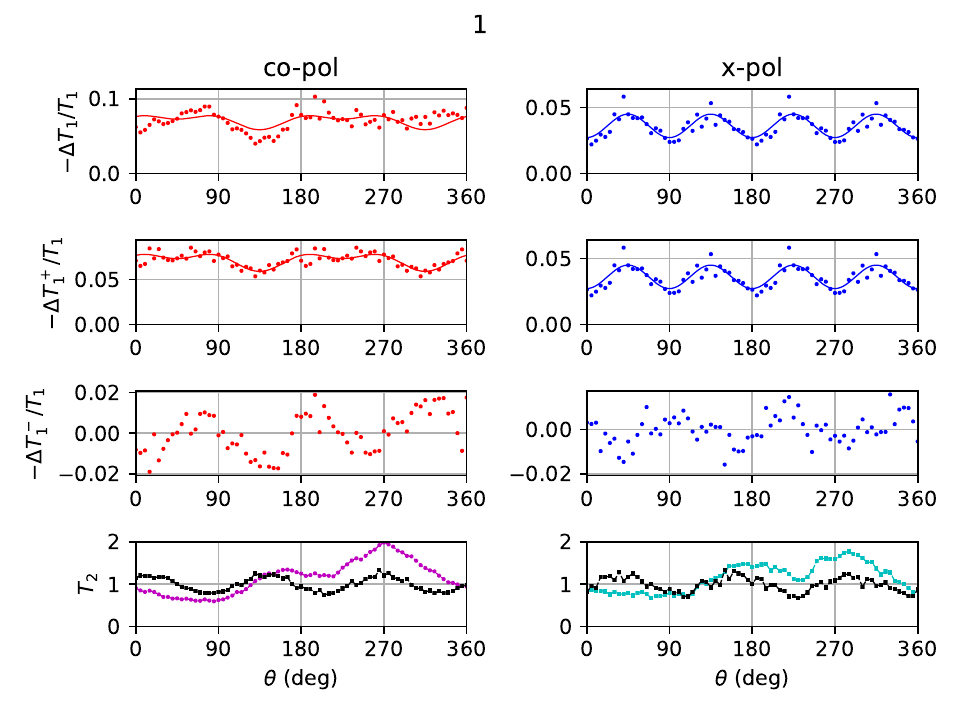}
\caption{\label{fig:rot_gap_700_run_1}Rotation scan data for GaP at $\lambda = 700$ nm for Run 1.}
\end{centering}
\end{figure}

\begin{figure}[ht]
\begin{centering}
\includegraphics[scale=0.75]{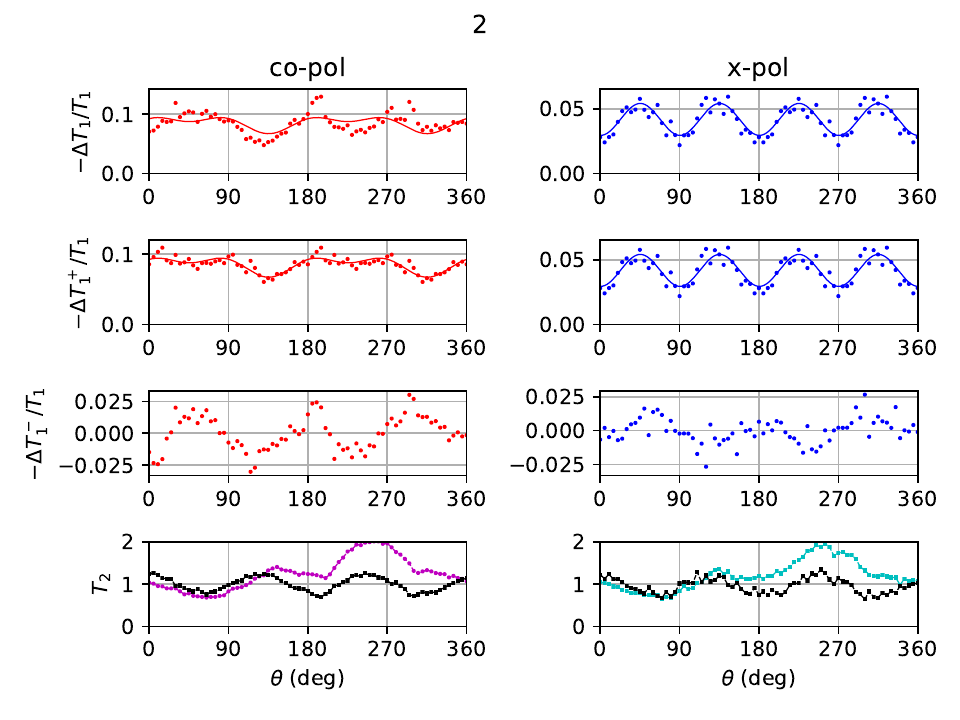}
\caption{\label{fig:rot_gap_700_run_2}Rotation scan data for GaP at $\lambda = 700$ nm for Run 2.}
\end{centering}
\end{figure}

\begin{figure}[ht]
\begin{centering}
\includegraphics[scale=0.75]{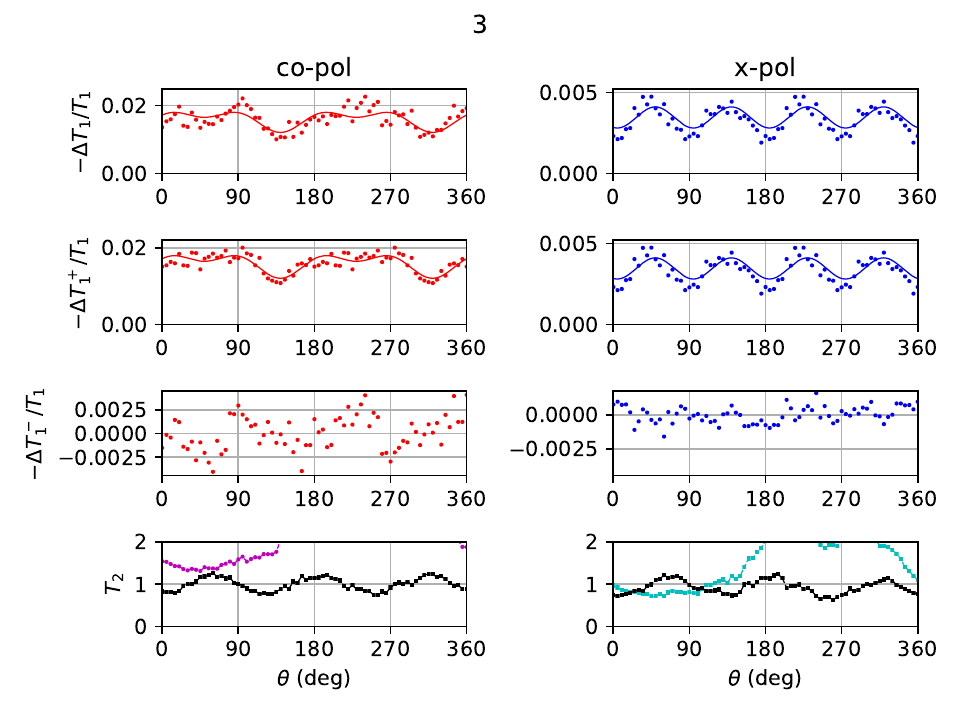}
\caption{\label{fig:rot_gap_700_run_3}Rotation scan data for GaP at $\lambda = 700$ nm for Run 3.}
\end{centering}
\end{figure}

\begin{figure}[ht]
\begin{centering}
\includegraphics[scale=0.75]{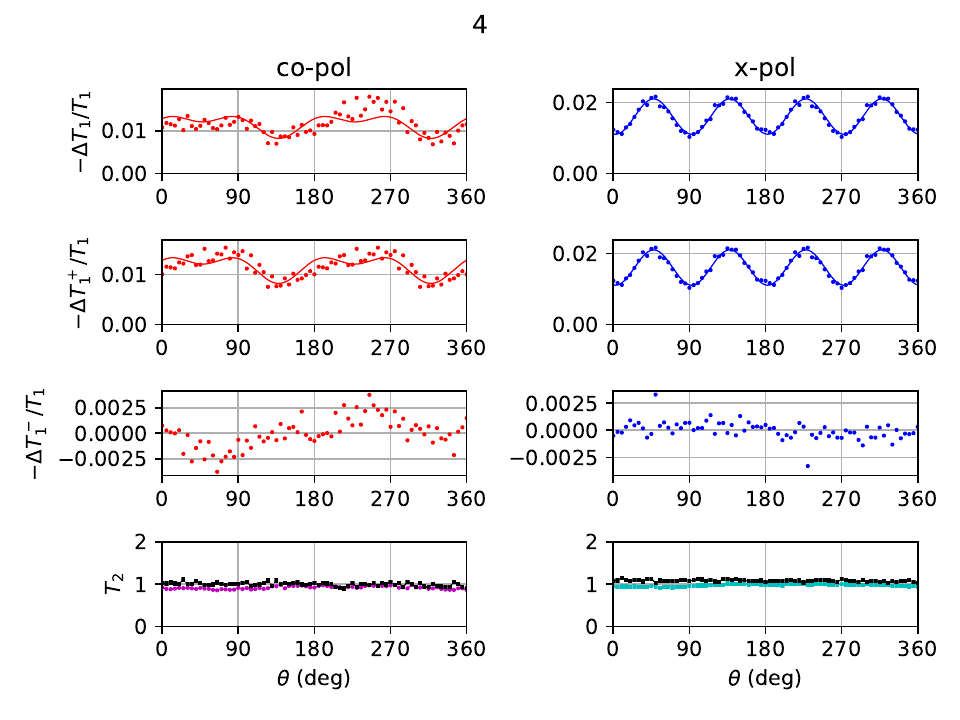}
\caption{\label{fig:rot_gap_800_run_4}Rotation scan data for GaP at $\lambda = 800$ nm for Run 4.}
\end{centering}
\end{figure}

\begin{figure}[ht]
\begin{centering}
\includegraphics[scale=0.75]{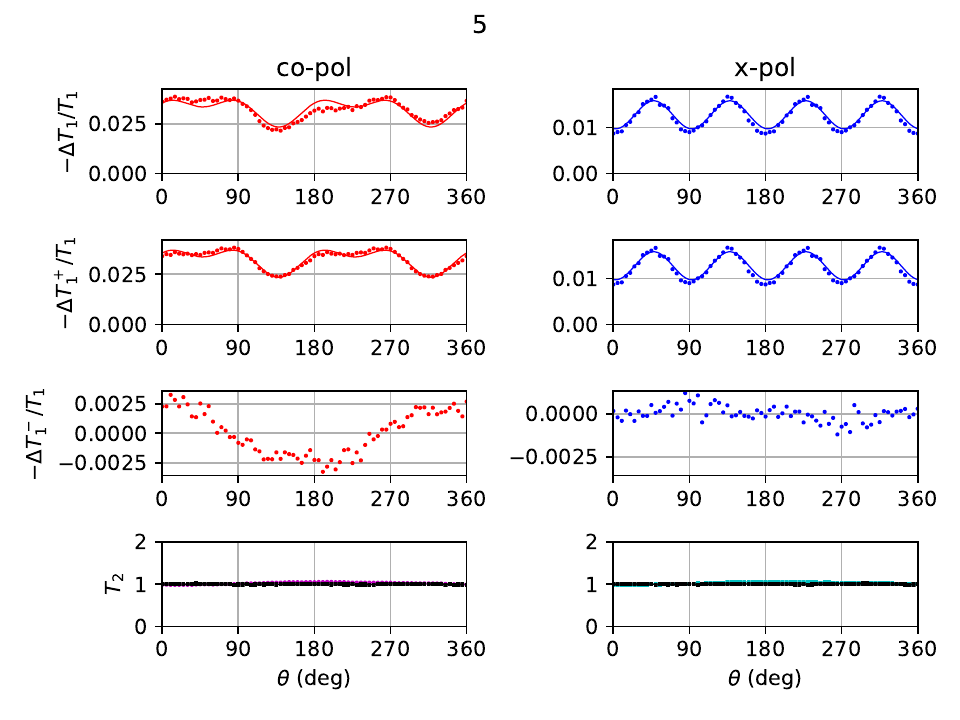}
\caption{\label{fig:rot_gap_800_run_5}Rotation scan data for GaP at $\lambda = 800$ nm for Run 5.}
\end{centering}
\end{figure}

\begin{figure}[ht]
\begin{centering}
\includegraphics[scale=0.75]{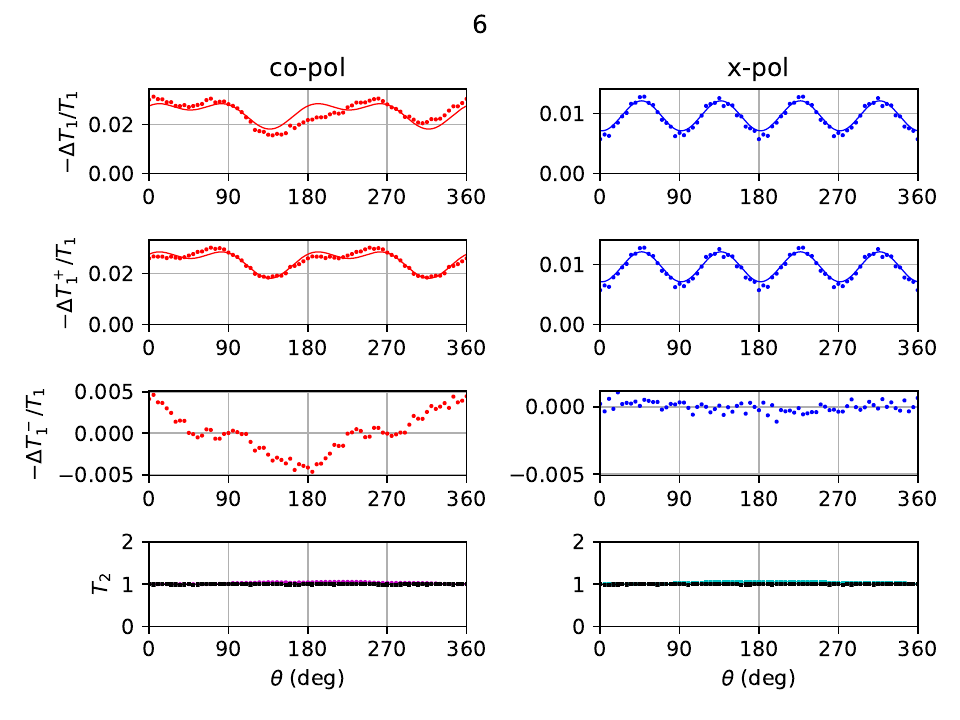}
\caption{\label{fig:rot_gap_800_run_6}Rotation scan data for GaP at $\lambda = 800$ nm for Run 6.}
\end{centering}
\end{figure}

\begin{figure}[ht]
\begin{centering}
\includegraphics[scale=0.75]{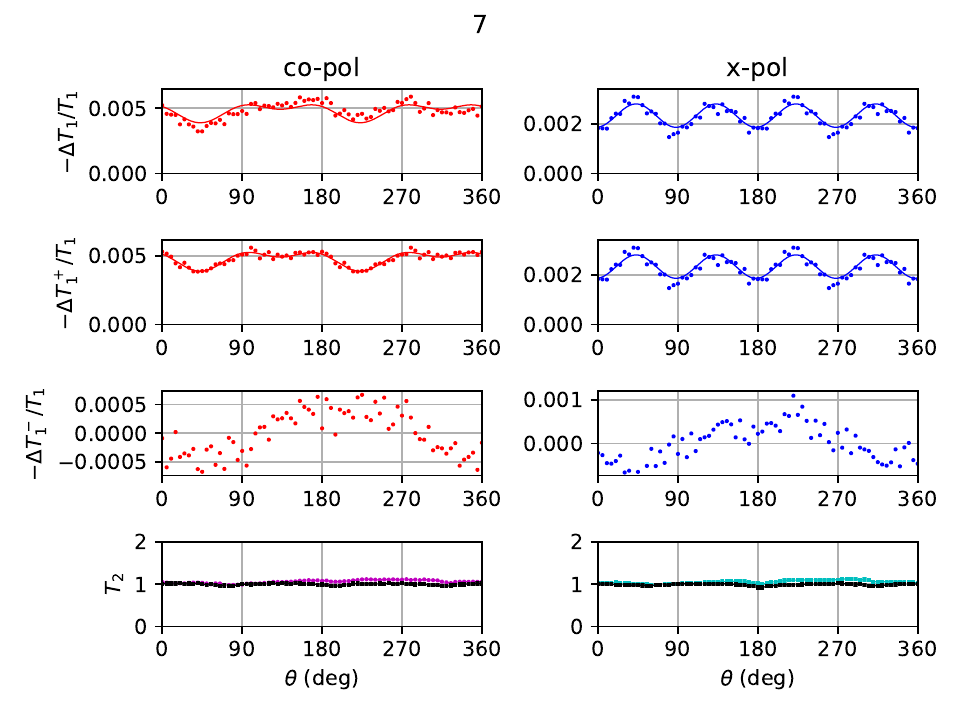}
\caption{\label{fig:rot_gap_900_run_7}Rotation scan data for GaP at $\lambda = 900$ nm for Run 7.}
\end{centering}
\end{figure}

\FloatBarrier

\subsubsection{\label{sec:rotsi}Si}

\begin{table}[ht]
\centering
\caption{\label{tab:si_2pa_rot}Rotation scan experimental and fit parameters for Si.}
\begin{tabular}{c c c c c c c c c}
\hline
Run& $\lambda$& Pol.& $\theta_0$& $\delta \theta_0$& $\sigma$& $\delta \sigma$& $\eta$& $\delta \eta$\\
 &\begin{tiny}(nm)\end{tiny}& & \begin{tiny}(deg)\end{tiny}& \begin{tiny}(deg)\end{tiny}& & & & \\
\hline
1&1200&s&76.9&3.3&-0.345&0.076&0.00&0.63\\
2&1200&s&80.5&3.8&-0.249&0.066&0.19&0.48\\
3&1200&s&81.2&1.5&-0.340&0.037&0.30&0.18\\
4&1200&s&81.1&1.9&-0.241&0.037&0.53&0.13\\
5&1300&s&0.0&2.4&-0.000&0.040&1.00&0.07\\
6&1300&s&83.1&1.3&-0.290&0.043&0.68&0.07\\
7&1400&s&0.0&1.5&-0.068&0.032&0.92&0.04\\
8&1400&s&0.0&2.4&-0.002&0.035&1.00&0.06\\
9&1600&s&91.8&2.0&-0.562&0.088&0.33&0.26\\
10&1600&s&93.3&2.0&-0.543&0.086&0.45&0.21\\
\hline
\end{tabular}
\end{table}

\FloatBarrier

\begin{figure}[ht]
\begin{centering}
\includegraphics[scale=0.75]{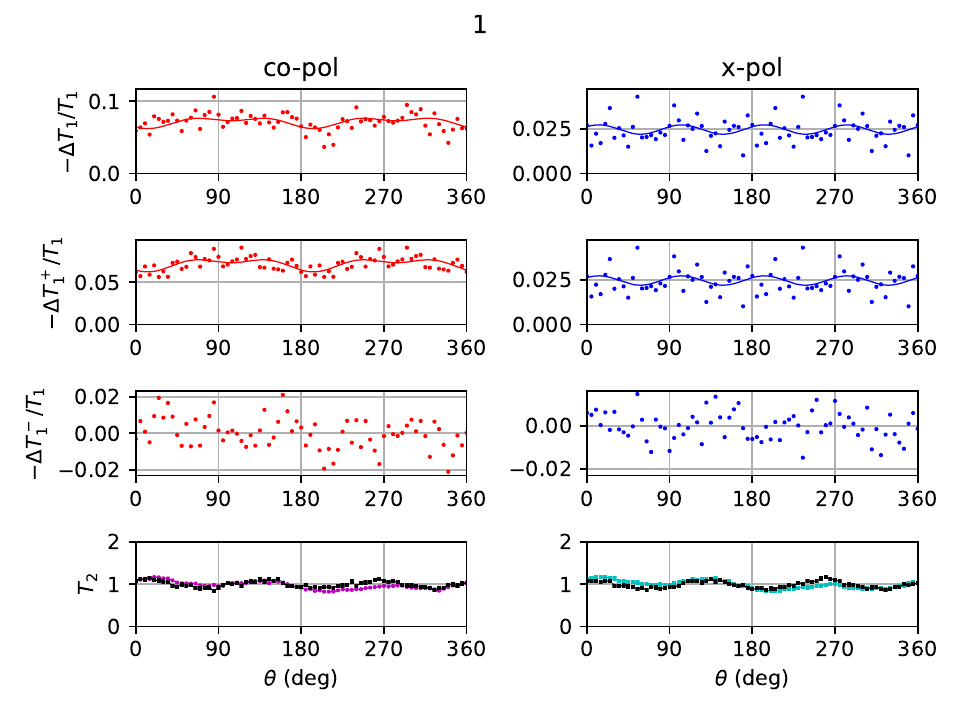}
\caption{\label{fig:rot_si_1200_run_1}Rotation scan data for Si at $\lambda = 1200$ nm for Run 1.}
\end{centering}
\end{figure}

\begin{figure}[ht]
\begin{centering}
\includegraphics[scale=0.75]{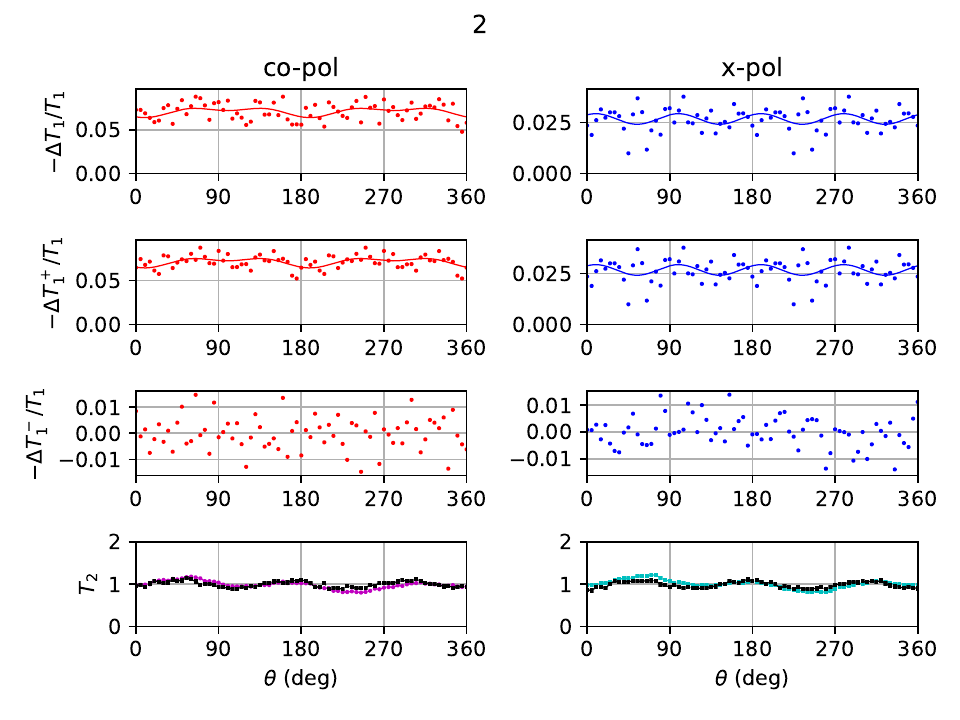}
\caption{\label{fig:rot_si_1200_run_2}Rotation scan data for Si at $\lambda = 1200$ nm for Run 2.}
\end{centering}
\end{figure}

\begin{figure}[ht]
\begin{centering}
\includegraphics[scale=0.75]{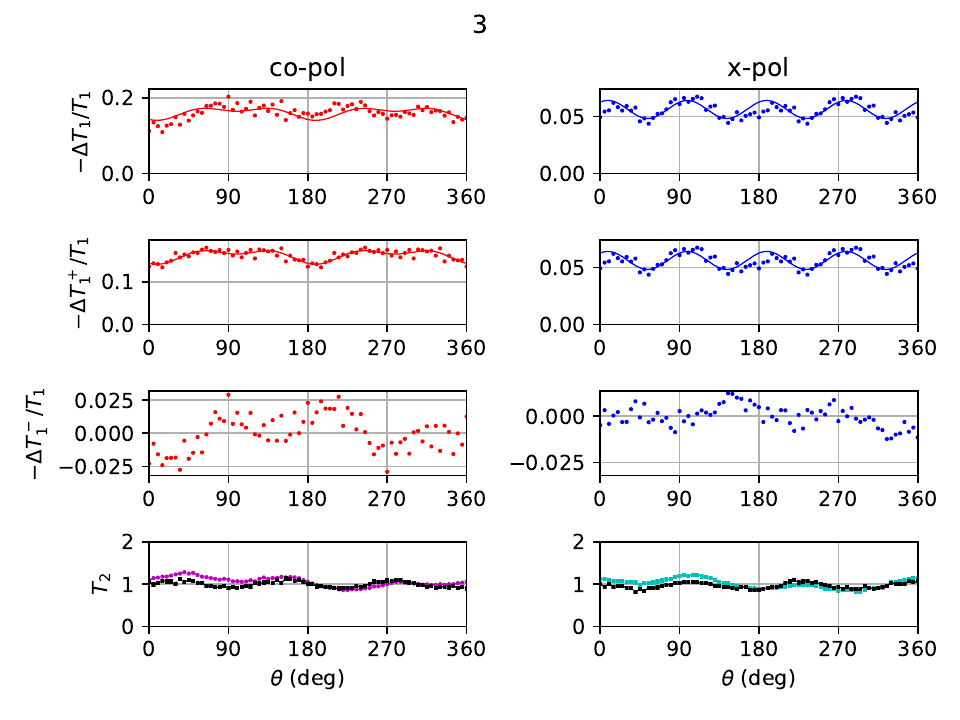}
\caption{\label{fig:rot_si_1200_run_3}Rotation scan data for Si at $\lambda = 1200$ nm for Run 3.}
\end{centering}
\end{figure}

\begin{figure}[ht]
\begin{centering}
\includegraphics[scale=0.75]{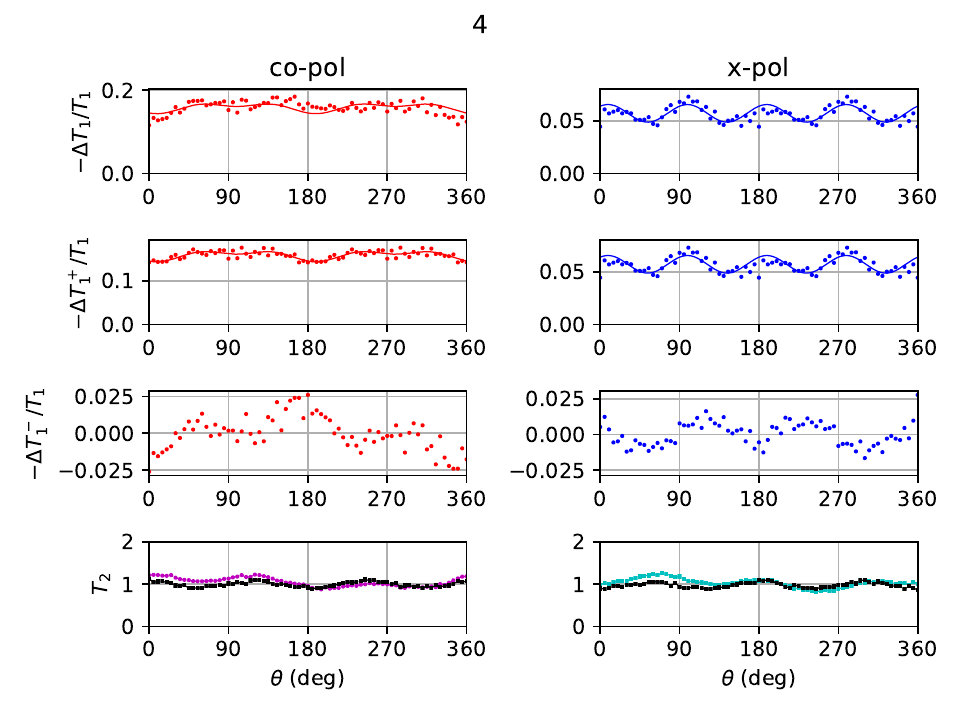}
\caption{\label{fig:rot_si_1200_run_4}Rotation scan data for Si at $\lambda = 1200$ nm for Run 4.}
\end{centering}
\end{figure}

\begin{figure}[ht]
\begin{centering}
\includegraphics[scale=0.75]{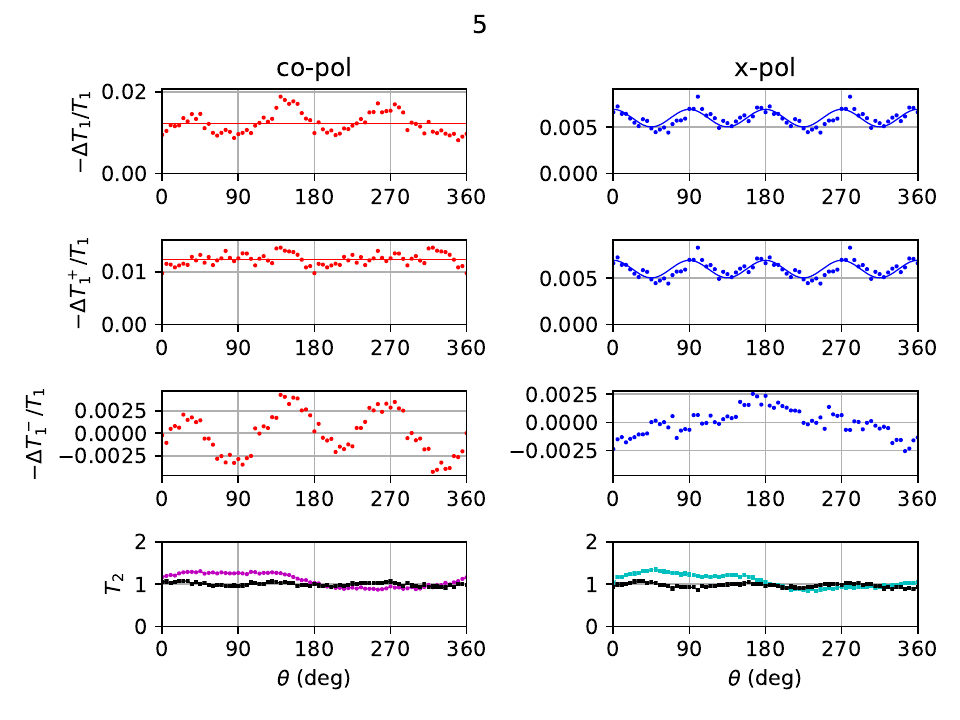}
\caption{\label{fig:rot_si_1300_run_5}Rotation scan data for Si at $\lambda = 1300$ nm for Run 5.}
\end{centering}
\end{figure}

\begin{figure}[ht]
\begin{centering}
\includegraphics[scale=0.75]{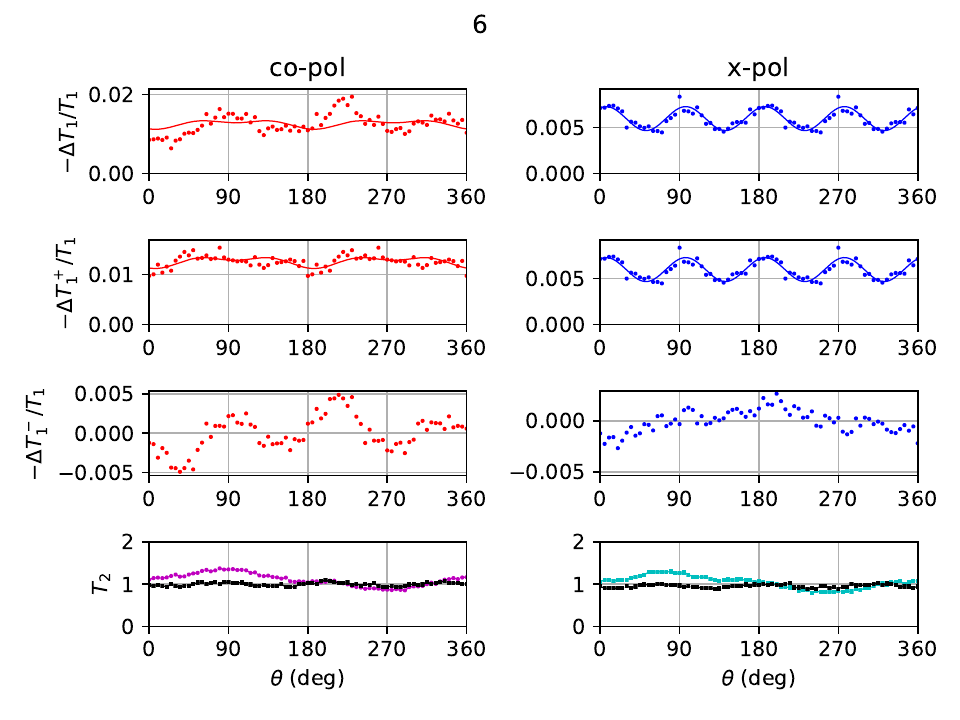}
\caption{\label{fig:rot_si_1300_run_6}Rotation scan data for Si at $\lambda = 1300$ nm for Run 6.}
\end{centering}
\end{figure}

\begin{figure}[ht]
\begin{centering}
\includegraphics[scale=0.75]{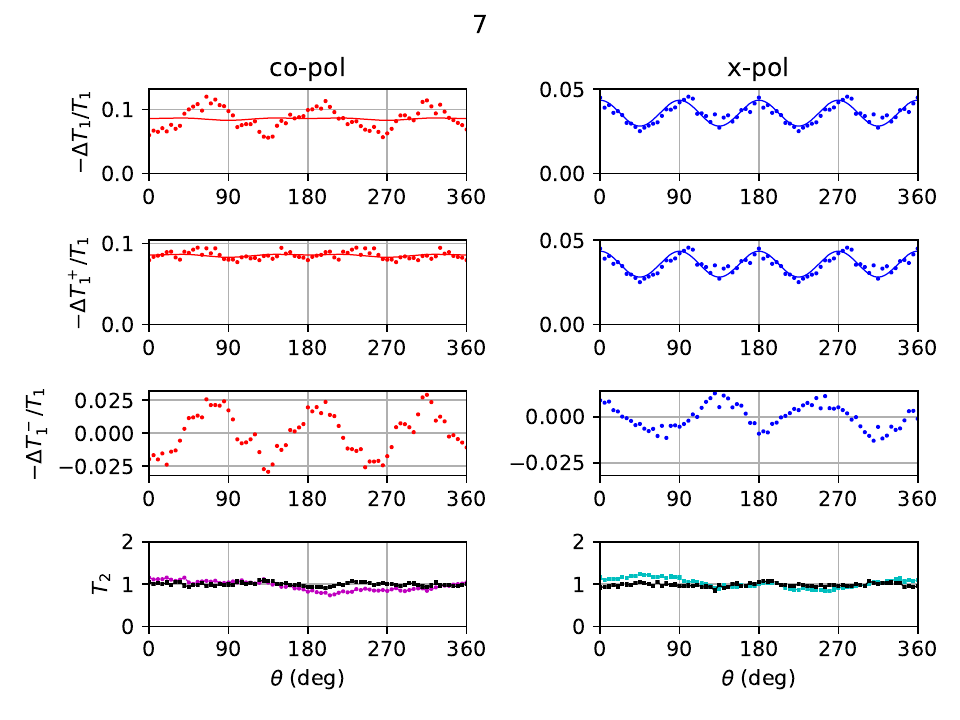}
\caption{\label{fig:rot_si_1400_run_7}Rotation scan data for Si at $\lambda = 1400$ nm for Run 7.}
\end{centering}
\end{figure}

\begin{figure}[ht]
\begin{centering}
\includegraphics[scale=0.75]{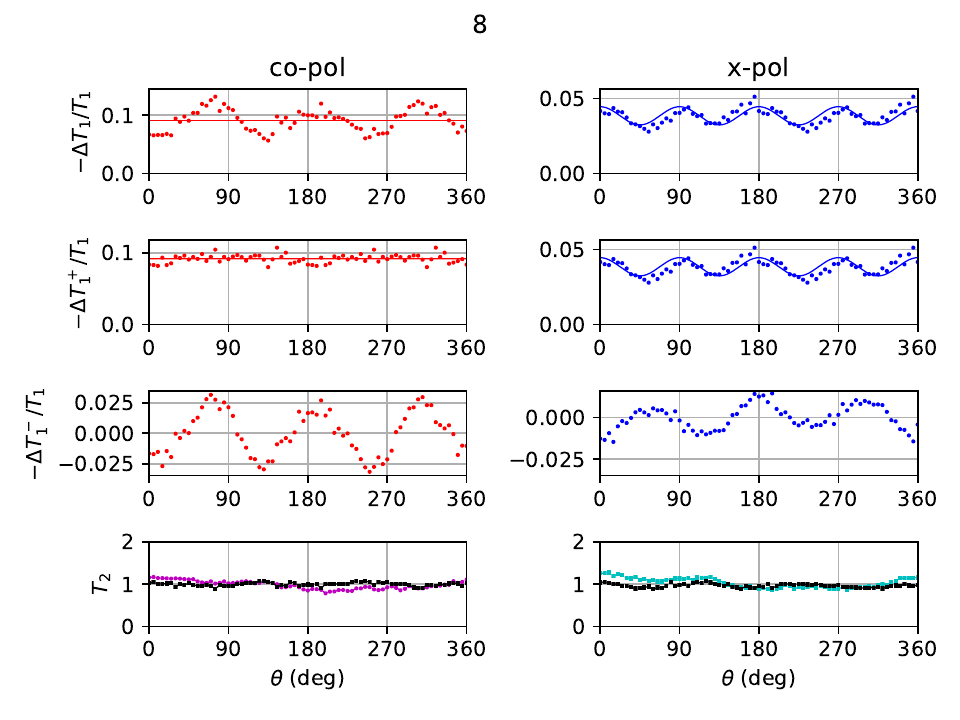}
\caption{\label{fig:rot_si_1400_run_8}Rotation scan data for Si at $\lambda = 1400$ nm for Run 8.}
\end{centering}
\end{figure}

\begin{figure}[ht]
\begin{centering}
\includegraphics[scale=0.75]{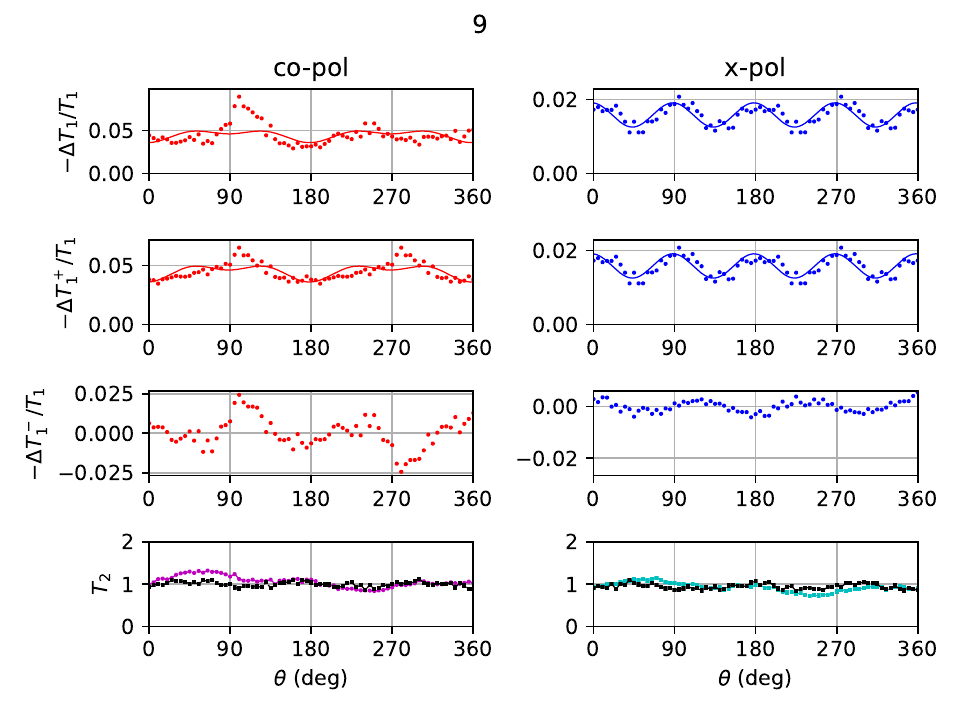}
\caption{\label{fig:rot_si_1600_run_9}Rotation scan data for Si at $\lambda = 1600$ nm for Run 9.}
\end{centering}
\end{figure}

\begin{figure}[ht]
\begin{centering}
\includegraphics[scale=0.75]{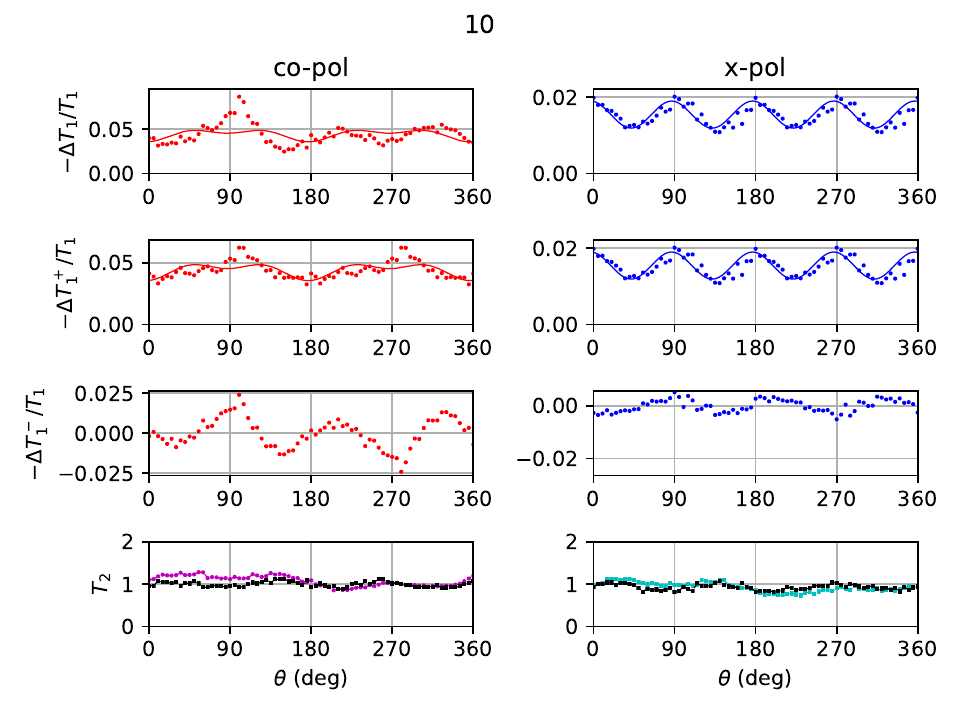}
\caption{\label{fig:rot_si_1600_run_10}Rotation scan data for Si at $\lambda = 1600$ nm for Run 10.}
\end{centering}
\end{figure}

\FloatBarrier

\section{\label{sec:anomalies}Anomalies}

\subsection{\label{sec:saturation}Possible saturation effect in time delay scans and alternate fitting procedure}

For a small number of time delay scans, we observed a 2PA response slightly wider than the SHG autocorrelation. For these cases, we considered the possiblity of saturation effect (either in sample or electronics) and this led to a widening of error bars in reported spectra. The fitting procedure was applied again only to the wings of the time delay scan data, omitting approximately the bottom half of the peaks. This fitting procedure resulted in a higher 2PA coefficient when the FWHM was wider than expected from SHG, but gave the same result for the 2PA coefficient if the FWHM was in agreement with that expected from SHG. The values of $\beta$ reported are for the fits to the complete data sets, but the reported uncertainties in the summary tables and plots in the article include the maximum range covered by uncertainties in $\beta$ for both fitting procedures.

\begin{figure}[ht]
\begin{centering}
\includegraphics[trim=0 0 0 30,clip,scale=0.65]{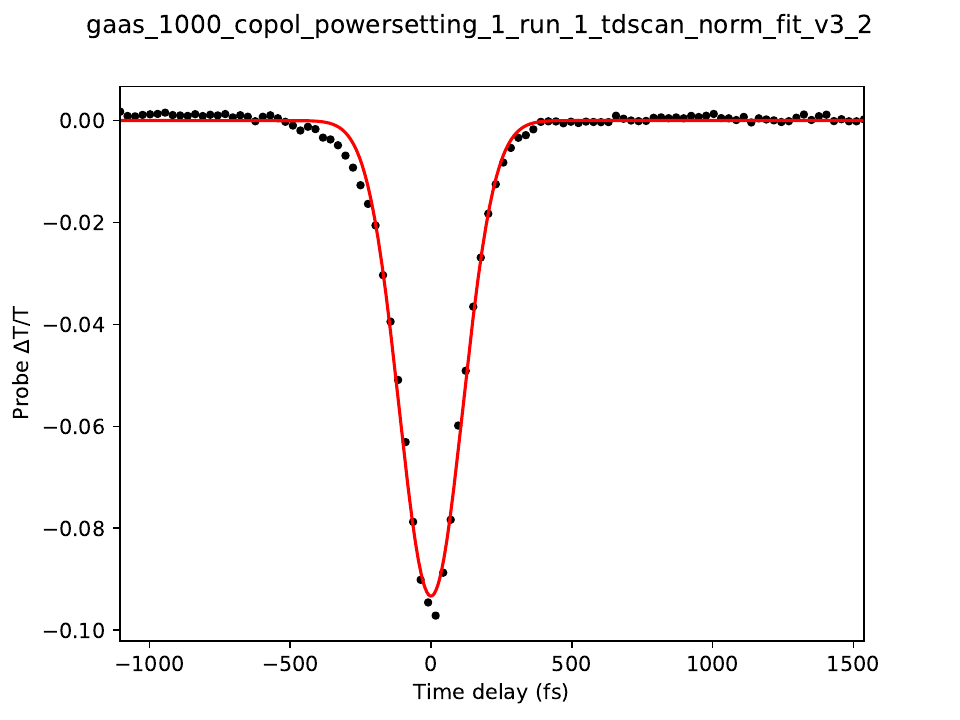}
\caption{\label{fig:td_sat_proc_gaas_1}Complete set of time delay scan data for GaAs at $\lambda = 1000$ nm for Run 1 (circles) and fit (red curve). The value of obtained from the fit was $\beta_{12}^{\parallel} = 16 \pm 3$ cm GW$^{-1}$.}
\end{centering}
\end{figure}

\begin{figure}[ht]
\begin{centering}
\includegraphics[trim=0 0 0 30,clip,scale=0.65]{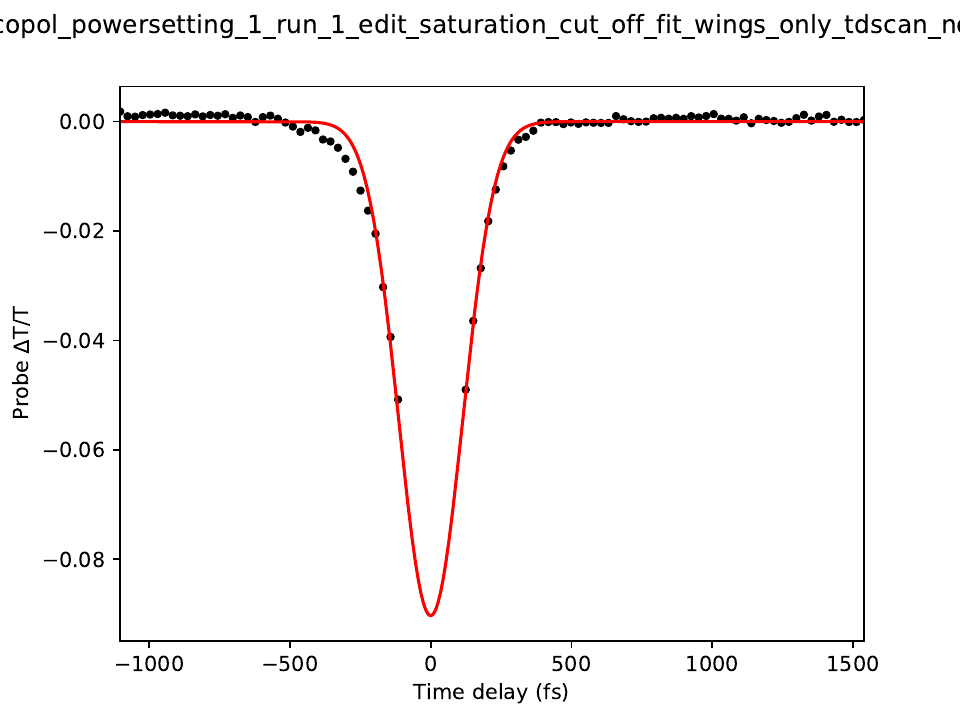}
\caption{\label{fig:td_sat_proc_gaas_2}Modified set of time delay scan data for GaAs at $\lambda = 1000$ nm for Run 1 for $|\Delta T_1/T_1| < 0.5\ \max |\Delta T_1/T_1|$ (circles) and fit (red curve). The value of obtained from the fit of this modified data was nearly the same as the complete set with $\beta_{12}^{\parallel} = 15 \pm 3$ cm GW$^{-1}$.}
\end{centering}
\end{figure}

\begin{figure}[ht]
\begin{centering}
\includegraphics[trim=0 0 0 30,clip,scale=0.65]{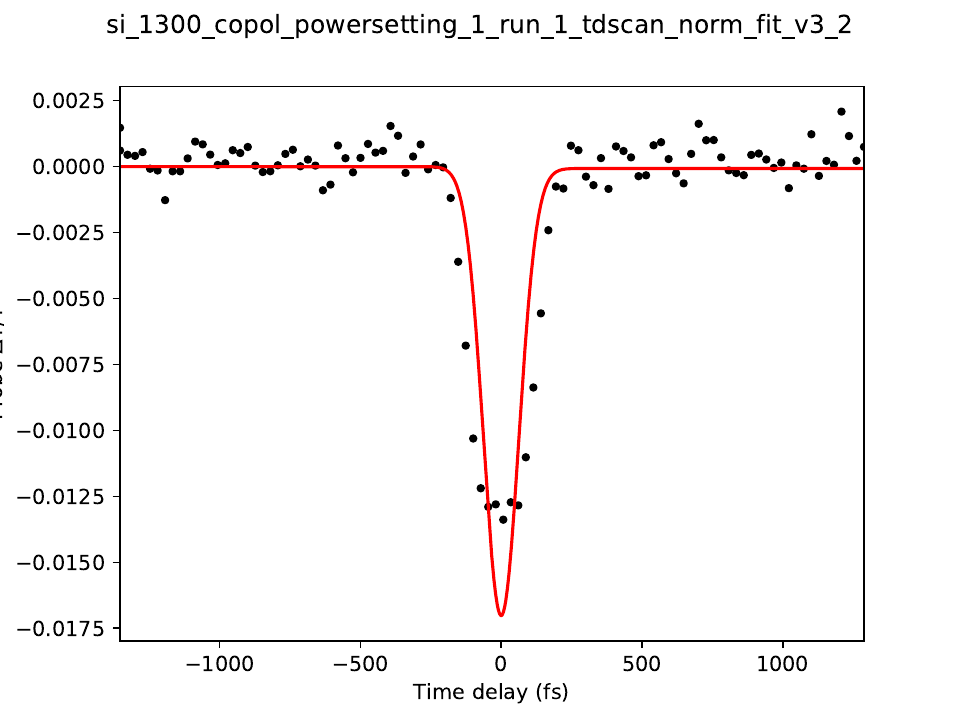}
\caption{\label{fig:td_sat_proc_si_1}Complete set of time delay scan data for Si at $\lambda = 1300$ nm for Run 5 (circles) and fit (red curve). The value of obtained from the fit was $\beta_{12}^{\parallel} = 0.39 \pm 0.08$ cm GW$^{-1}$.}
\end{centering}
\end{figure}

\begin{figure}[ht]
\begin{centering}
\includegraphics[trim=0 0 0 30,clip,scale=0.65]{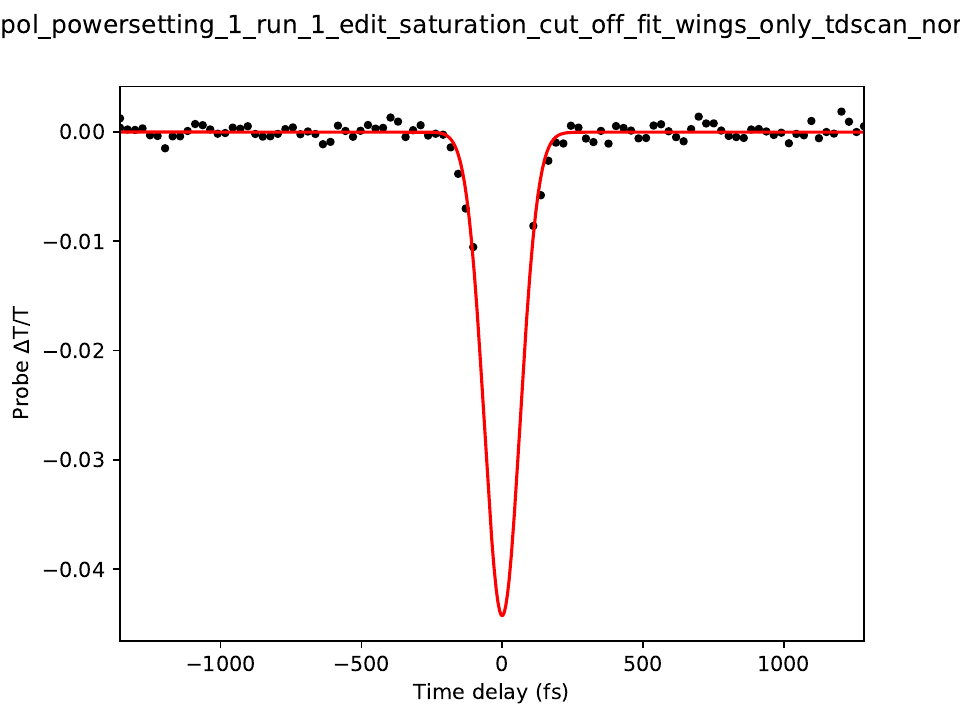}
\caption{\label{fig:td_sat_proc_si_2}Modified set of time delay scan data for Si at $\lambda = 1300$ nm for Run 5 for $|\Delta T_1/T_1| < 0.5\ \max |\Delta T_1/T_1|$ (circles) and fit (red curve). The value of obtained from the fit of this modified data was considerably larger than for the complete set with $\beta_{12}^{\parallel} = 1.0 \pm 0.2$ cm GW$^{-1}$.}
\end{centering}
\end{figure}

Figs. \ref{fig:td_sat_proc_gaas_1} -- \ref{fig:td_sat_proc_gaas_2} show an application of this procedure to a case where the possible saturation effect was small and the change in the value of $\beta$ obtained from fitting the complete vs. the modified set of data was small. On the other hand, Figs. \ref{fig:td_sat_proc_si_1} -- \ref{fig:td_sat_proc_si_2} show an application of this procedure to a case where the possible saturation effect was large and the change in the value of $\beta$ obtained from fitting the complete vs. the modified set of data was also large. The runs for which this procedure was applied are listed in Table \ref{tab:sat_alt_fit_proc}.

\begin{table}[ht]
\centering
 \caption{\label{tab:sat_alt_fit_proc}Run numbers of time delay scans for which the alternative fitting procedure described in this section was applied to account for a possible saturation effect for estimating a possible systematic uncertainty in $\beta$.}
  \begin{tabular}{l l}
  \hline
  Sample& Run numbers\\
  \hline
   GaAs& 1 2 3 4 5 7 9 10 11 12 13 14 15\\
   GaP& 1 2 3 4 5 6 7 8 9 10 11 12 13 14 15\\
   Si& 5 7 8 9 10 11\\
   \hline
  \end{tabular}
\end{table}

\FloatBarrier

\subsection{\label{sec:hwhianomaly}Temporally-broadened asymmetric time delay scan anomaly near $E_g$}

\begin{figure}[ht]
\begin{centering}
\includegraphics[scale=1.0]{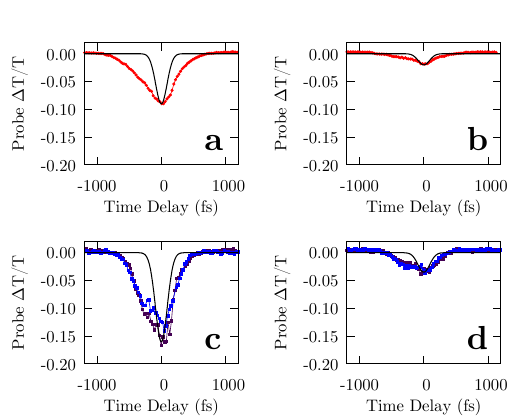}
\caption{\label{fig:anomaly_eg}Time delay scan anomalies near the band gap of each material with time delay scan data (colored points and lines) and normalized 2PA autocorrelation curves expected from SHG autocorrelation (black curves). a) GaAs at $\hbar \omega = 1.38$ eV in co-polarized geometry. b) GaAs at $\hbar \omega = 1.38$ eV in cross-polarized geometry. c) GaP at $\hbar \omega = 1.91$ eV in co-polarized geometry. d) GaP at $\hbar \omega = 1.91$ eV in cross-polarized geometry.}
\end{centering}
\end{figure}

The temporal width of the nonlinear loss was anomalously broadened in GaAs and GaP as the excitation neared their respective band gaps in both polarization geometries. Fig. \ref{fig:anomaly_eg} shows the anomaly observed in GaAs and GaP in both polarization geometries. The observed temporal FWHM in the nonlinear loss was a factor of 2.3 -- 3 greater than expected for 2PA alone. Asymmetry in the time delay scans also support the presence of other effects not accounted for in our model. Both of these features are obvious against the second-order autocorrelation signal expected for 2PA (as measured by SHG) which is superimposed on the data. The possible origins of these anomalies are discussed in the article.

\FloatBarrier

\subsection{\label{sec:hwloanomaly}Antisymmetric time delay scan anomaly near $E_g/2$}

\begin{figure}[ht]
\begin{centering}
\includegraphics[scale=1.0]{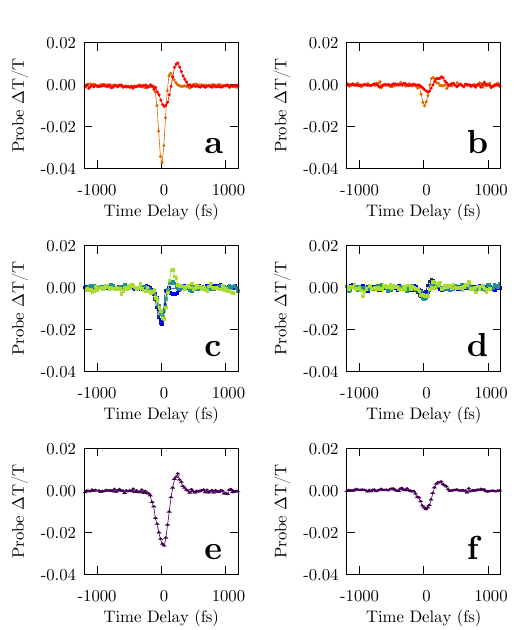}
\caption{\label{fig:anomaly_eg2}Time delay scan anomalies near the two-photon band edge of each material. a) GaAs at $\hbar \omega = 0.62$ eV (red) and $\hbar \omega = 0.69$ eV (orange) in co-polarized geometry. b) GaAs in cross-polarized geometry. c) GaP at $\hbar \omega = 0.77$ eV (lime green), $\hbar \omega = 0.89$ eV (teal), $\hbar \omega = 1.03$ eV (blue), and $\hbar \omega = 1.24$ eV (dark blue) in co-polarized geometry. d) GaP in cross-polarized geometry. e) Si at $\hbar \omega = 0.62$ eV (violet) in co-polarized geometry. f) Si in cross-polarized geometry.}
\end{centering}
\end{figure}

Below the two-photon absorption band edge of GaAs and GaP, nonlinear loss was very small except at high intensities (a factor of 1 -- 20 greater than intensities at excitation energies $\hbar \omega > E_g/2$). This is consistent with 2PA being energetically disallowed but a time delay scan anomaly was still observed near zero time delay in both polarization geometries. A time delay scan anomaly was also observed in Si near to but slightly above $E_g/2$ at $\hbar \omega = 0.62$ eV, where the magnitude of 2PA was also quite low. Fig. \ref{fig:anomaly_eg2} shows the anomaly observed in all three samples in both polarization geometries. This anomaly was asymmetric and notably included an antisymmetric component (about $\Delta t =0$) where $\Delta T_1/T_1 > 0$ briefly for positive time delays. The possible origins of these anomalies are discussed in the article.

\FloatBarrier

\section{\label{sec:additionalsi}Additional supporting information}

Additional information that support the findings of this study are available from the corresponding author upon reasonable request.

\bibliography{supplement}